\documentclass[11pt]{article}
\usepackage{graphicx,epsfig}
\usepackage{graphics,amssymb,float}
\usepackage{cite,float}
\usepackage[usenames,dvips]{color}
\usepackage{amsmath, mathbbol}

\usepackage{multirow}
\makeatletter

\@addtoreset{equation}{section}
\makeatother

\def\lsim{\;\raise0.3ex\hbox{$<$\kern-0.75em\raise-1.1ex\hbox{$\sim$}}\;}
\def\gsim{\;\raise0.3ex\hbox{$>$\kern-0.75em\raise-1.1ex\hbox{$\sim$}}\;}

\def\ben{\begin{enumerate}}  \def\een{\end{enumerate}}
\def\bit{\begin{itemize}}    \def\eit{\end{itemize}}
\def\beq{\begin{equation}}   \def\eeq{\end{equation}}
\def\ba{\begin{array}}       \def\ea{\end{array}}
\def\bea{\begin{eqnarray}}   \def\eea{\end{eqnarray}}

\begin{document}

\setcounter{footnote}{0}

\vspace*{-2cm}
\begin{flushright}
LPT Orsay 12-17 \\
CFTP 12-006 \\
PCCF RI 12-04\\

\vspace*{2mm}
\today
\end{flushright}
\begin{center}
\vspace*{15mm}
{\Large\bf Lepton flavour violation: physics potential of a Linear Collider} \\
\vspace{1cm}
{\bf A. Abada$^{a}$, A. J. R. Figueiredo$^{b,c}$, J. C. Rom\~ao$^{b}$ and 
A. M. Teixeira$^{c}$
}

 \vspace*{.5cm} 
$^{a}$ Laboratoire de Physique Th\'eorique, CNRS -- UMR 8627, \\
Universit\'e de Paris-Sud 11, F-91405 Orsay Cedex, France

\vspace*{.2cm} 
$^{b}$ Centro de F\'{\i}sica Te\'orica de Part\'{\i}culas, 
Instituto Superior T\'ecnico, \\ Av. Rovisco Pais 1, 
1049-001 Lisboa, Portugal

\vspace*{.2cm} 
$^{c}$ Laboratoire de Physique Corpusculaire, CNRS/IN2P3 -- UMR 6533,\\ 
Campus des C\'ezeaux, 24 Av. des Landais, F-63177 Aubi\`ere Cedex, France

\end{center}

\vspace*{10mm}
\begin{abstract}
We revisit the potential of a Linear Collider concerning the study
of lepton flavour violation, in view of new LHC bounds and of 
the (very) recent developments in lepton
physics. Working in the framework of a type I 
supersymmetric seesaw, we evaluate the prospects of observing 
seesaw-induced lepton flavour violating final states of the type 
$e\,\mu$ plus missing energy, arising from $e^+ e^-$ and $e^- e^-$ 
collisions. In both cases we address the potential 
background from standard model and supersymmetric charged
currents. We also explore  the possibility of electron and positron beam
polarisation. 
The statistical significance of the signal,
  even in the absence of kinematical and/or detector cuts, renders the
observation of such flavour violating events feasible over large
regions of the parameter space.
We further consider the $\mu^-\mu^- + E^T_\text{miss}$ final
state in the $e^-e^-$ beam option finding that, due to a very
suppressed background, this process turns out to be a truly clear probe of a supersymmetric
seesaw, assuming the latter to be  the unique source of lepton flavour violation. 
\end{abstract}

\section{Introduction}
Although the Large Hadron Collider (LHC) is yet to discover
supersymmetry (SUSY), this class of models remains one of the most
appealing extensions of the Standard Model (SM).  
A very
attractive explanation for the smallness of neutrino masses, $m_\nu$, and their
mixings~\cite{neutrino:data} is then to embed a seesaw mechanism 
in the framework of SUSY models.
A high scale (close to the grand unification - GUT - scale) type I seesaw~\cite{seesaw:I}
can account for $m_\nu$ and the observed pattern of the mixing angles,
for naturally large values of the Yukawa  
couplings, $Y^\nu$. 
Such scenarios are also very appealing as they can  
explain the observed baryon asymmetry of the Universe (BAU) via 
leptogenesis~\cite{Davidson:2008bu} (through the CP and lepton number
violating decays of the heavy mediators 
responsible for the smallness of neutrino masses, i.e., the right-handed neutrinos). 
Furthermore, when realised at such high scales, the new dynamics 
has only a marginal effect on electroweak (EW)
precision measurements and observables~\cite{Abada:2007ux}, contrarily to other seesaw
realisations as, for instance, in the case of a type II seesaw~\cite{seesaw:II}.
By itself, and independently of the scale at which it is realised, 
the type I seesaw is very hard to probe - 
directly or indirectly -  as the mediators are very difficult to  
produce at colliders (in particular when heavier than the TeV scale) 
and/or their effects at low-energies are extremely
small (typically in association with small couplings). 
However, when embedded in the framework of SUSY models - 
the so-called SUSY seesaw - the type I seesaw offers 
the interesting feature 
that flavour violation in the slepton sector is radiatively induced~\cite{Borzumati:1986qx}
and, at low-energies, the 
new contributions to a large array of charged 
lepton flavour violation (cLFV) observables 
might provide a unique probe of this mass generation mechanism. 

In the framework of a type I SUSY seesaw, 
low-energy cLFV observables such as $\ell_i\to\ell_j\gamma$,
$\ell_i\to 3 \ell_j$, $\mu - e$ conversion in nuclei, etc.,  
have been extensively 
addressed~\cite{Hisano:1995cp, Hisano:1995nq, Hisano:1998fj,
  Buchmuller:1999gd, Kuno:1999jp, Ellis:1999uq, Casas:2001sr, Lavignac:2001vp,
  Bi:2001tb, Ellis:2002fe,Deppisch:2002vz, Fukuyama:2003hn,
  Brignole:2004ah,  Masiero:2004js, Fukuyama:2005bh,
  Petcov:2005jh,Arganda:2005ji, Deppisch:2005rv, Yaguna:2005qn,
  Calibbi:2006nq,Antusch:2006vw,Arganda:2007jw, Arganda:2008jj}. 
Moreover, cLFV has also been studied in high-energy observables~\cite{Arkanihamed:1996au,Hinchliffe:2000np,Carvalho:2002jg,Buckley:2006nv,Hirsch:2008dy,Carquin:2008gv,Buras:2009sg,Abada:2010kj,Abada:2011mg,Calibbi:2011dn,Calibbi:2011vi,Galon:2011wh,Arbelaez:2011bb}, as those that can be probed
at the LHC as, for instance, lepton flavour violating neutralino
decays and slepton mass 
splittings. 
The impact of a type I seesaw on the SUSY Higgs spectrum has been recently discussed
in~\cite{Herrero:2009tm}.  
Furthermore, under the assumption of a unique source of 
Lepton Flavour Violation (LFV), one can rely on the
synergy of high- and low-energy observables and devise powerful probes
(using data from neutrino experiments, rare lepton decays, and collider
searches), which allow to shed light on the fundamental nature of LFV.

At hadron colliders (such as the Tevatron and the LHC) sleptons are in general only
produced in the decay chains of a strongly interacting 
sparticle, i.e. squarks or gluinos; if these states are heavy (as
suggested by the negative searches at the LHC), then slepton
production, even if the latter are comparatively light, 
is extremely difficult (occurring only via EW gaugino production, 
which is typically subdominant). Nevertheless, numerous
slepton-dedicated studies have been conducted  
for the LHC, in particular focusing on the case in which sleptons are 
produced from gaugino decays. Hence,  
the studies of cLFV observables are necessarily associated with specific decay chains that only
occur in (reduced) regions of the SUSY parameter space: in general, 
cLFV is studied through dilepton mass distributions ($m_{\ell \ell}$)
of $\chi_2^0 \to \ell_i \ell_j \chi_1^0$, via observables such as
additional edges in $m_{\ell \ell}$ and slepton mass splittings (see, for 
example~\cite{Carvalho:2002jg,Hirsch:2008dy,Carquin:2008gv,Buras:2009sg,Abada:2010kj,Abada:2011mg,Calibbi:2011dn,Calibbi:2011vi}).

A high-energy lepton collider, such as ILC~\cite{:2007sg,:2010zzd},
CLIC~\cite{CLIC:CDR} or a muon collider~\cite{Palmer:2011zz}, 
will allow to extend and to complement LHC searches. Following a
potential discovery of new physics at the LHC,  
linear colliders offer numerous advantages for precision studies, in particular 
the possibility of  probing the properties of the new states.  
Linear lepton Colliders (LC) constitute an excellent laboratory to study
the slepton sector (see, e.g.~\cite{Blaising:2012vd,slepton:lepton:colliders}). 
In contrast to hadron colliders, 
the exact nature of the colliding particles is
known in a Linear Collider - valid for electron and muon colliders - and the polarisation
of the beams can be known with high precision. 
In fact, the possibility of
beam polarisation is instrumental to either suppress or enhance SM
contributions, so that the latter can be differentiated 
from new physics processes. 
It has been extensively pointed out that the ability of adjusting
the polarisation of each beam independently (and simultaneously)
opens unique possibilities to test the properties of the produced
particles: their quantum numbers and ``chiral'' couplings
can thus be directly probed under a minimal set of 
assumptions~\cite{MoortgatPick:2010zza}. 

In the case of supersymmetric models, a LC will complete the LHC exploration of 
SUSY~\cite{Weiglein:2004hn}, possibly resolving
particularly difficult cases where LHC measurements fail to clearly
identify SUSY particles, and determining the properties of the new particles 
(for instance, CP properties~\cite{ee:CP}, 
the Majorana nature of neutralinos~\cite{ee:majorana} and of other 
new states~\cite{Cannoni:2002ny}, etc.).
Concerning the slepton sector, and despite a possibly heavy
squark and gluino sectors, lighter sleptons can be directly produced
at a Linear Collider, operating at a comparatively smaller centre of
mass (c.o.m.) energy, $\sqrt s$. 
In addition to very precise measurements of slepton parameters, 
such a clean environment (with a low hadronic background when compared
to the LHC) further allows dedicated  
studies of rare processes, as is the case of flavour violation. 

 SUSY cLFV  processes at a Linear Collider, 
such as  for example $e^{\pm} e^- \to \ell_i^\pm \ell_j^- + E^T_\text{miss}
(+\text{jets})$ and 
$e^{\pm} e^- \to \ell_i^\pm \ell_j^-$,
have been previously addressed in the literature. 
Early analyses have treated flavour violation in the slepton sector using a
sneutrino mixing (oscillation) based
approach~\cite{ArkaniHamed:1997km,Hisano:1998wn,Guchait:2001us,Cannoni:2003my},
while recent studies considered an effective parametrization of the
amount of LFV~\cite{Carquin:2011rg}, in both cases without focusing on a specific
mechanism of flavour violation in the lepton sector. Other studies
have specifically considered type I seesaw-inspired
models~\cite{Deppisch:2003wt,Deppisch:2004pc}. 
A Linear Collider could also offer the possibility of working in the
so called $e \gamma$ or $\gamma \gamma$ modes - thus becoming a  
very high energy photon collider, with the option of polarised photon
beams~\cite{PC}. cLFV has also been considered in $\gamma \gamma \to
\ell_i \ell_j$ collisions using a phenomenological
effective approach~\cite{Cannoni:2005gy}.

In recent years, not only neutrino parameters have been better
constrained~\cite{Schwetz:2011zk}, but crucial data has just been
obtained by several collaborations, with very precise measurements of  
the Chooz angle~\cite{Abe:2011fz,Hartz:2012pr,Adamson:2012rm,An:2012eh,Ahn:2012nd}.
Furthermore, negative LHC SUSY searches
have also suggested that the SUSY spectrum might be far heavier than what
had previously been considered~\cite{LHC:2011}. 
Moreover, the sensitivity of the
experiments dedicated to the observation of low-energy cLFV
transitions has also significantly improved~\cite{PDG}, especially
in what concerns $\mu \to e \gamma$ decays~\cite{arXiv:1107.5547} and  
$\mu-e$ conversion in nuclei~\cite{Glenzinski:2010zz,Cui:2009zz}. 

Parallel to these achievements, recent  technological developments
potentially suggest that the future Linear Collider will be an
ambitious project concerning the expected detector efficiency,
as well as electron and positron beam polarisations. It is also expected 
that such a collider will run at (very) large 
centre of mass energies, leading to a very high luminosity - see
e.g.~\cite{:2010zzd,CLIC:CDR}. In view of all the above, in this study 
we revisit the potential manifestations of cLFV in a future LC assuming a
type I SUSY seesaw as a unique source of LFV, discussing how the possible
direct signals and their synergy with other strongly correlated cLFV
observables can contribute to probe the
mechanism of neutrino mass generation.  

In this work we study cLFV in association to slepton and gaugino pair
production in both $e^+ e^-$ and $e^- e^-$ collisions.  
In view of the (recent) stringent experimental bounds on flavour violation in the
$\mu-e$ sector, we only consider final states involving muons and/or
electrons, further allowing to maximally benefit from the
efficiency of the muon detectors.
Let us recall that the reconstruction of high-energy
electrons and muons is a key feature of a LC - for instance in the
case of CLIC~\cite{CLIC:CDR,Blaising:2012vd},
the muon and electron average identification efficiencies are 
expected to be as high as 99\% and 96\%, respectively. 
However, we do allow for
(intermediate) tau production, focusing exclusively on its purely 
leptonic decays, $\tau \to \nu_\tau \,\ell \,\bar \nu_\ell$.
We thus consider all processes leading to final states involving two
charged leptons ($\mu, e$) and missing energy. In the presence of
flavour violation in the neutrino sector, as confirmed by neutrino
data, potential LFV backgrounds arise from charged lepton 
currents: even in the absence of a seesaw mechanism, these backgrounds
are present both in the SM as well as in its Minimal
Supersymmetric extension (MSSM), once neutrino 
oscillations are accommodated via 
the leptonic mixing matrix, $U_\text{MNS}$~\cite{PMNS}. 
Thus, in addition to the pure SUSY seesaw signal, both the SM and
SUSY charged current interactions provide important contributions to
$e^{\pm} e^- \to \ell_i^\pm \ell_j^- + E^T_\text{miss}$ ($i \neq j$). 
While in the "signal" $E^T_\text{miss}$ is strictly composed of SUSY
neutral particles escaping the detector (the lightest SUSY 
particle (LSP), which is the lightest neutralino, $\chi_1^0$), 
in the SM (MSSM) $E^T_\text{miss}$ is in the form of neutrinos
(neutrinos + neutralinos). 

In addition to different flavour final states,
we also discuss the unique potential of the 
$e^- e^- \to \mu^-\mu^- + E^T_\text{miss}$ channel, since in this case
the background is considerably suppressed - in this case 
the SM contributions to the background are essentially
non-existent. In view of such a clean signal,  
this process deserves a careful analysis since, in addition to
directly probing the Majorana nature of the fermionic neutral
superparticles, it also allows to relate  the neutrino mass generation
mechanism to a cLFV high-energy observable. 

The purpose
of the present study is to emphasise the potential of a future Linear
Collider in disentangling a SUSY seesaw induced Lepton Flavour Violation 
from the potential SM and SUSY backgrounds. 
We assume centre of mass energies between 500 GeV and 3 TeV
and integrated luminosities of 0.5 and 3 ab$^{-1}$, and we  
explore the possibility of electron and positron beam
polarisation, comparing the significance of the signal for different beam
configurations.
Although a full detector simulation
as well as a thorough  study of relevant kinematical cuts lies beyond
the scope of this study, we nevertheless
suggest several possibilities for kinematical cuts. 

Instead of scanning over the parameter space of minimal supergravity
(mSUGRA) inspired constrained SUSY models 
we illustrate our study via four representative benchmark
points, both in the low and high $\tan \beta$ 
regime\footnote{Most of the  chosen benchmark points comply with 
recent LHCb bounds on BR($B_s \to \mu^+ \mu^-$)~\cite{Aaij:2012ac}.}. 
In addition to
complying with current LHC bounds, the chosen benchmark points exhibit
a viable dark matter relic density~\cite{Larson:2010gs}.
Our analysis reveals that, despite the SM and
SUSY backgrounds, the expected number of signal events should allow to
probe cLFV in extensive regions of the SUSY seesaw parameter space.

The paper is organised as follows: in Section~\ref{sec:susyseesaw}, we
briefly review Lepton Flavour Violation  in the type I SUSY seesaw;
Sections~\ref{sec:lfv:lc:epem} and~\ref{sec:lfv:lc:emem}  are respectively devoted
to the discussion of a charged lepton flavour violation signal (and
corresponding background) at a Linear Collider for the $e^+e^-$ and
$e^-e^-$ beam options. In Section~\ref{sec:res}, we
collect the relevant numerical results and discuss the potential
observation of a SUSY seesaw cLFV signature at a Linear Collider, 
considering different beam configurations/polarisations, 
always stressing the correlation with the corresponding relevant
low-energy observables. Our final remarks are given in
Section~\ref{sec:concs}.  Technical details concerning the computation
are summarised in the Appendix.

\section{LFV in the SUSY seesaw}\label{sec:susyseesaw}
Irrespective of the Dirac/Majorana nature of the neutrinos, and
of the underlying mechanism of $\nu$-mass generation, charged lepton
currents (i.e., $W^\pm \, \bar \ell_i \, \nu_j$) do violate lepton flavour,
with a strength given by the relevant entry of the $U_\text{MNS}$ leptonic mixing
matrix, $U_{ij}$. Under the standard parametrization (and omitting Majorana CP
violating phases for simplicity), the $U_\text{MNS}$ is given by
\begin{equation}
U_\text{MNS}=
\left( 
\begin{array}{ccc} 
c_{12} \,c_{13} & s_{12} \,c_{13} & s_{13} \, e^{-i \delta} \\ 
-s_{12}\, c_{23}\,-\,c_{12}\,s_{23}\,s_{13}\,e^{i \delta} 
& c_{12} \,c_{23}\,-\,s_{12}\,s_{23}\,s_{13}\,e^{i \delta} 
& s_{23}\,c_{13} \\ 
s_{12}\, s_{23}\,-\,c_{12}\,c_{23}\,s_{13}\,e^{i \delta} 
& -c_{12}\, s_{23}\,-\,s_{12}\,c_{23}\,s_{13}\,e^{i \delta} 
& c_{23}\,c_{13}
\end{array} \right) \,,
\label{Umns}
\end{equation}
where  $c_{ij} \equiv \cos \theta_{ij}$, $s_{ij} \equiv \sin
\theta_{ij}$,  with $\theta_{ij}$ the leptonic mixing angles and $\delta$ the Dirac CP phase. 

Depending on the hypothesised mechanism of neutrino
mass generation, and on the specific framework onto which it is
embedded, additional sources of
charged lepton flavour violation might emerge, leading to potentially
important contributions to a number of cLFV observables. 

Here we consider a type I seesaw embedded into the constrained MSSM (cMSSM), which is
thus extended by
three right-handed neutrino superfields. The model is defined by its
superpotential and soft-SUSY breaking Lagrangian, whose leptonic parts
we detail here:
\begin{equation}\label{eq:Wlepton:def}
\mathcal{W}^\text{lepton}\,=\,\hat N^c\,Y^\nu\,\hat L \, \hat H_2 \,+\,
\hat E^c\,Y^l\,\hat L \, \hat H_1 \,+\,
\frac{1}{2}\,\hat N^c\,M_N\,\hat N^c\,,
\end{equation}
where, and without any loss of generality, 
we work in a basis where both $Y^l$ and $M_N$ are diagonal, 
$Y^l =\operatorname{diag}  (Y^e, Y^\mu,Y^\tau)$, 
$ M_N=\operatorname{diag}  (M_{N_1},M_{N_2},M_{N_3})$. The relevant 
slepton soft-breaking terms are then
\begin{align}\label{eq:Lsoftslepton:def}
\mathcal{V}_\text{soft}^\text{slepton}\,=-\mathcal{L}^\text{slepton}\,&=
m_{\tilde L}^2 \, \tilde \ell_L \,\tilde 
\ell_L^*\,+\,m_{\tilde E}^2 \tilde \ell_R \,\tilde \ell_R^*
\,+\,m^2_{\tilde \nu_R}\,\tilde \nu_R\,\tilde \nu_R^*\,+
\nonumber \\
& +\, \left(
A_l\, 
H_1\,\tilde \ell_L \,\tilde \ell_R^*
\,+\,A_\nu\, 
H_2\,\tilde \nu_L \,\tilde \nu_R^* 
\,+\, B_{\nu}\,\tilde \nu_R\,\tilde \nu_R\, +\text{H.c.}
\right)\,.
\end{align}
Further assuming a flavour blind mechanism of SUSY breaking (for
instance minimal supergravity inspired), the soft breaking parameters
obey universality conditions at some high-energy
scale, which we choose to be the gauge coupling unification
scale $M_\text{GUT} \sim 10^{16}$ GeV: 
\begin{align}\label{eq:cMSSM:univcond}
&
\left(m_{\tilde L}\right)^2_{ij}\,=\,
\left(m_{\tilde E}\right)^2_{ij}\,=\,
\left(m_{\widetilde {\nu}_R}\right)^2_{ij}\,=\, m_0^2\,\delta_{ij}\,,\,\,
\nonumber \\
& \left(A_l\right)_{ij}\,=\, A_0 \, \left(Y^l \right)_{ij},\,\,
\left(A_\nu \right)_{ij}\,=\, A_0 \, \left(Y^\nu \right)_{ij},
\end{align}
where $m_0$ and $A_0$ are the universal scalar soft-breaking mass and
trilinear couplings of the cMSSM, and $i,j$ denote lepton flavour
indices ($i,j=1,2,3$).  

After electroweak symmetry breaking (EWSB), and assuming the seesaw
limit (i.e. $M_{N_i}\,\gg\,M_{_\text{EW}}$), one recovers the 
usual seesaw equation for the light
neutrino mass matrix,
\begin{equation}\label{eq:seesaw:light}
m_\nu\,=\, - {m_D^\nu}^T M_N^{-1} m_D^\nu  \,, 
\end{equation}
with $m_D^\nu=Y^\nu\,v_2$ ($v_i$ being the vacuum expectation values (VEVs) of the neutral
Higgs scalars, $v_{1(2)}= \,v\,\cos (\sin) \beta$, with $v=174$
GeV), and where $M_{N_i}$ corresponds to the masses of the heavy right-handed neutrino
eigenstates. The light neutrino matrix  $m_\nu$ is diagonalized by
the $U_\text{MNS}$ as $m_{\nu}^\text{diag}
={U_\text{MNS}}^T\,m_{\nu} \,U_\text{MNS}$. 

A convenient means of parametrizing the neutrino Yukawa couplings,
while at the same time allowing to accommodate the experimental data,
is given by the Casas-Ibarra parametrization~\cite{Casas:2001sr},
which reads at the seesaw scale, $M_N$,
\begin{equation}\label{eq:seesaw:casas}
Y^\nu v_2=m_D^\nu \,=\, i \sqrt{M^\text{diag}_N}\, R \,
\sqrt{m^\text{diag}_\nu}\,  {U_\text{MNS}}^{\dagger}\,.
\end{equation}
In the above, $R$ is a complex orthogonal $3 \times 3$ matrix that
encodes the possible mixings involving the right-handed neutrinos, 
and which can be parameterized in terms of three 
complex angles $\theta_i$ $(i=1,2,3)$.

The non-trivial flavour
structure of $Y^\nu$ at the GUT scale will induce (through the running
from $M_\text{GUT}$ down to the seesaw scale) flavour mixing in the
otherwise approximately flavour conserving soft-SUSY breaking terms.
In particular, there will be radiatively induced flavour mixing in the slepton
mass matrices, manifest in the $LL$ and $LR$ blocks of the
$6\times 6$ slepton mass matrix; an analytical estimation using the
leading order (LLog) approximation leads to the following corrections
to the slepton mass terms: 
\begin{align}\label{eq:LFV:LLog}
(\Delta m_{\tilde{L}}^2)_{_{ij}}&\,=
\,- \frac{1}{8\, \pi^2}\,  (3\, m_0^2+ A_0^2)\, ({Y^{\nu}}^\dagger\, 
L\, Y^{\nu})_{ij} 
\,,
\nonumber \\
(\Delta A_l)_{_{ij}}&\,=
\,
- \frac{3}{16 \,\pi^2}\, A_0\, Y^l_{ij}\, ({Y^{\nu}}^\dagger\, L\, Y^{\nu})_{ij}
\,,
\nonumber \\
(\Delta m_{\tilde{E}}^2)_{_{ij}}&\,\simeq
\,
0\,\,;\, L_{kl}\, \equiv \,\log \left( \frac{M_\text{GUT}}{M_{N_k}}\right) \,
\delta_{kl}\,.
\end{align}
The amount of flavour violation is
encoded in the matrix elements $({Y^\nu}^\dagger L Y^\nu)_{ij}$ of
Eq.~(\ref{eq:LFV:LLog}), which can be related to high- and 
low-energy neutrino parameters using Eq.~(\ref{eq:seesaw:casas}).

These renormalisation group equation (RGE) 
induced flavour violating corrections have an
impact regarding both flavour non-universality and flavour violation
in the charged slepton sector, leading to high- and low-energy cLFV
observables that can be experimentally probed. 

At low-energies, there will be important contributions to radiative
and three-body lepton decays, as well as to $\mu-e$ conversion in
nuclei (a detailed discussion, and the corresponding expressions can
be found, for example, in~\cite{Raidal:2008jk}).
In particular, and considering a seesaw scale close to the GUT scale 
(i.e. $M_N \sim \mathcal{O}(10^{12-15}\text{ GeV})$), so that $Y^\nu$
is large ($Y^\nu \sim \mathcal{O}(1)$),
one expects values of  BR($\mu \to e\gamma$),
BR($\tau \to \mu \gamma$) and CR($\mu-e$, N) to be well within the
reach of current dedicated experiments. 

At the LHC, cLFV can also be studied, for instance in 
relation with the ${\chi}^0_2 \to {\chi}^0_1\,
\ell^{\pm}\,{\ell^{\mp}}$ decay chains. 
In scenarios where the
${\chi}^0_2$ is sufficiently heavy to decay via a real (on-shell)
slepton, several cLFV observables can be manifest:
(i) sizable widths for  cLFV decay processes like
$\chi_2^0 \to \chi_1^0\ \ell_i^\pm\, \ell_j^\mp$~\cite{Arkanihamed:1996au,Hinchliffe:2000np, Carvalho:2002jg, Hirsch:2008dy, Carquin:2008gv};
(ii) multiple edges in di-lepton invariant mass
distributions $\chi_2^0 \to \chi_1^0 \ \ell_i^{\pm} \ell_i^{\mp}$,
arising from the 
exchange of a different flavour slepton $\tilde \ell_j$ 
(in addition to the left- and right-handed sleptons, $\tilde \ell_{_{L,R}}^i$);
(iii) flavoured slepton mass splittings. 

As discussed, for instance 
in~\cite{Buras:2009sg,Abada:2010kj,Calibbi:2011dn,Calibbi:2011vi}, the interplay
of low- and high-energy observables might provide valuable
insight into the underlying mechanism of neutrino mass generation.

\section{LFV at a Linear Collider: $e^+ e^-$ beam option}\label{sec:lfv:lc:epem}
In this work we focus our attention on possible 
signals of cLFV in processes of the type $e^+ e^- \to \ell_i^+
\ell^-_j$ + missing energy. 
It is important to emphasise here that one must distinguish LFV as
originating from charged current interactions, and whose source is
strictly related to the $U_\text{MNS}$ matrix (in a model-independent
way), 
from LFV originating from neutral currents involving sleptons. 
The latter, although also
associated to the $U_\text{MNS}$, can be related to the implementation
of a specific mechanism for cLFV: in~\cite{Carquin:2011rg}, 
effective flavour violation entries
in the slepton mass matrices were considered; 
in~\cite{Deppisch:2003wt,Deppisch:2004pc}, 
lepton flavour violation has been addressed for a Linear Collider in
the framework of a type I SUSY seesaw. Here, and following the latter
approach, we revisit cLFV  arising from a type I SUSY seesaw at a
Linear Collider in the light of recent neutrino data, low-energy cLFV
bounds and, more importantly, for a comparatively heavy SUSY spectrum
as suggested by recent LHC searches.   

In our work, we will thus study signals of SUSY seesaw cLFV as 
\begin{equation}
e^+\, e^- \,\to \,\ell_i^+\,\ell^-_j + 2\,\chi_1^0 \,,
\end{equation}
further focusing on final states of the type $e^+\mu^-$, 
so to explore the maximal efficiency of the muon 
detector~\cite{:2010zzd,CLIC:CDR,Blaising:2012vd}. We recall that 
we do allow for (intermediate) tau production, but consider only
its purely leptonic decays, $\tau \to \nu_\tau\, \ell \,\bar \nu_\ell$. 
Clearly, under the
assumption of leptonic mixing, as parametrized by the $U_\text{MNS}$,
both the SM and the MSSM provide important, 
if not dominant, contributions. In particular, one
expects missing energy signatures such as neutrinos pairs (in the case
of the SM) and/or pairs of neutrinos + 2 neutralinos
in the case of the MSSM. 

Before proceeding, let us mention that
LFV can also be manifest in more involved sparticle production and
decays: depending on the specific SUSY spectrum, other
processes might be present, leading to more complex final states (e.g.,
with four charged leptons, mesons, and missing energy).

\bigskip
\noindent In this study we focus on the following processes 
\begin{equation}\label{eq:epem:ABC}
e^+\,e^-\, \to \,
\left\{
\begin{array}{l l}
e^+\,\mu^- + 2\,\chi_1^0 \quad & \text{(A)}\\ 
e^+\,\mu^- + 2\,\chi_1^0 + {(2,4)}\,\nu\quad &\text{(B)}\\
e^+\,\mu^- + {(2,4)}\,\nu\quad &\text{(C)}
\end{array}
\right.
\end{equation}
as well as the charge-conjugated final states. 

\begin{figure}[ht!]
\begin{center}
\begin{tabular}{ccc}
\raisebox{-2mm}{\epsfig{file=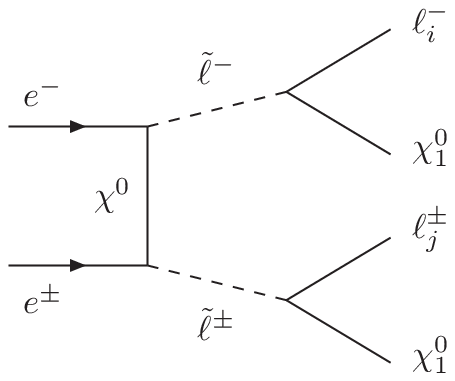, clip=, angle=0, width=37mm}}
\hspace*{4mm}&\hspace*{4mm}
\epsfig{file=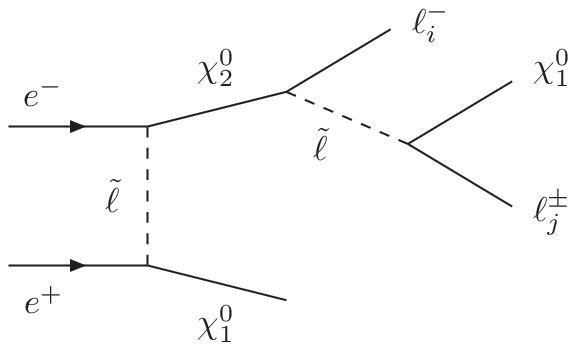, clip=, angle=0, width=50mm} 
\hspace*{4mm}&\hspace*{0mm}
\raisebox{-5mm}{\epsfig{file=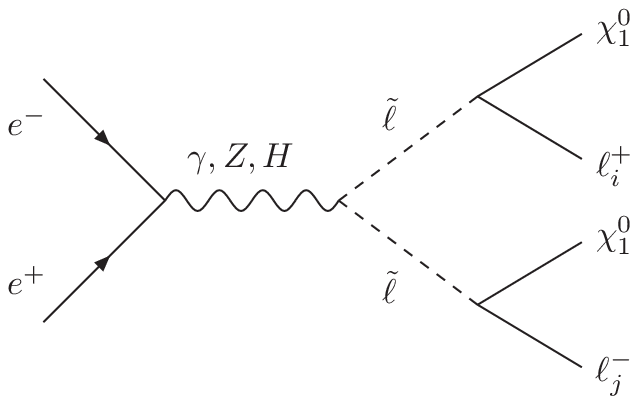, clip=, angle=0, width=50mm}}
\end{tabular}
\caption{Illustrative examples of processes contributing to signal
  (A), for both $e^+e^-$ and $e^-e^-$ beam
  options.}
\label{fig:diagA}
\end{center}
\end{figure}

The process (A), which corresponds to the signal,  will arise from both $s$- and $t$-channel
processes. In the former case, one can have photon-, $Z$, and Higgs
mediated production of slepton pairs (whose flavour content is
decorrelated from the initial state electrons) 
and mixed neutralino
($\chi^0_{i}\,\chi^0_{j}$) pair production. In the latter case, the
$t$-channel exchange of neutralinos 
gives rise to slepton production.
Although in most cases the event is expected to have a symmetric
topology, it is important to notice that, if allowed by the SUSY spectrum, one
might be led to asymmetric cases. 
Some illustrative examples of  diagrams leading to final state of type
(A) can be found in Figure~\ref{fig:diagA}.  

\noindent (B) encompasses most of the processes leading to what we denote
``charged current SUSY background'', in other words cLFV processes
that occur via SUSY charged currents. In this case, flavour violation reflects only
the existence of low-energy leptonic mixings, and could even be
accommodated with massive Dirac neutrinos. As an example, one can have 
mixed chargino ($\chi^\pm_{1,2} \, \chi^\mp_{1,2}$) $s$-channel production,
while gauginos can also be produced via the $t$-channel exchange of 
a neutral or charged slepton.  Moreover,
the decays of a would-be final state tau (i.e., an intermediate tau
state) can account for additional neutrino pairs.

Finally, the SM background (processes of type  (C)) corresponds to $s$-channel
photon exchange, and/or $t$-channel lepton exchange 
(Dirac or Majorana neutrinos). Like for the SUSY background,
``$W$-strahlung'' can lead to sizable contributions.
Different flavour
opposite-sign final state leptons would arise from charged current
interactions and, as in processes of type (B), 
would only reflect the existence of low-energy
mixings in the lepton sector. Multiple final state neutrino pairs can arise, 
for instance, in association to intermediate tau decays (as in (B)).  

\begin{figure}[ht!]
\begin{center}
\begin{tabular}{cc}
\epsfig{file=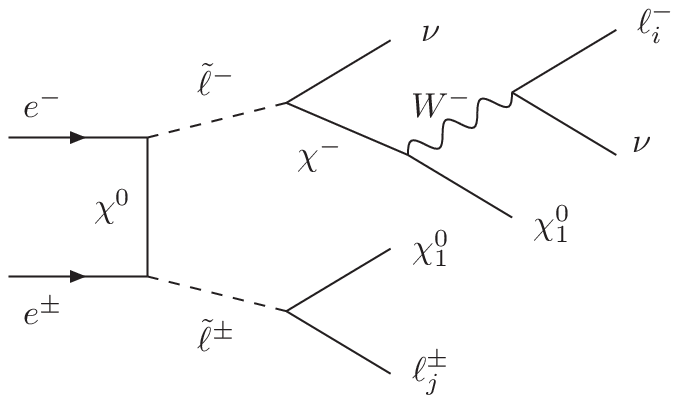, clip=, angle=0, width=50mm}
\hspace*{10mm}&\hspace*{10mm}
\epsfig{file=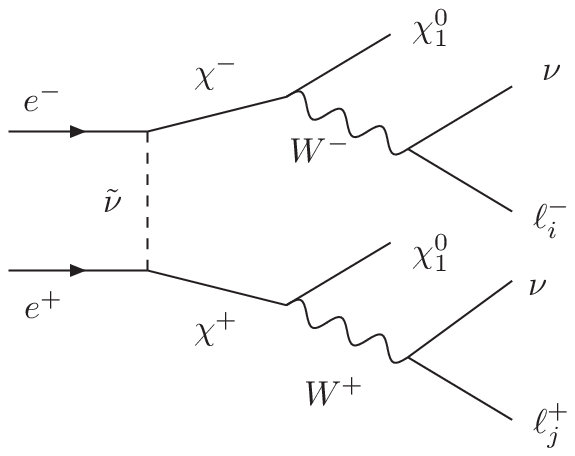, clip=, angle=0, width=50mm} 
\vspace*{5mm}
\\
\raisebox{0mm}{\epsfig{file=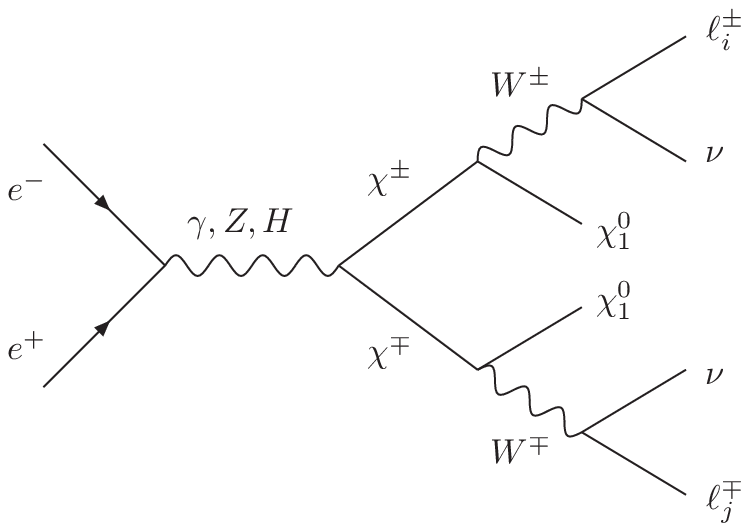, clip=, angle=0, width=50mm}}
\hspace*{10mm}&\hspace*{10mm}
\epsfig{file=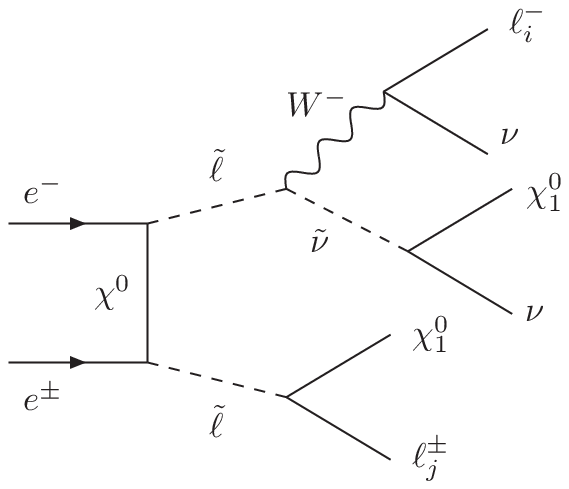, clip=, angle=0, width=50mm} 
\end{tabular}
\caption{Examples of processes contributing to the SUSY
  background (B), for both $e^+e^-$ and $e^-e^-$ beam options. }
\label{fig:diagB}
\end{center}

\end{figure}

\begin{figure}[ht!]
\begin{center}
\begin{tabular}{ccc}
\epsfig{file=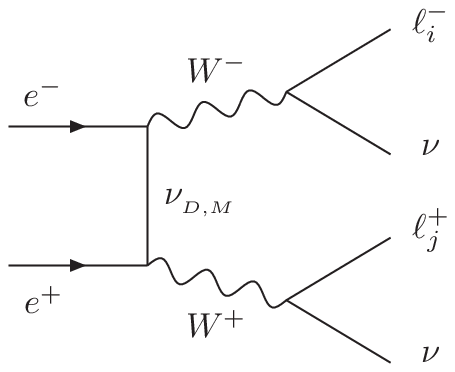, clip=, angle=0, width=40mm}
\hspace*{4mm}&\hspace*{4mm}
\epsfig{file=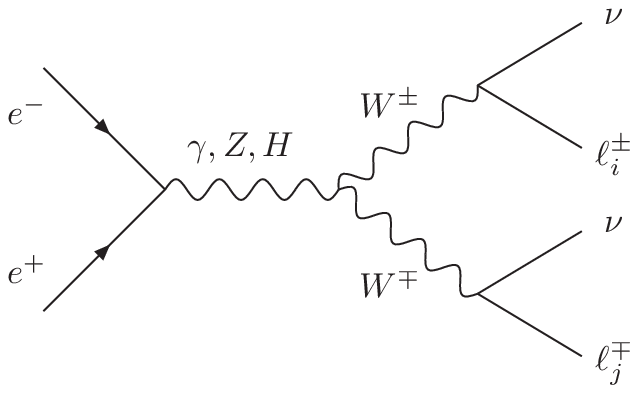, clip=, angle=0, width=53mm} 
&
\vspace*{5mm}
\\
\raisebox{0mm}{\epsfig{file=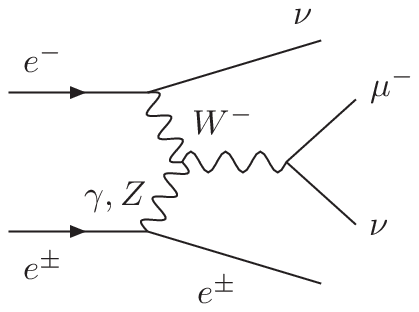, clip=, angle=0, width=42mm}}
\hspace*{4mm}&\hspace*{4mm}
\raisebox{0mm}{\epsfig{file=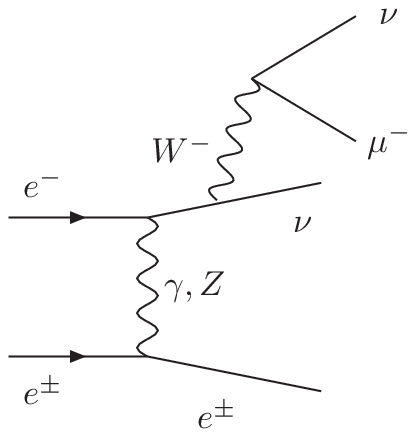, clip=, angle=0, width=40mm}}
\hspace*{4mm}&\hspace*{4mm}
\epsfig{file=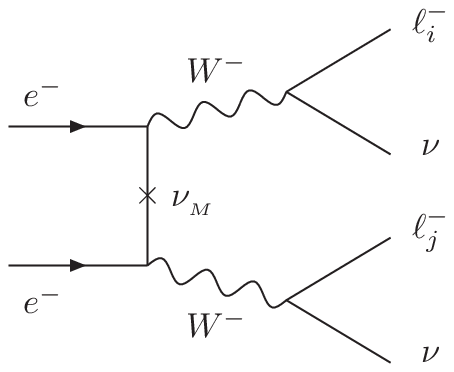, clip=, angle=0, width=45mm} 
\end{tabular}
\caption{Examples of processes contributing to the SM
  background (C), for both $e^+e^-$ and $e^-e^-$ beam options. 
}
\label{fig:diagC}
\end{center}
\end{figure}

Examples of background processes of type (B) and (C)  can 
be found in Figures~\ref{fig:diagB} and~\ref{fig:diagC}. 
Notice that 
some of the processes displayed are due to the nature of the light
neutrinos (Dirac vs. Majorana), while the viability 
of others is dictated by the specific SUSY spectra.

We notice that the distinct topology of the different processes,
as well as specific dedicated cuts,
should already allow a first selection of the ``signal'' events (A).

The possibility of electron and positron beam polarisation is also
instrumental in disentangling the above 
processes~\cite{MoortgatPick:2010zza}: in the ideal case
of 100\% polarised $e^+$ and $e^-$ beams, that is an LL polarisation
of $(P_{e^-},P_{e^+})=(-100\%,-100\%)$,
and with the exception of the $W$-strahlung processes, 
the other SM contributions (C) would
vanish, as well as most of those arising from pure SUSY LFV 
charged currents (B).  
Under this ``ideal'' beam configuration, and via some 
kinematical cuts, a cLFV signal 
(A) would clearly point towards flavour mixing in the slepton sector
(for instance due to a SUSY seesaw), manifest in neutral currents.
 
 Other sources of background have also been considered.  
For instance, the Higgs-mediated contributions have been
taken into account, but their contributions to final states of the
type 
$e^+\,\mu^- + (0\ \text{or }4) \chi_1^0 + {(2,\ 4 \text{ or }6)}\,\nu$
are small, especially compared
to the previously discussed processes (C). 
Likewise, three body  processes such as $e^+ e^- \to \chi^0 \,\ell^\pm \,\tilde \ell^\mp, \,
\chi^\pm \,W^\mp \,\chi^0$ and $\tilde \ell^\pm \,W^\mp \,\tilde
\nu$ are found to be subdominant and have thus been neglected.

A final remark, concerning the possibility of 
same-flavour final states (such as $\mu^+ \mu^-$) is still in order. 
Although same-flavour final state leptons will be
explored in the case of the $e^- e^-$ beam option 
(in Sections \ref{sec:lfv:lc:emem} and  \ref{golden}), we
will not consider them here, since such a signal would be dominantly generated
from flavour conserving interactions.

\section{LFV at a Linear Collider: $e^- e^-$ beam option}\label{sec:lfv:lc:emem}
In addition to its many potentialities to probe the Majorana nature of
neutral fermions which implies the violation of total lepton number, 
$e^- e^-$ collisions provide one of the cleanest experimental 
environments to test lepton flavour violation in the slepton 
sector~\cite{Deppisch:2003wt}.  
In the SM and the MSSM (and their type I 
seesaw extensions), $s$-channel 
processes are excluded due to the absence of doubly charged particles
(for instance, as those present in a type II seesaw~\cite{seesaw:II}). 
As done in the previous section, we again focus on $e^- e^- \to
\ell_i^- \ell^-_j + E^T_\text{miss}$ (at least $i$ or $j \neq e$).
As for the $e^+ e^-$ case, 
we include  processes involving
intermediate $\tau$s, which decay only leptonically, as background contributions.
 
In particular, we study the following possibilities:
\begin{equation}\label{eq:emem:ABC}
e^-\,e^-\, \to \,
\left\{
\begin{array}{l l}
e^-\,\mu^-
+ 2\,\chi_1^0 \quad & \text{(A)}\\ 
e^-\,\mu^-
+ 2\,\chi_1^0 +{(2, 4)}\,\nu\quad &\text{(B)}\\
e^-\,\mu^-
+ {(2, 4)}\,\nu\quad &\text{(C)}
\end{array}
\right.
\end{equation}
Some diagrams corresponding to these processes were illustrated in
Figures~\ref{fig:diagA},~\ref{fig:diagB} and~\ref{fig:diagC}.

In the processes of type (A) - signal -, charged slepton production occurs via $t$-channel
neutralino exchange. 
For SUSY spectra in which $m_{\tilde \ell} \lesssim
m_{\chi^\pm_1}$, this is typically the dominant channel leading to 
$\ell^-_i \ell^-_j + 2\chi_1^0$ final states. 

Other than the contributions 
coming from single $W$-production ($W$-strahlung), we notice that additional processes
might contribute to the background. For example, if the spectrum is
such that $m_{\tilde \ell} \gtrsim m_{\chi^\pm_1}$, one
of the sleptons decays via $\tilde \ell^- \to  \nu \, \chi^-_1
\to \nu \, \chi_1^0 \, W^- \to 2 \nu \, \chi^0_1 \, \ell^-$. 
We stress that in this case, polarising the beams has little effect on
disentangling the signal (A) from the background (B), 
as both processes occur via identical slepton
production ($t$-channel neutralino exchange) mechanisms.

The SM background (C) would arise from $t$-channel single
$W$-production, as well as from the
exchange of a Majorana neutrino; notice that the latter cross section
(and similar to what occurs for neutrinoless double beta decays
 - $0\nu2\beta$ -) is
extremely tiny, at most $\sigma(e^- e^- \to W^- W^-) \sim 
\mathcal{O}(10^{-22}$ fb), due to the
smallness of the light neutrino masses.

\bigskip
\textcolor{black}{In addition to studying 
the prospects for $e^-\mu^-$ final states, a 
$\mu^-\mu^-$ + missing energy final state is a truly ``golden
channel'' for cLFV at a Linear Collider. In this case, the SM
background (C) is negligible, and the SUSY background processes are
always subdominant. Furthermore, in addition to probing cLFV in neutral
currents in a remarkably clean way, this channel
also allows to fully benefit from the LC muon detector efficiency.}

\section{LC potential for LFV discovery}\label{sec:res}
We now proceed to address the different cLFV processes discussed in the
previous sections (Eqs.~(\ref{eq:epem:ABC}, \ref{eq:emem:ABC})),
considering that the unique source of LFV is a type I seesaw mechanism. 

As discussed in Section~\ref{sec:susyseesaw}, we embed
a type I seesaw into the framework of the cMSSM. At the GUT scale, the
neutrino Yukawa couplings are parametrized as in
Eq.~(\ref{eq:seesaw:casas}), and we adopt a conservative approach where
no mixings, other than those associated to the $U_\text{MNS}$, are
present\footnote{In general, the limit $R=1$
translates into a ``conservative'' limit for flavour violation:
apart from possible cancellations, and for a fixed
seesaw scale, this limit typically provides
a lower bound for the amount of the generated LFV.} (i.e. $R=1$). 
Concerning low-energy lepton
data, we take the following (best-fit) values for the neutrino mixing
angles (solar and atmospheric)~\cite{GonzalezGarcia:2010er,Schwetz:2011zk},
\begin{align}\label{eq:mixingangles:data}
& 
\sin^2\theta_{12}\,=\ 0.31^{+0.017}_{-0.015}, 
\quad 
\sin^2\theta_{23}\,=\ 0.52^{+0.06}_{-0.07}.
\end{align}
In what concerns the Chooz angle, given the very recent results from
Double-Chooz~\cite{Abe:2011fz}, 
T2K~\cite{Hartz:2012pr}, MINOS~\cite{Adamson:2012rm}, Daya-Bay~\cite{An:2012eh}
and RENO~\cite{Ahn:2012nd}, we choose a representative 
value for this experimentally favoured large $\theta_{13}$ regime: 
\begin{equation}
\theta_{13}\,= 10^\circ\,.
\end{equation}
Finally, the light neutrino spectrum is specified by
the following intervals for the mass-squared differences
\begin{align}\label{eq:lightmasses:data}
& 
\Delta\, m^2_\text{12} \,=\,(7.6\, \pm 0.2)\,\times 10^{-5}\,\,\text{eV}^2\,,
\quad 
\Delta \, m^2_\text{13} \,=\left\{ \begin{array}{l} \,(-2.40\, \pm \
    0.09)\,\times 10^{-3}\,\,\text{eV}^2\,\\  
\,(+2.50\, \pm \ 0.13)\,\times 10^{-3}\,\,\text{eV}^2\ 
 \end{array}\right. \,,
\end{align}
where the two ranges for $\Delta \, m^2_\text{13}$ correspond to
normal and inverted neutrino spectrum~\cite{Schwetz:2011zk}.

Although in our study we
considered both hierarchical and degenerate right-handed neutrino
spectra, in the following  discussion we will present the results for the case of  
a degenerate right-handed neutrino spectrum.

\medskip
In our numerical analysis, and to illustrate the LFV potential of a
Linear Collider, we will consider two sets of cMSSM (mSUGRA-like)
points\footnote{In the final stages of this work, a new ILC-dedicated
  benchmark proposal has appeared~\cite{LCbenchmark2012}.}, 
defined in Table~\ref{table:points}, 
representative of regimes of low and (very) large $\tan \beta$. 
For each regime, we further present two possibilities, one with a
``light'' spectrum (i.e. with a gluino mass just above the LHC exclusion bound) 
and the other with a comparatively heavier one (associated with
$m_{\tilde g}=2$ TeV).   
The points with low $\tan \beta$ have a phenomenology similar to that
of the points along the 10.1.n  slope of the recent LHC benchmark
proposal~\cite{AbdusSalam:2011fc}. Moreover, the relatively low-values of $m_0$ (when
compared to $M_{1/2}$) lead to a lighter slepton spectrum, that can be
easily produced at a LC (while avoiding the LHC bounds on the
strongly interacting sector).
We notice that in this case, the correct dark matter relic density
(computed using MicrOmegas~\cite{Belanger:2008sj}) is 
obtained from 
neutralino-stau co-annihilation, and these points are thus characterised
by a very small NLSP-LSP mass difference.
The very large value of $\tan \beta=52$
chosen for the second set of points is also due to our willingness to 
have the LSP as a viable dark matter candidate (in this case we are in
the so-called ``Higgs funnel'' region). 

In Table~\ref{table:spectrum}, 
we collect the most relevant information
regarding the spectrum of the chosen benchmark points. 
The low-energy sparticle masses and mixings 
have been numerically evaluated
with the SPheno public code~\cite{Porod:2003um}. LHC bounds on the
SUSY spectrum~\cite{LHC:2011}
have been applied; in all cases, the lightest Higgs mass is around 
117 GeV, still marginally compatible with the new
LHC data~\cite{LHC:Higgs:2012}.
Notice that the  non-minimal SUSY scenarios can still be considered 
to accommodate such data, with a minimal impact concerning
the present lepton flavour dedicated analysis. This is the case of
deviations from strict mSUGRA inspired universality in the Higgs
sector - the so-called Non-Universal Higgs Mass
models -, or third generation non-universality, which could increase the value of 
$m_h$ (especially given the already heavy stop sector)~\cite{Arbey:2011ab}.

We have taken into account all relevant 
constraints from low-energy cLFV dedicated
experiments~\cite{PDG}. In particular, regarding the bound on BR($\mu \to
e \gamma$), which will be the most constraining one for our study, 
we have considered the most recent MEG results~\cite{arXiv:1107.5547}, 
\begin{equation}\label{eq:MEG:2011}
\text{BR}(\mu \to e \gamma)\, < 2.4 \times 10^{-12}\,,\quad 90\%\text{
  C.L.}\,.
\end{equation}  
\begin{table}
\centering
\begin{tabular}{|c|c|c|c|c|}
\hline
&  C-light & C-heavy& F-light & F-heavy  \\ \hline
$m_0$ (GeV)   & 150  & 200 & 600  & 750  \\ \hline
$M_{1/2}$ (GeV)  & 727.9 & 949.2 & 667.0  & 872.1  \\ \hline
$\tan\beta$ & 10 & 10& 52 & 52  \\ \hline
$A_0$  (GeV)  & 0  & 0 & 0 & 0 \\ \hline
sign($\mu$) & 1 & 1  & 1 & 1 \\ \hline
\end{tabular}
\caption{Representative points used in the numerical analysis.}\label{table:points}
\end{table}
A comment is still in order concerning the recent LHCb bounds of 
BR($B_s \to \mu^+\mu^-$)~\cite{Aaij:2012ac}:
points F (in the large $\tan \beta$ regime) are associated with 
large contributions to the latter observable. In particular due to its lighter spectrum, F-light 
is clearly ruled out based on this bounds, while F-heavy  would be marginally compatible at 
 3-$\sigma$. 
Both points C are in good agreement with these bounds, and in  the
subsequent discussion we will mostly illustrate  
our results focusing on points C-light and C-heavy.

\begin{table}
	\centering
	\begin{tabular}{|c|c|c|c|c|}
	\hline
&  C-light & C-heavy	& F-light & F-heavy  \\ \hline
	$\chi^0_1$ & 303.7   & 401.1& 279.9   & 370.6         \\ \hline
	$\chi^0_2$  & 574.2   & 756.9& 530.3   & 700.9        \\ \hline
	$\chi^0_3$ & -882.3  & -1114.7& -786.5  & -993.9       \\ \hline
	$\chi^0_4$ & 893.5   & 1124.4& 797.5   & 1003.8        \\ \hline
	$\chi^{\pm}_1$   &574.4  & 757.1  & 530.5  & 701.1          \\ \hline
	$\chi^{\pm}_2$    & 893.8  & 1124.5    & 798.3  & 1004.2         \\ \hline
	$(\tilde{\nu}_e)_L$  & 504.2  & 657.9& 742.4  & 943.3         \\ \hline
	$(\tilde{\nu}_{\mu})_L$     & 503.9   & 657.6   & 651.0   & 830.0         \\ \hline
	$(\tilde{\nu}_{\tau})_L$    & 502.1   & 655.2   & 741.8   & 368.93        \\ \hline
	$\tilde{e}_R$ & 314.4 & 409.7& 650.3 & 817.6        \\ \hline
	$\tilde{e}_L$ & 510.7   & 662.9& 747.0 & 946.9        \\ \hline
	$\tilde{\mu}_R$    & 314.4  & 409.7    & 649.5 & 816.6    \\ \hline
	$\tilde{\mu}_L$    & 509.9  & 662.6    & 746.5 & 946.3    \\ \hline
	$\tilde{\tau}_1$    & 306.5  & 401.1  & 383.6 & 495.8     \\ \hline
	$\tilde{\tau}_2$     & 510.6  & 661.4   & 672.8 & 847.4   \\ \hline
	\end{tabular}
	\caption{Slepton, neutralino and chargino mass spectrum (in GeV) 
for the points of Table~\ref{table:points}.}\label{table:spectrum}
\end{table}

In what follows, we
separately discuss the prospects for $e^+e^-$ and $e^-e^-$, and in
both cases the effects of polarising the beams. 
In our analysis, we have  taken into account all
contributions to the different final states, allowing for
the production and exchange of all possible 
(s-)states\footnote{Although in the present analysis we only collect the
results regarding the leading-order LFV processes mentioned above, 
we have also investigated more complex production and decay processes,
with additional LFV vertices and longer decay chains, leading
to final states involving a larger number of particles.}. Our study
is based on an algorithmic calculation of the possible production and
decay modes, considering that the majority of the events proceeds 
from an on-shell primary production (so that there are no interference 
effects between the different contributions), with subsequent
two-body cascade decays (the exception being 3-body decays of the $\tau$). 
For SUSY $e^+ e^-$ cross-sections and branching fractions, we have used
SPheno~\cite{Porod:2003um}, while for SUSY $e^- e^-$ and SM cross
sections we have developed dedicated routines.
For each given final state, we have generated all possible allowed
production and decay chains, arising from
each of the considered primary production modes (more detailed
information can be found in the Appendix).

\bigskip
Regarding the performance of the Linear Collider, we will assume
values for the c.o.m. energy in the interval 
$500$ GeV $\lesssim \sqrt s \lesssim 3$ TeV, and benchmark values for
the integrated luminosity of $0.5$ and 3 ab$^{-1}$. In our
analysis, we adopt the following definition of the significance of the cLFV signal:
\begin{equation}\label{eq:sigma:def}
\mathcal{S}\,=\,\frac{\mathcal{N}(\text{signal})}{\sqrt{\mathcal{N}(\text{signal}
+\text{background})}}\,,
\end{equation}
where $\mathcal{N}$ denotes the number of events and ``background'' will be
identified with (B) and/or (B+C), see Eqs.~(\ref{eq:epem:ABC},
\ref{eq:emem:ABC}). 

In this study we will only conduct a phenomenological
analysis, i.e., focusing only on the theoretical estimations of the
potential number of events at a LC operating at a given 
$\sqrt s$, for possibly polarised beams. 
Although Linear Collider CRDs do not consider cLFV slepton production and/or decays,
there are dedicated studies addressing, for example in the case of
$e^+ e^- \to \tilde \ell^+ \tilde \ell^- \to  \ell_i^+ \ell^-_i
\chi_1^0 \chi_1^0$, event reconstruction methods and detector
performances, as well as proposals for several cuts that allow to optimise the
significance. The different cuts are based on the key characteristics
of the signal events which, in addition to the different flavour final
state leptons, are missing energy, missing transverse momentum and
acoplanarity. The analysis procedure proposed in~\cite{LCD-Note-2011-018} makes use
 of the following discriminating variables to separate the signal from
the large backgrounds: dilepton energy $(E_{\ell_i}+E_{\ell_j})$; 
vector and algebraic sums of the leptons' momenta $(p^T_{\ell_i}, p^T_{\ell_j})$;
dilepton invariant mass $m_{{\ell_i}{\ell_j}}$ and velocity
$\beta_{{\ell_i}{\ell_j}}$; dilepton's missing momenta angle
$\theta(\vec p^T_{\ell_i}, \vec p^T_{\ell_j})$;
dilepton acolinearity and acoplanarity; 
dilepton energy imbalance $\Delta=|E_{\ell_i}-E_{\ell_j}|/|E_{\ell_i}+E_{\ell_j}|$.
The values of the above cuts are chosen to optimise 
significance versus signal efficiency.
As discussed in~\cite{CLIC:CDR}, selection efficiencies
can be as high as 97\% for di-muon and 94\% for di-electron final
states. Therefore based on the similarities between the flavour conserving processes discussed in ~\cite{CLIC:CDR,LCD-Note-2011-018} with the cLFV signals studied here,  we expect that a dedicated parallel analysis with the introduction of appropriate cuts will allow to optimise the statistical significance in the cLFV case as well. The LC appears thus a perfect facility for  the
study of $e^{\pm} e^- \to e^{\pm}
\mu^- + E^T_\text{miss}$ and $e^- e^- \to \mu^-
\mu^- + E^T_\text{miss}$ processes.

\subsection{$e^+e^-$ beam option}\label{sec:res:epem}
\begin{figure}[ht!]
\begin{center}
\begin{tabular}{cc}
\epsfig{file=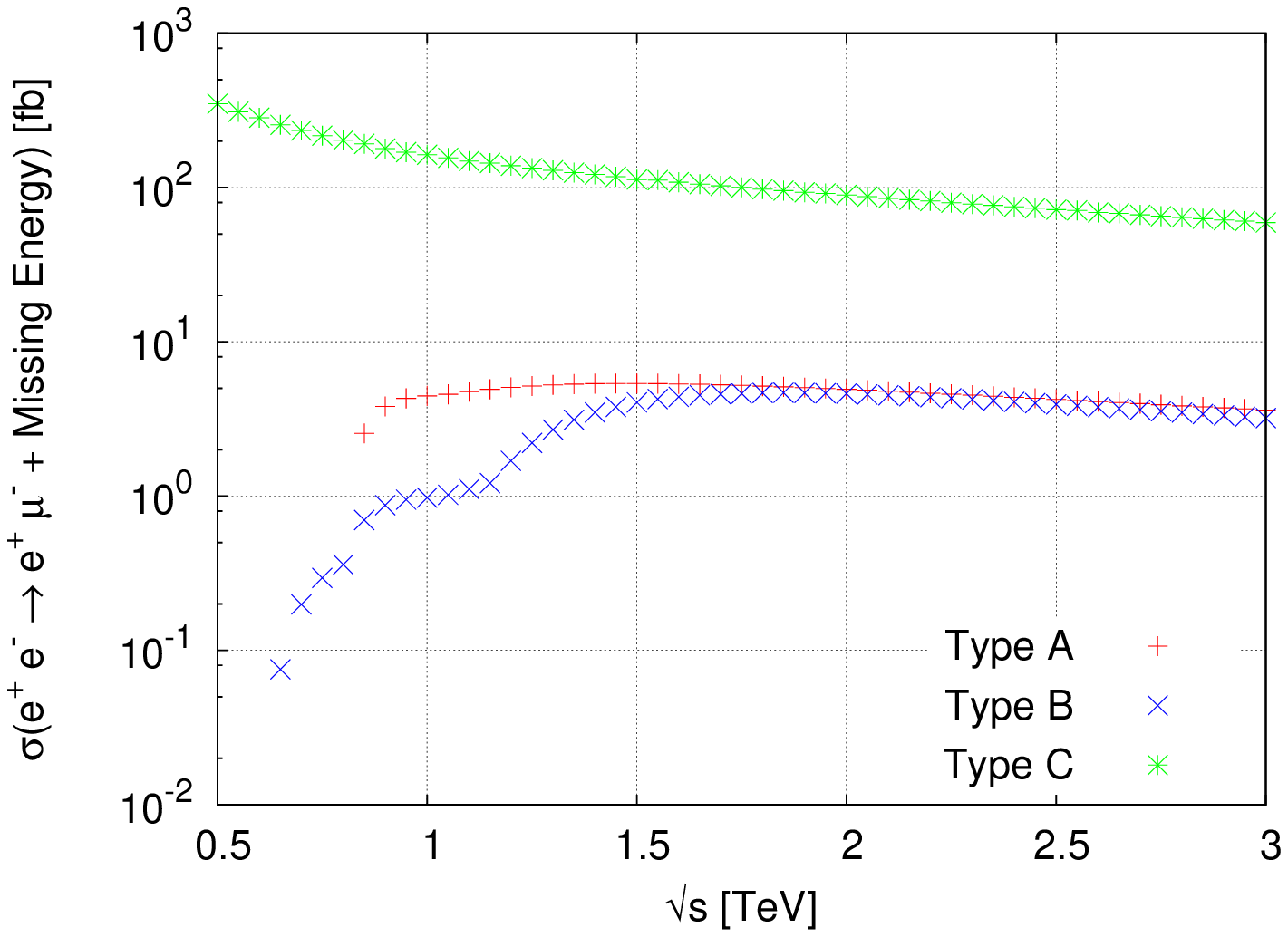, 
clip=, angle=0, width=75mm}
&
\epsfig{file=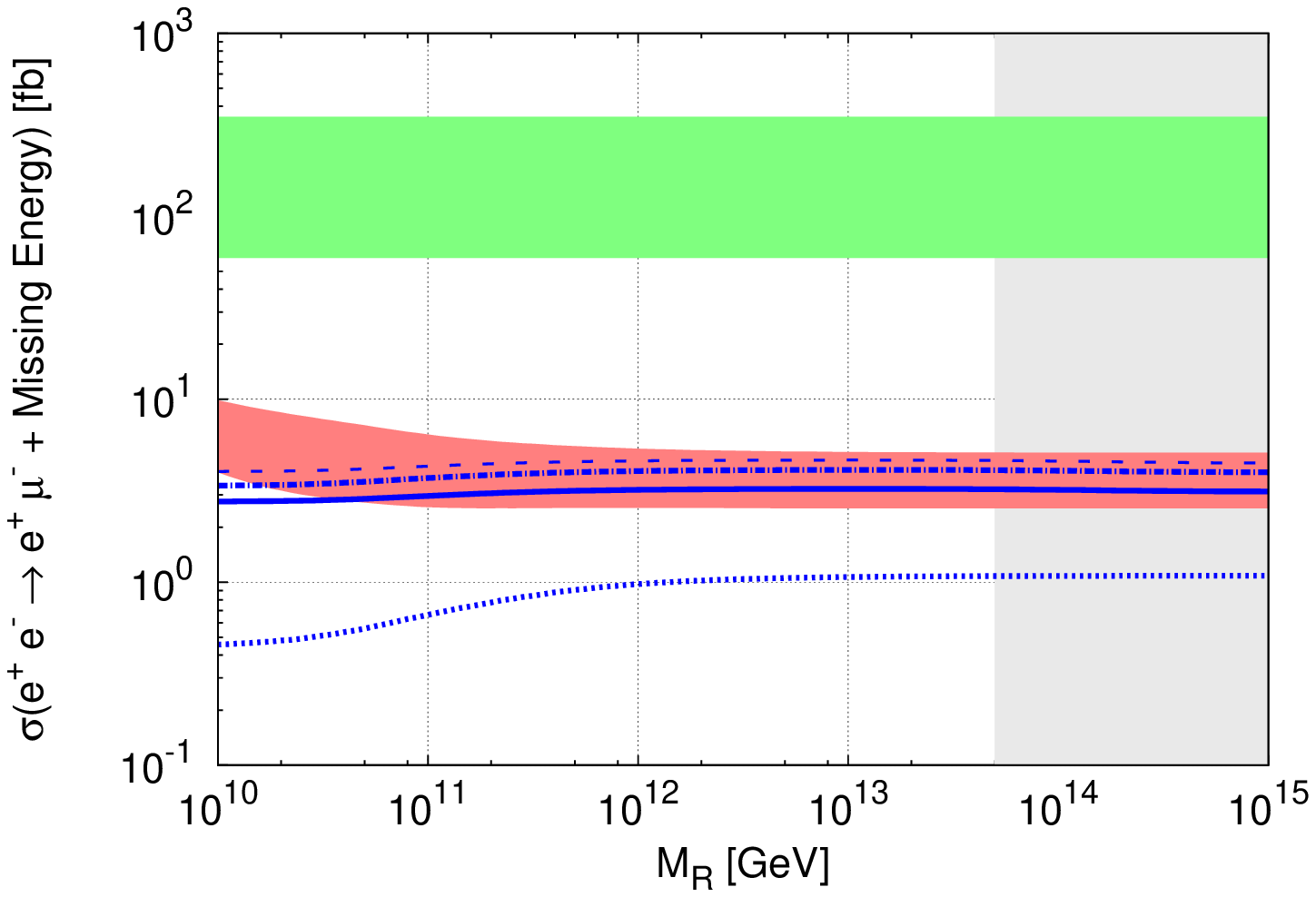, 
clip=, angle=0, width=75mm} 
\\
\epsfig{file=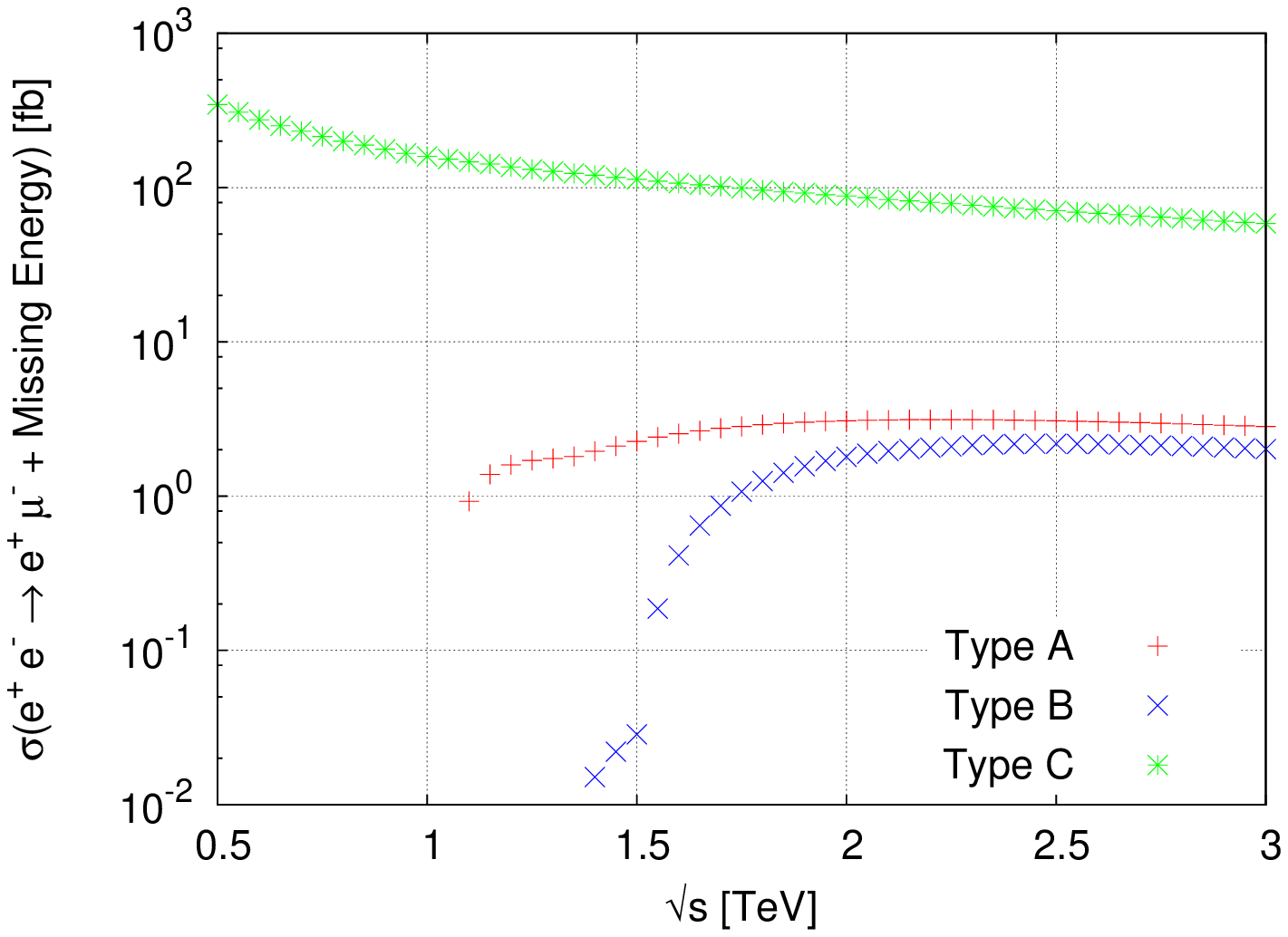, 
clip=, angle=0, width=75mm}
&
\epsfig{file=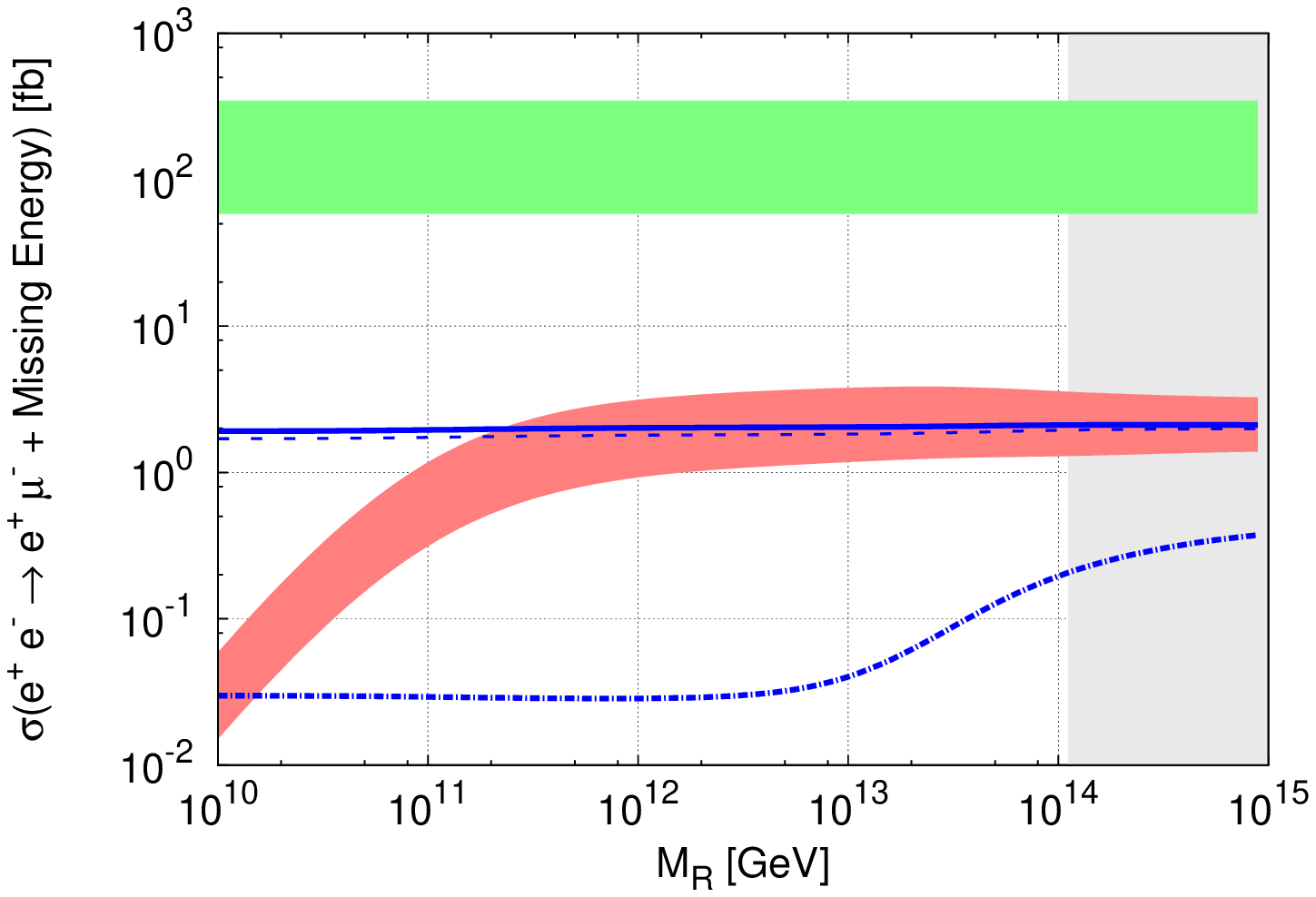, 
clip=, angle=0, width=75mm}
\end{tabular}
\caption{On the left, cross section for $e^+ e^- \to e^+
\mu^- + E^T_\text{miss}$ {(with $E^T_\text{miss}=2 \chi_1^0, 2 \chi_1^0
+  (2,4)\nu, (2,4) \nu$)}, for points C-light and C-heavy (upper and lower
panels, respectively), as a function of the  centre of mass energy,
$\sqrt{s}$. We fix $M_R = 10^{12}$ GeV, and denote 
the signal (A) with red crosses, the SUSY charged current background
(B) with blue times, and the SM charged current background (C) by 
green asterisks. On the right, cross section for $e^+ e^- \to e^+
\mu^- + E^T_\text{miss}$ (with $E^T_\text{miss}=2 \chi_1^0, 2 \chi_1^0
+  (2,4)\nu, (2,4) \nu$), for points C-light and C-heavy (upper and lower
panels, respectively), as a function of the right-handed
neutrino mass ($M_R$).  A green band denotes the SM background, while
the red one corresponds to the signal (A). Blue lines denote the SUSY
backgrounds, computed for $\sqrt{s}=$ 1, 1.5, 2 and 3 TeV
(respectively corresponding to dotted, dashed-dotted, dashed and full
lines).  The (grey) shaded region corresponds to 
values of $M_R$ already excluded by the present bound on BR($\mu \to e
\gamma$). 
In all cases we have taken a degenerate right-handed neutrino spectrum,
and set $\theta_{13}=10^\circ$. }
\label{fig:Clh:Msqrts}
\end{center}
\end{figure}

We begin by studying the processes 
$e^+ e^- \to e^+
\mu^- + E^T_\text{miss}$, with $E^T_\text{miss}=2 \chi_1^0, 2 \chi_1^0
+   (2,4) \nu, (2,4) \nu$, for the four representative points described in 
Table~\ref{table:points} 
(C-light, C-heavy, F-light and F-heavy). 
We display in Fig.~\ref{fig:Clh:Msqrts} these cross section for 
points C-light and C-heavy, assuming a
degenerate right-handed neutrino spectrum. 

As can be easily seen from the left panels of 
Fig.~\ref{fig:Clh:Msqrts}, the SM background (in particular due to the
$W$-strahlung processes)
clearly dominates over any SUSY contribution, for all values of the
c.o.m. energy, $\sqrt s$. 
However, due to the
different topology of the SM background (C), when compared to the SUSY
events (A) and (B),  we expect (in view of what was previously argued) that these can  be successfully
disentangled (see for instance the several cuts above,
  as well as those proposed in~\cite{CLIC:CDR}. 
  Due to the very light spectrum of both C-points (see
Tables~\ref{table:points} and~\ref{table:spectrum}), the signal is
associated to a slightly larger cross section,  
and for the case of C-heavy, it could
be the only SUSY signal for low values of $\sqrt{s}$.
Although not displayed here, we have also estimated the contributions
of Higgs-mediated processes (via intermediate $h\, Z$, $H\, A$ or $H^+
\,H^-$ production), finding that these are very subdominant, with 
associated cross sections typically
below $10^{-1}$ fb.  Moreover, we notice that the contribution from
intermediate $\tau$ decays accounts for about 
about 90\% of the SUSY background in the case of points 
F, and circa 50\% (100\% for LL polarization) in the case of points C, provided the c.o.m.
energy is above the $\tilde\ell_L \,\tilde\ell_R$ production
threshold; otherwise it is about 100\% for both cases.

Additional information, especially on the seesaw origin of the signal,
can be found on the right-handside plots, where we display its dependence
 on the degenerate right-handed neutrino mass, 
$M_R$. For point C-heavy, the seesaw origin of the signal is manifest,
increasing with $M_R$ as would be expected from the LLog approximation
of Eq.~(\ref{eq:LFV:LLog}), thus revealing that such an   
$e^+ e^- \to e^+ \mu^- + 2 \chi_1^0$ cross section is clearly due to
the working hypothesis of a type I SUSY seesaw. 
For sufficiently large values of the seesaw scale, 
the signal (A) would even dominate over the SUSY background (B), 
but such regions become eventually excluded, as
the associated amount of flavour violation in the $e-\mu$ sector would
induce excessively large BR($\mu \to e \gamma$). 
 The peculiar spectrum of point C-light, with near-degenerate $\tilde
e_L, \tilde \mu_L$ states, is such that even for a very small $LL$ 
slepton mixing (i.e. $(m^{\tilde \ell}_{LL})^2_{12} \ll (m^{\tilde
  \ell}_{LL})^2_{11,22}$), sufficiently small to avoid $\mu \to e
\gamma$ constraints, a significant $\tilde
e_L - \tilde \mu_L$ mixing can indeed occur due to tiny $LR$ effects.
This accounts for the apparent ``flatness'' exhibited by the signal $\sigma$
versus $M_R$ curves. 

Assuming an integrated luminosity of 0.5 (3) ab$^{-1}$, the
expected number of signal events would 
 be $\mathcal{O}(5\times10^{3} (
10^4))$ for points C-light and $\mathcal{O}(10^{3} (10^4))$ for
C-heavy. In the former, the maximal number of events is associated
with a lower regime of $M_R$, while in the latter, one finds the
opposite behaviour (in agreement with Fig.~\ref{fig:Clh:Msqrts}).
 
\begin{figure}[ht!]
\begin{center}
\begin{tabular}{cc}
\epsfig{file=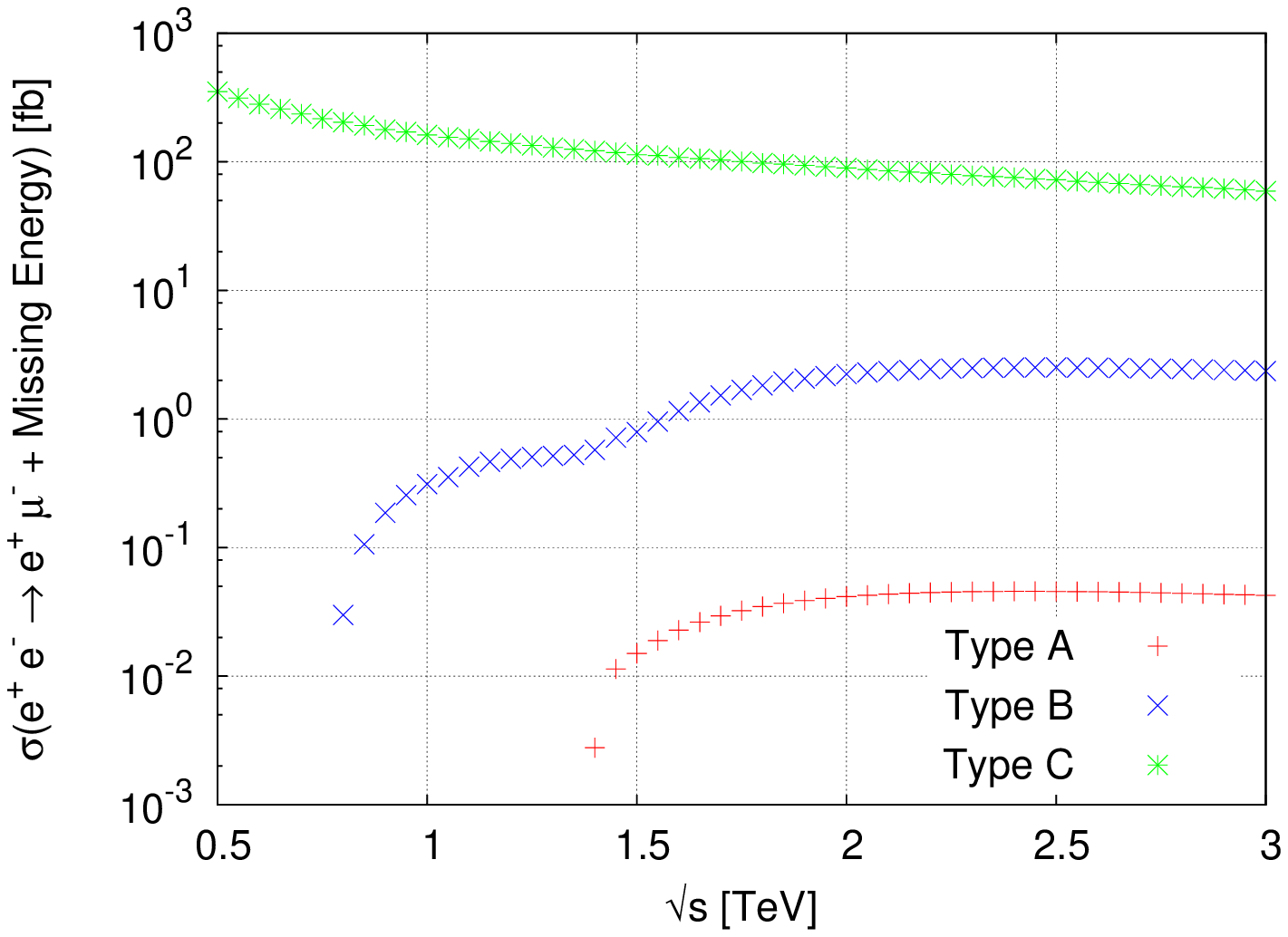, 
clip=, angle=0, width=75mm}
&
\epsfig{file=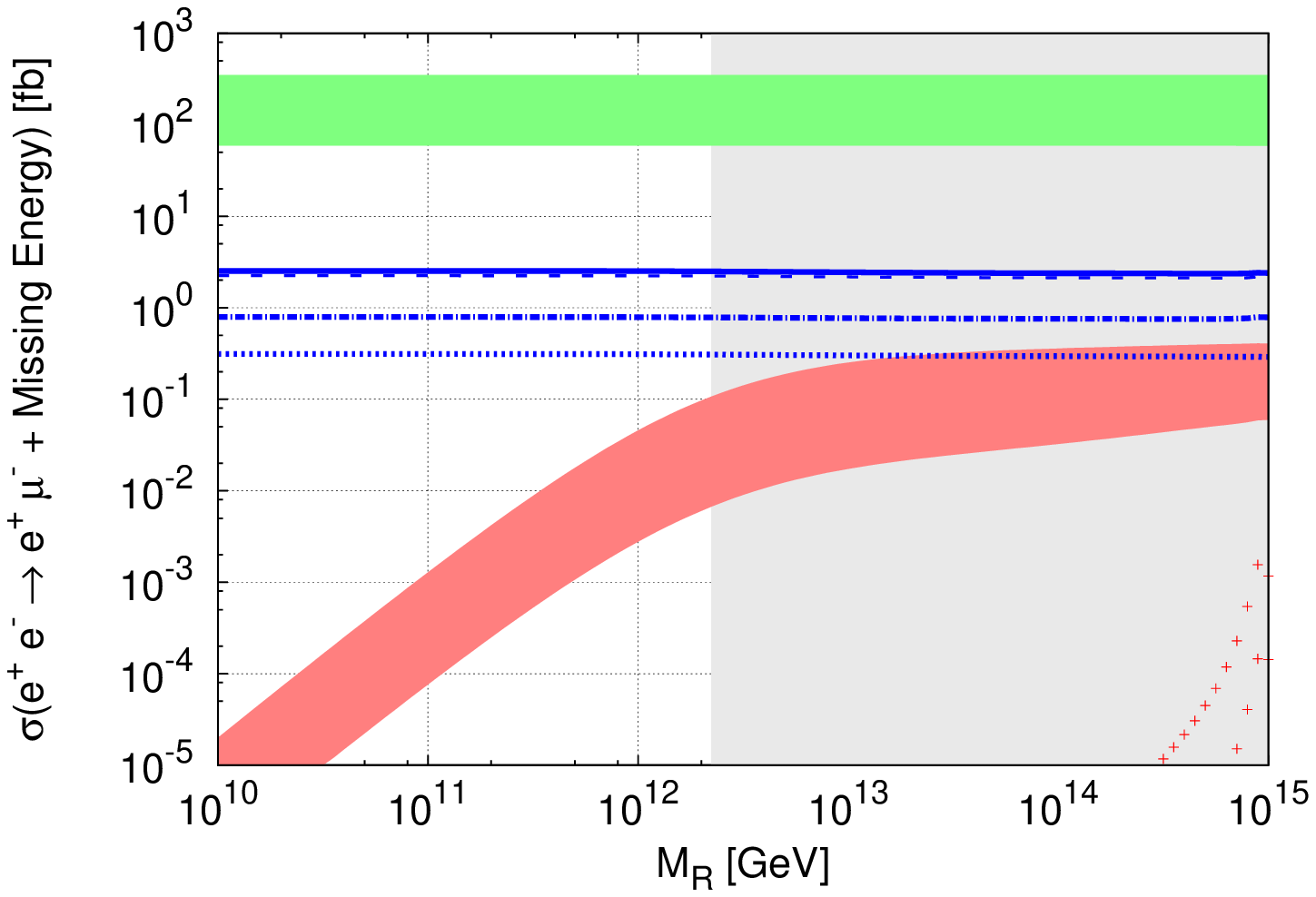, 
clip=, angle=0, width=75mm} 
\\
\epsfig{file=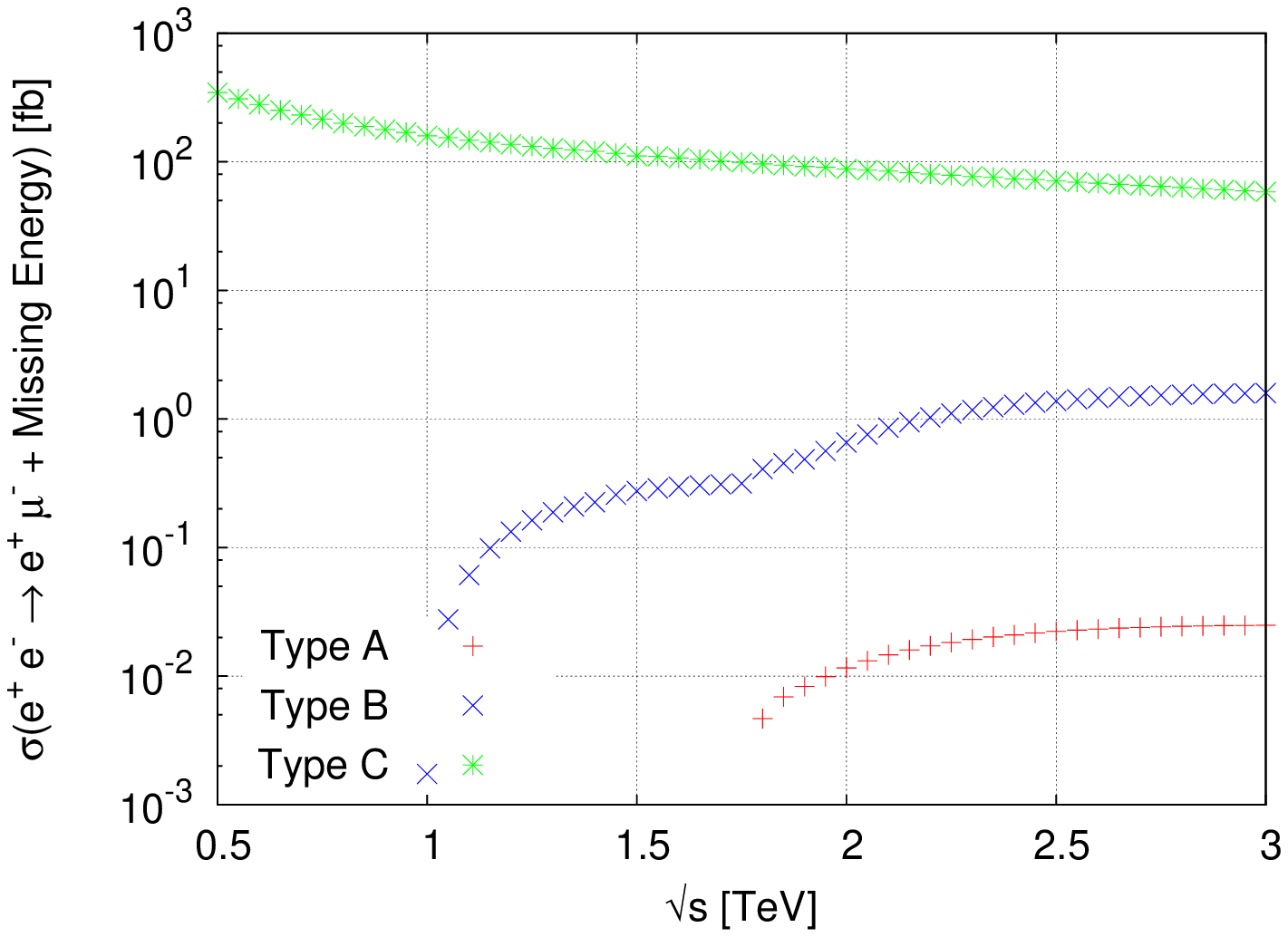, 
clip=, angle=0, width=75mm}
&
\epsfig{file=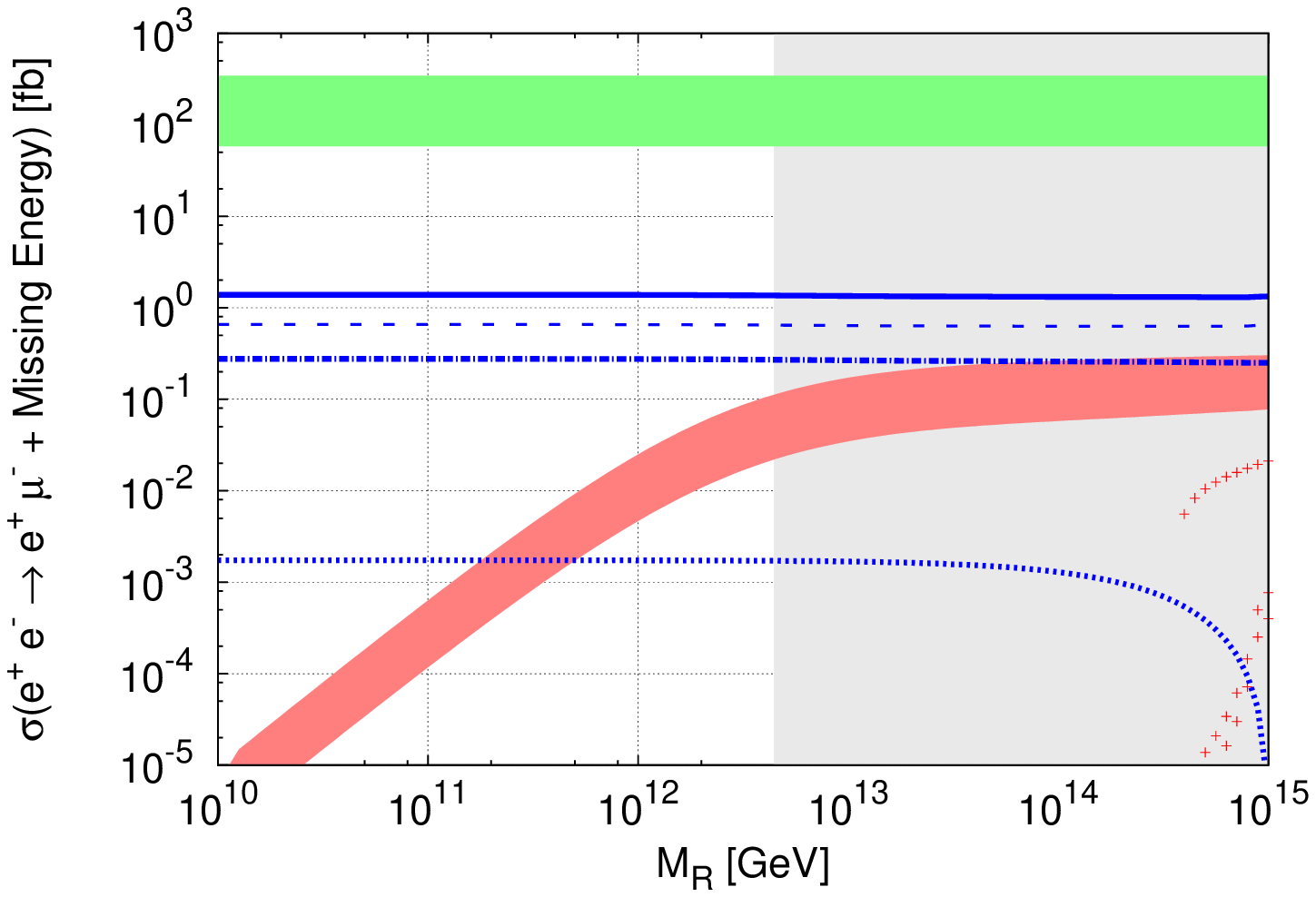, 
clip=, angle=0, width=75mm}
\end{tabular}
\caption{Same as in Fig.~\ref{fig:Clh:Msqrts}, but for points 
F-light and F-heavy (upper and lower
panels, respectively), the only difference being that a full line is
now associated to $\sqrt s=2.5$ TeV. }
\label{fig:Flh:Msqrts}
\end{center}
\end{figure}

In Fig.~\ref{fig:Flh:Msqrts}, we display analogous studies, but now
for points F-light and F-heavy.  
The prospects are very similar to
those of points C, the only relevant difference being that
for all experimentally compatible $M_R$ and all 
$\sqrt s$ regimes, the cross sections for the 
SUSY background (B) are much larger than those of the signal, (A). 
For the same nominal values of the integrated
luminosity, the
expected number of events would be however much smaller than for
points C:
$\mathcal{O}(50 (300))$ for F-light and $\mathcal{O}(10(100))$ for
F-heavy, for $\mathcal{L}=$ 0.5 (3)
ab$^{-1}$.
Since points F are associated with excessively large contributions to 
BR($B_s \to \mu^+ \mu^-$), and thus experimentally disfavoured, 
we will focus most of the subsequent analysis on points C-light and
C-heavy.

\begin{figure}[ht!]
\begin{center}
\begin{tabular}{cc}
\epsfig{file=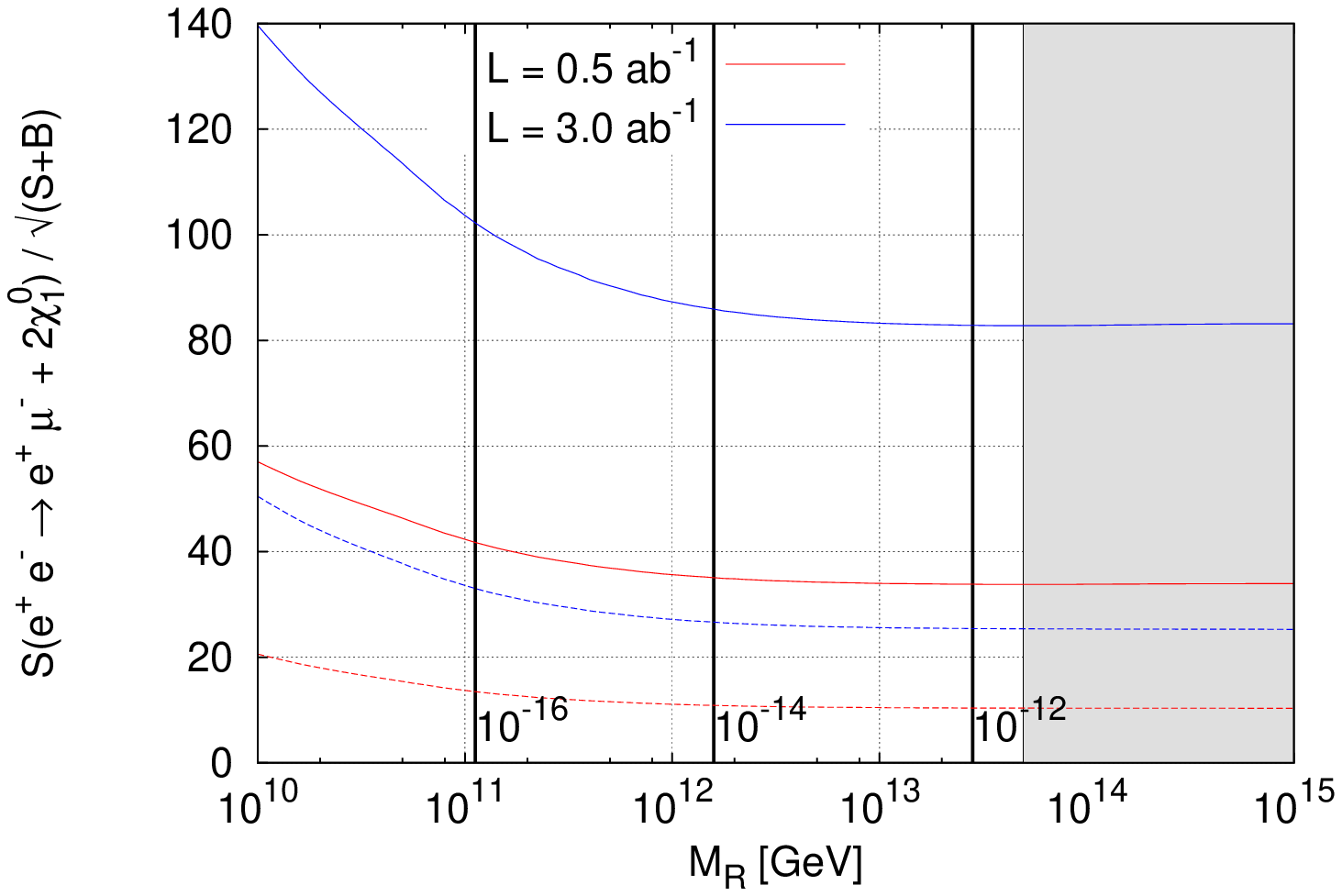, 
clip=, angle=0, width=75mm}
&
\epsfig{file=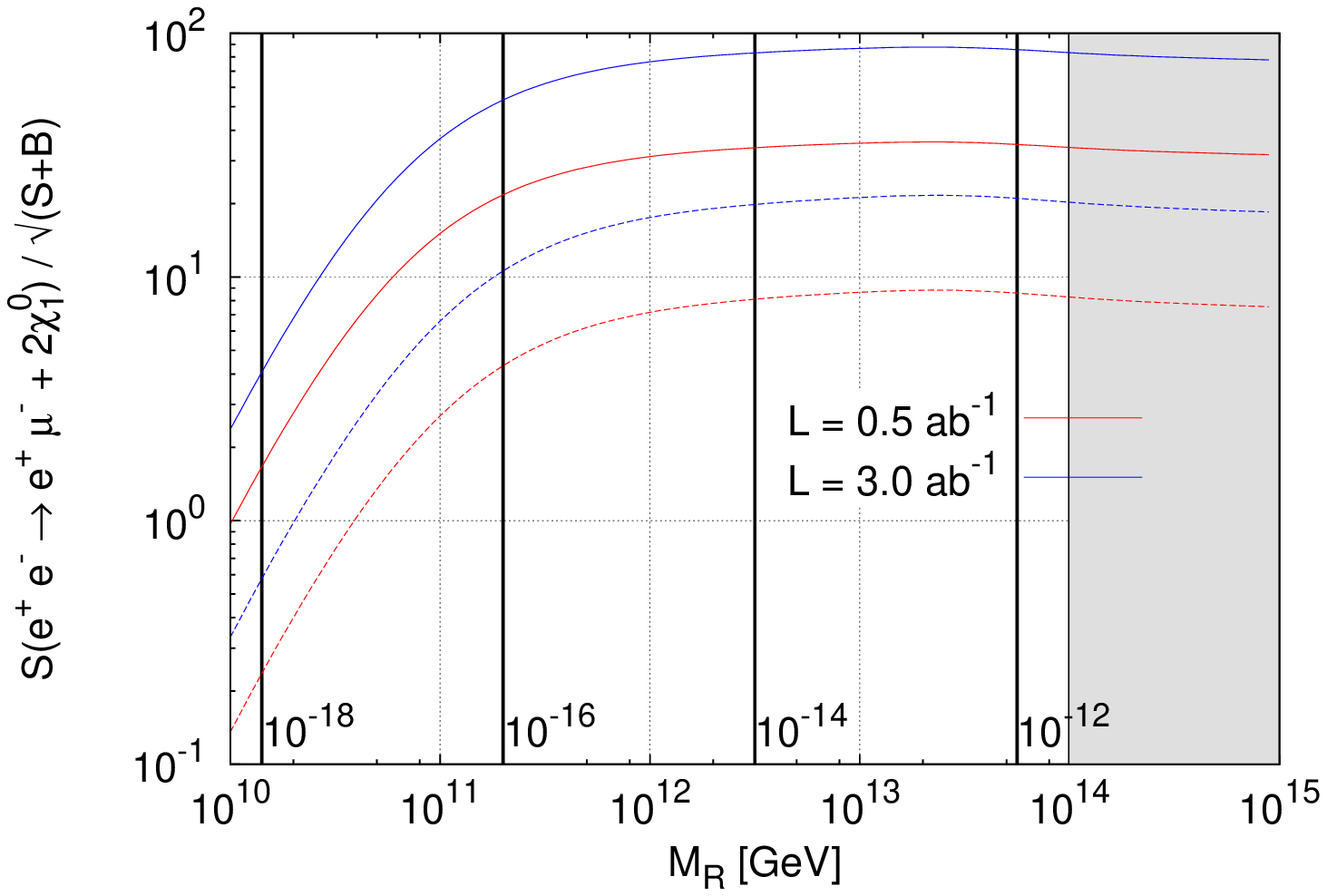, 
clip=, angle=0, width=75mm} 
\end{tabular}
\caption{Significance of the signal for points C-light (left) and
  C-heavy (right), as a function of the seesaw scale ($M_R$), 
  for $\sqrt s$=2 TeV. We assume nominal values for the integrated
  luminosity of $0.5$ and 3 ab$^{-1}$ (respectively red and blue
  curves). The full/dashed lines correspond to a SUSY/SM + SUSY background, i.e., 
  (B)/(B+C). Vertical lines denote the $M_R$-corresponding value of BR($\mu \to e \gamma$)
  while the (grey) shaded region represents values of $M_R$ 
  already excluded by the present experimental bound on BR($\mu \to e \gamma$). 
  Same assumptions for the remaining parameters as in Fig.~\ref{fig:Clh:Msqrts}.}
\label{fig:Clh:Sig}
\end{center}
\end{figure}

For completeness, we display in Fig.~\ref{fig:Clh:Sig} the
significance of the cLFV seesaw signal (cf. Eq.~(\ref{eq:sigma:def})) as
a function of the seesaw scale, for points C-light and C-heavy. For
illustrative purposes, we have chosen $\sqrt s=2$ TeV. 
As is manifest, for the case of point C-light,
one expects that the significance of the events should be no smaller
than 20 (10) for the case of luminosity of 3 (0.5) ab$^{-1}$,
throughout the considered interval for $M_R$. By itself, and in the
absence of dedicated cuts that would in principle allow to enhance the
significance, this is an extremely promising result, in the sense
that the observation of seesaw-induced cLFV events appears to be
potentially feasible at a Linear Collider. As expected, the
situation is slightly worse for a C-heavy spectrum;
nevertheless, a significance close to 10 is potentially within reach
for the considered integrated luminosities.

\bigskip
For both cases (points C and F),
considering a hierarchical RH neutrino spectrum does not have a
significant impact on the results displayed before, other than the
well-known effects on cLFV (see, e.g.~\cite{Raidal:2008jk}), such as a small
reduction of the amount of flavour violation in the $e-\mu$ for the
same choice of $M_{N_3}=M_R$. 

Let us recall here that these results correspond to the 
case where there are no sources of flavour violation other than the
$U_\text{MNS}$, i.e. the limit $R=1$. This indeed
corresponds to a ``conservative'' limit for flavour violation:
apart from possible cancellations, and for a fixed
seesaw scale, this limit typically provides
a lower bound for the amount of the generated LFV.
For the present analysis, taking generic (complex) values\footnote{We
  notice that for the case of degenerate right-handed neutrinos, any
  real $R$ matrix is equivalent to taking $R=1$, as can be seen from
  Eq.~(\ref{eq:seesaw:casas}).} of  
$R_{ij}$ would lead to an increase in 
the cross sections; however, this would also generate large
contributions to low-energy cLFV observables (namely $\mu \to e
\gamma$), strongly reducing the available parameter space.

\bigskip
We now consider how polarising the beams can help resolve the signal. 
As discussed before, if LFV charged currents - both arising from the SM
background (C) and SUSY background (B) - can be eliminated, any remaining 
cLFV signal such as those of (A) would clearly point towards 
flavour mixing in the slepton sector 
that would be manifest in neutral currents.

\begin{figure}[ht!]
\begin{center}
\begin{tabular}{cc}
\epsfig{file=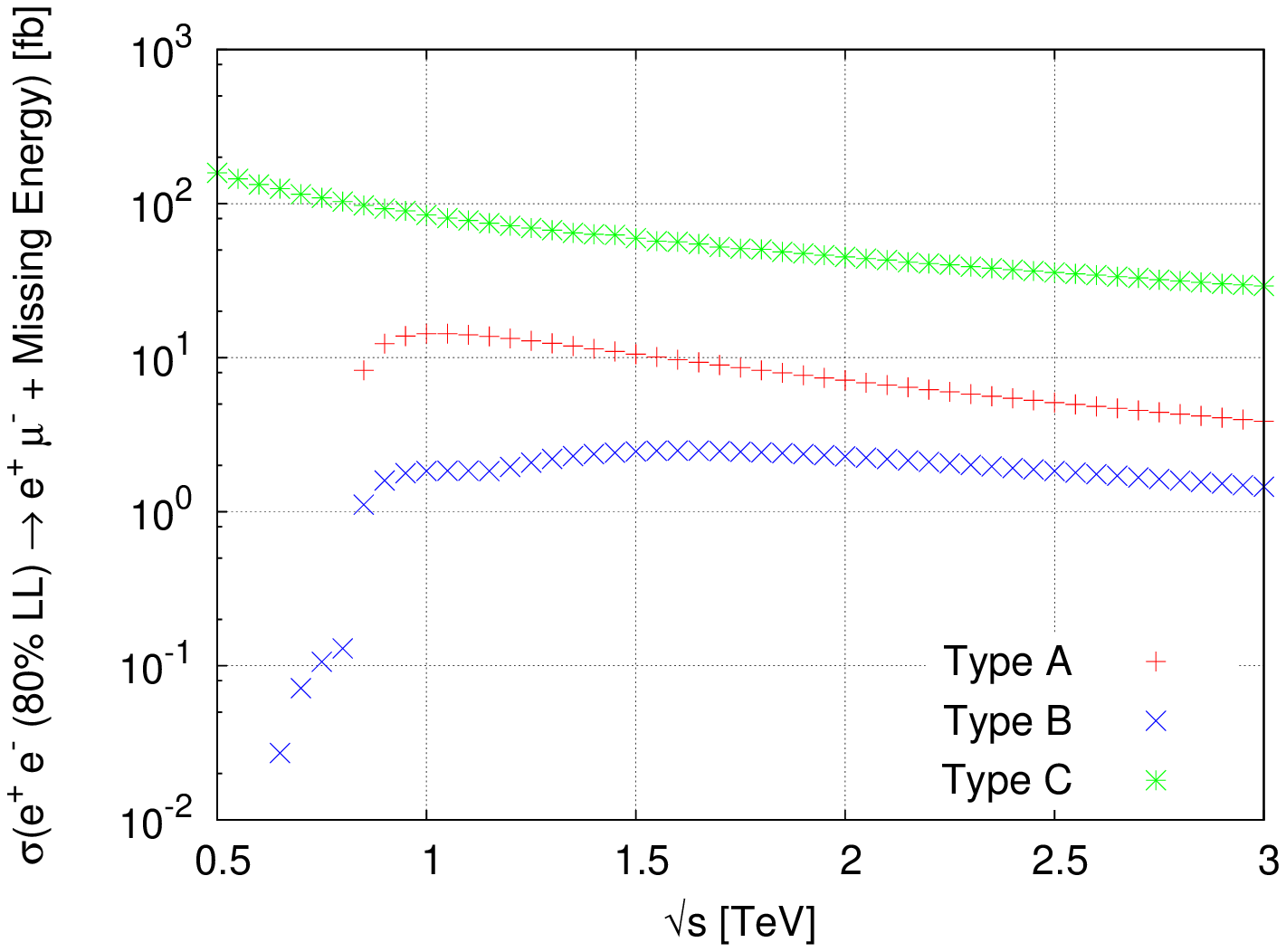, 
clip=, angle=0, width=75mm}
&
\epsfig{file=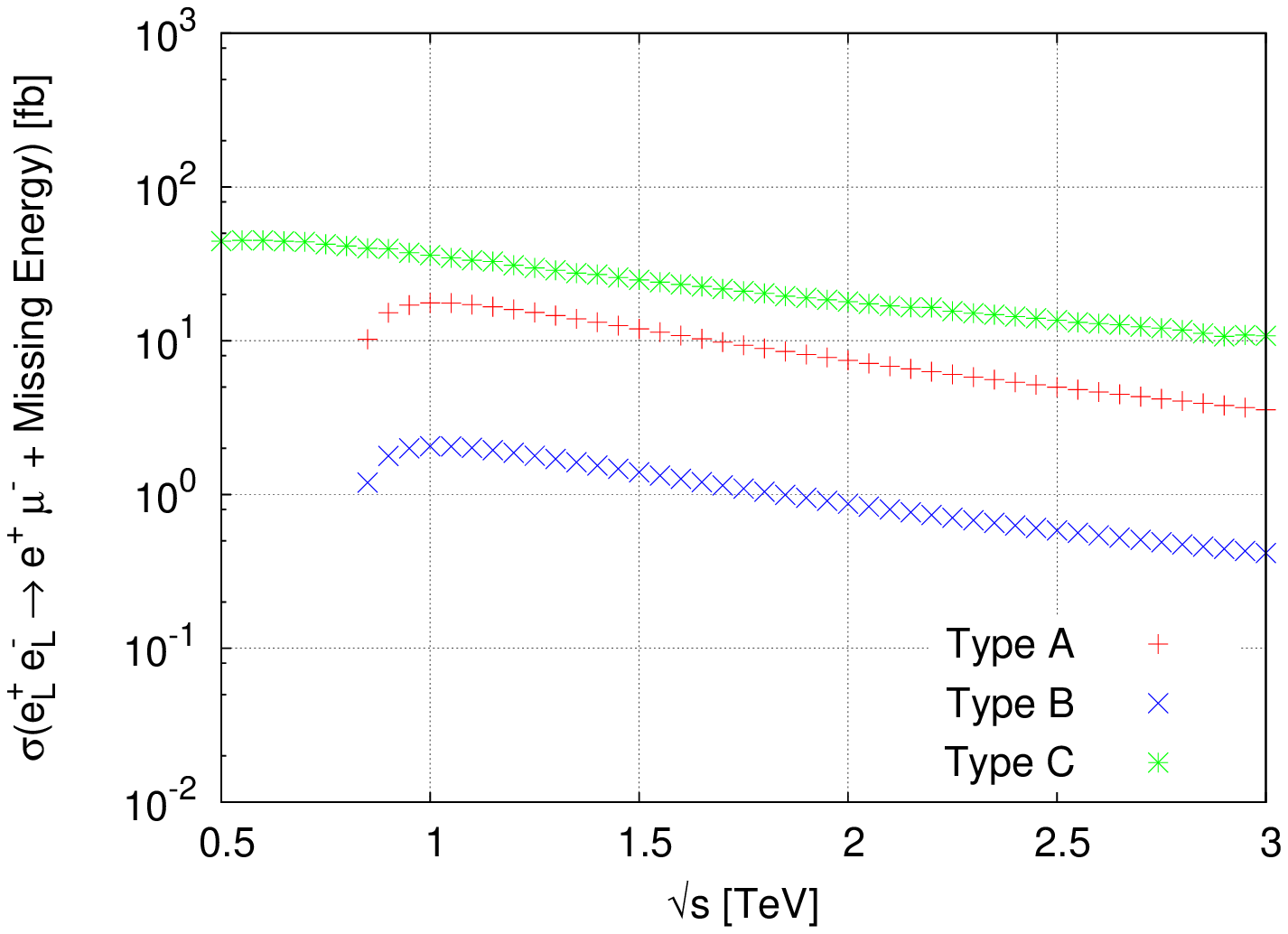, 
clip=, angle=0, width=75mm} 
\\
\epsfig{file=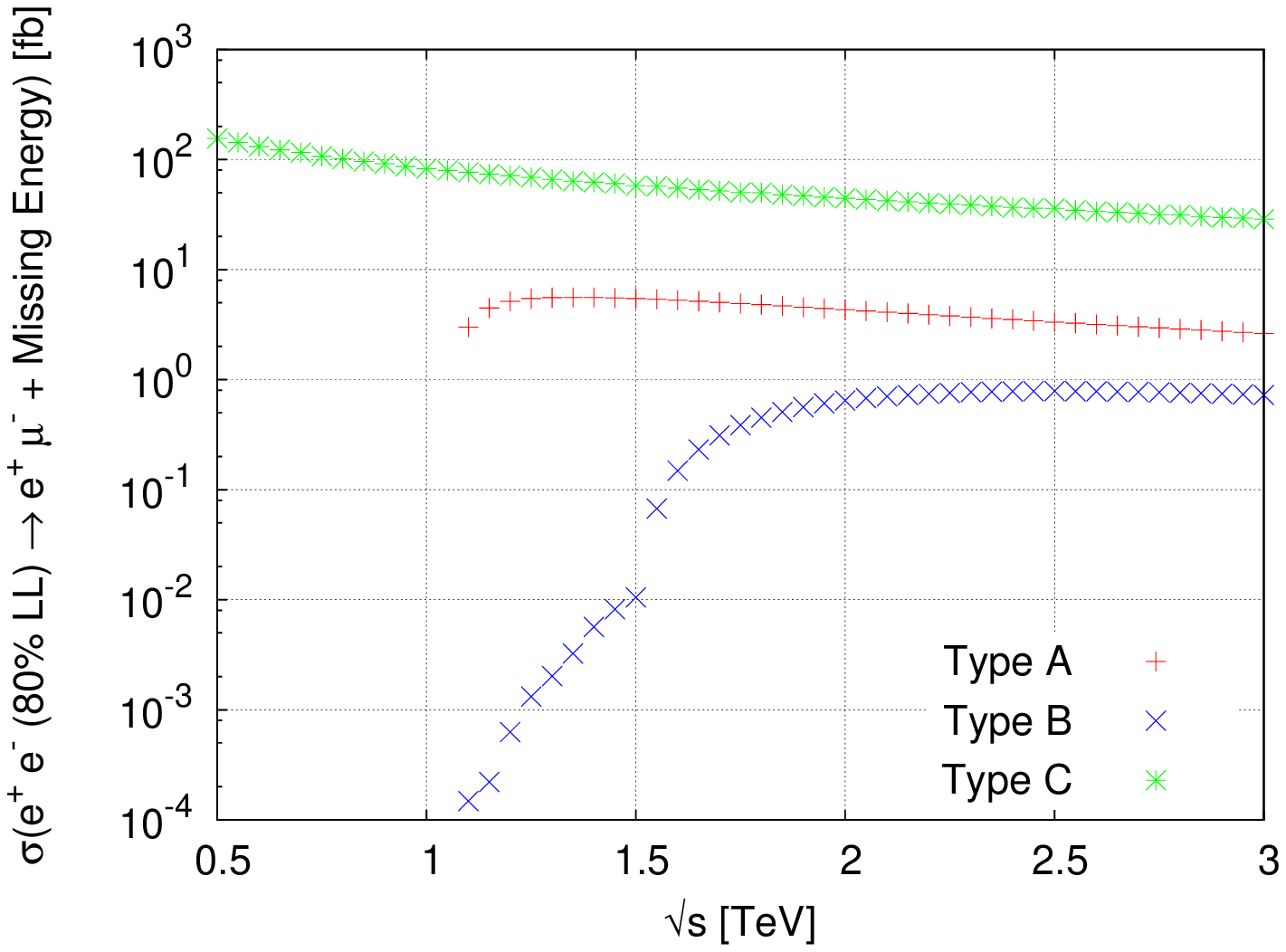, 
clip=, angle=0, width=75mm}
&
\epsfig{file=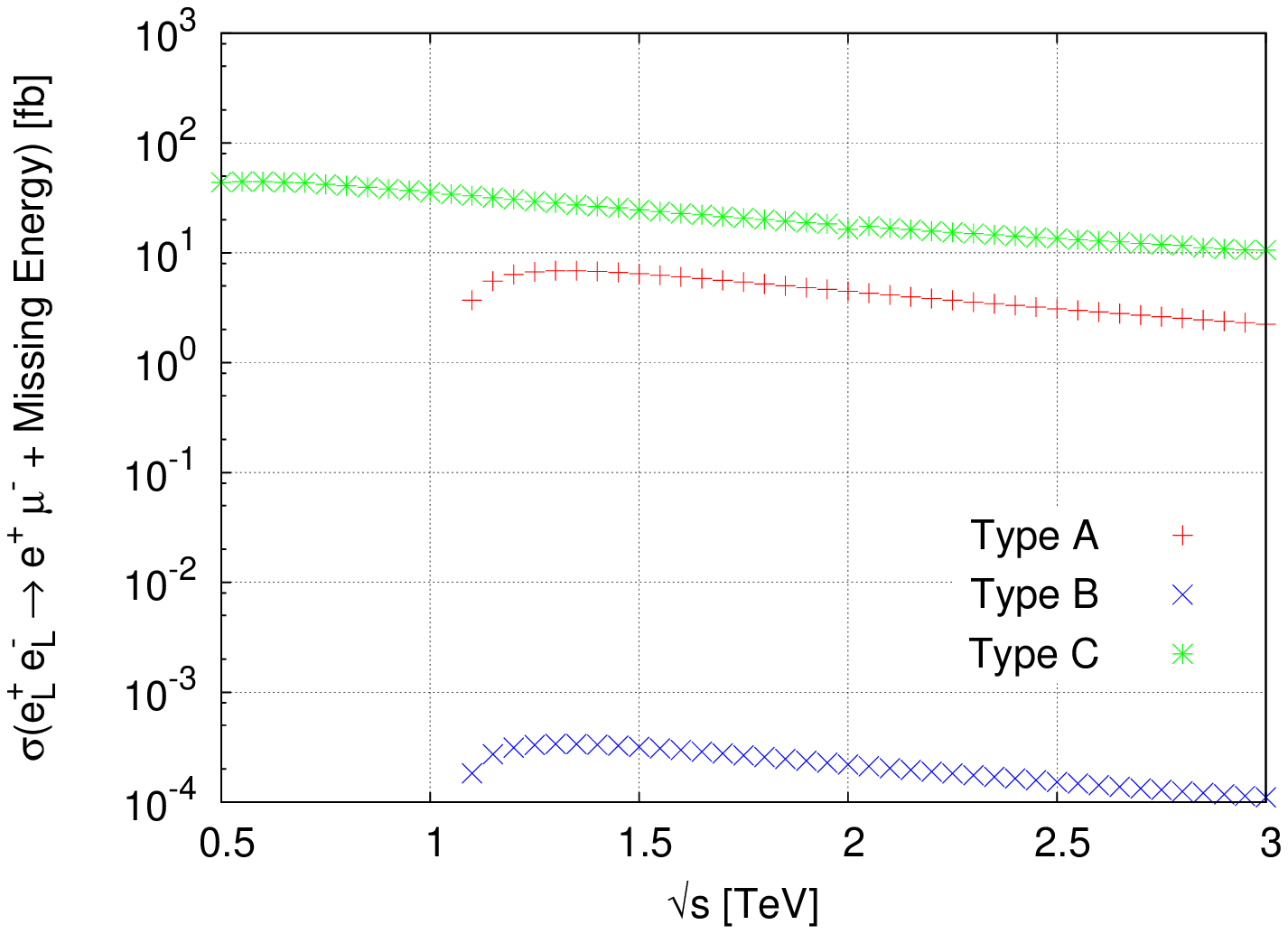, 
clip=, angle=0, width=75mm}
\end{tabular}
\caption{Cross section for $e^+ e^- \to e^+
\mu^- + E^T_\text{miss}$ {(with $E^T_\text{miss}=2 \chi_1^0, 2 \chi_1^0
+  (2,4)\nu, (2,4) \nu$)}, for points C-light and C-heavy (upper and lower
panels, respectively), as a function of the  centre of mass energy,
$\sqrt{s}$, for polarised beams. Left panels:
$(P_{e^+},P_{e^-})=(-80\%,-80\%)$; right panels
$(P_{e^+},P_{e^-})=(-100\%,-100\%)$. 
We fix $M_R = 10^{12}$ GeV, and again denote 
the signal (A) with red crosses, the SUSY charged current background
(B) with blue times, and the SM charged current background (C) by 
green asterisks.  
In both cases we have taken a degenerate right-handed neutrino spectrum,
and set $\theta_{13}=10^\circ$.}
\label{fig:Clh:Msqrts:pol80-100}
\end{center}
\end{figure}

In what follows, and in addition to the ``ideal'' polarisation
scenario, i.e. a full LL polarisation
of $(P_{e^+},P_{e^-})=(-100\%,-100\%)$, we consider a more realistic
(but still optimistic) case where one could have 80\% polarisation for
both beams, $(P_{e^+},P_{e^-})=(-80\%,-80\%)$. 
In Fig.~\ref{fig:Clh:Msqrts:pol80-100} we display the resulting cross
sections, for points C-light and C-heavy. Notice that 
due to the SU(2)$_L$ nature of the dominant $W$-strahlung background,
LL polarisation cannot fully remove these contributions. 
Otherwise, polarising the electron and positron beams 
would significantly enhance the
signal with respect to the remaining SM and other SUSY background. 
For instance, for point C-light, 80\% beam polarisation would 
already allow to have
a dominant cLFV signal, while in both C points, fully polarised beams
would reduce the background stemming from processes other than 
$W$-emission.
If dedicated detector simulations - with appropriated cuts - 
can indeed reduce the SM background (especially the dominant
contributions due to single- and double-$W$ emissions), and thus 
isolate the SUSY contributions, polarised electron and
positron beams would allow to obtain a signal of
cLFV, as possibly induced by a SUSY seesaw.
Regarding the expected number of events in the polarised cases, 
for $\sqrt s=$ 2 TeV, and 
$\mathcal{L}=0.5$ (3) ab$^{-1}$, one expects (in both
100\% and 80\% polarization cases),  
values of order of $5 \times 10^3$ ($3 \times 10^4$) for
C-light and $10^3$ ($10^4$) for point C-heavy.

The results displayed in Fig.~\ref{fig:Clh:Msqrts:pol80-100}
correspond to a fixed value of the (degenerate) right-handed neutrino
scale ($M_R = 10^{12}$ GeV). Conducting an analysis similar to that
leading to the right-handside 
panels of Figs.~\ref{fig:Clh:Msqrts} and~\ref{fig:Flh:Msqrts} 
(i.e., study of $e^+ e^- \to e^+
\mu^- + E^T_\text{miss}$ as a function of $M_R$), we have verified
that for sufficiently large values of $M_R$, which are still
compatible with the bounds from BR($\mu \to e \gamma$),
the signal cross section is larger than that of the
SUSY background; 
in particular, this occurs for spectra similar to
that of point C-heavy.
If, as discussed above, dedicated cuts allow to reduce the SM
backgrounds, then one could  also
infer some information concerning the scale of a possible underlying
type I seesaw. 

\begin{figure}[ht!]
\begin{center}
\begin{tabular}{cc}
\epsfig{file=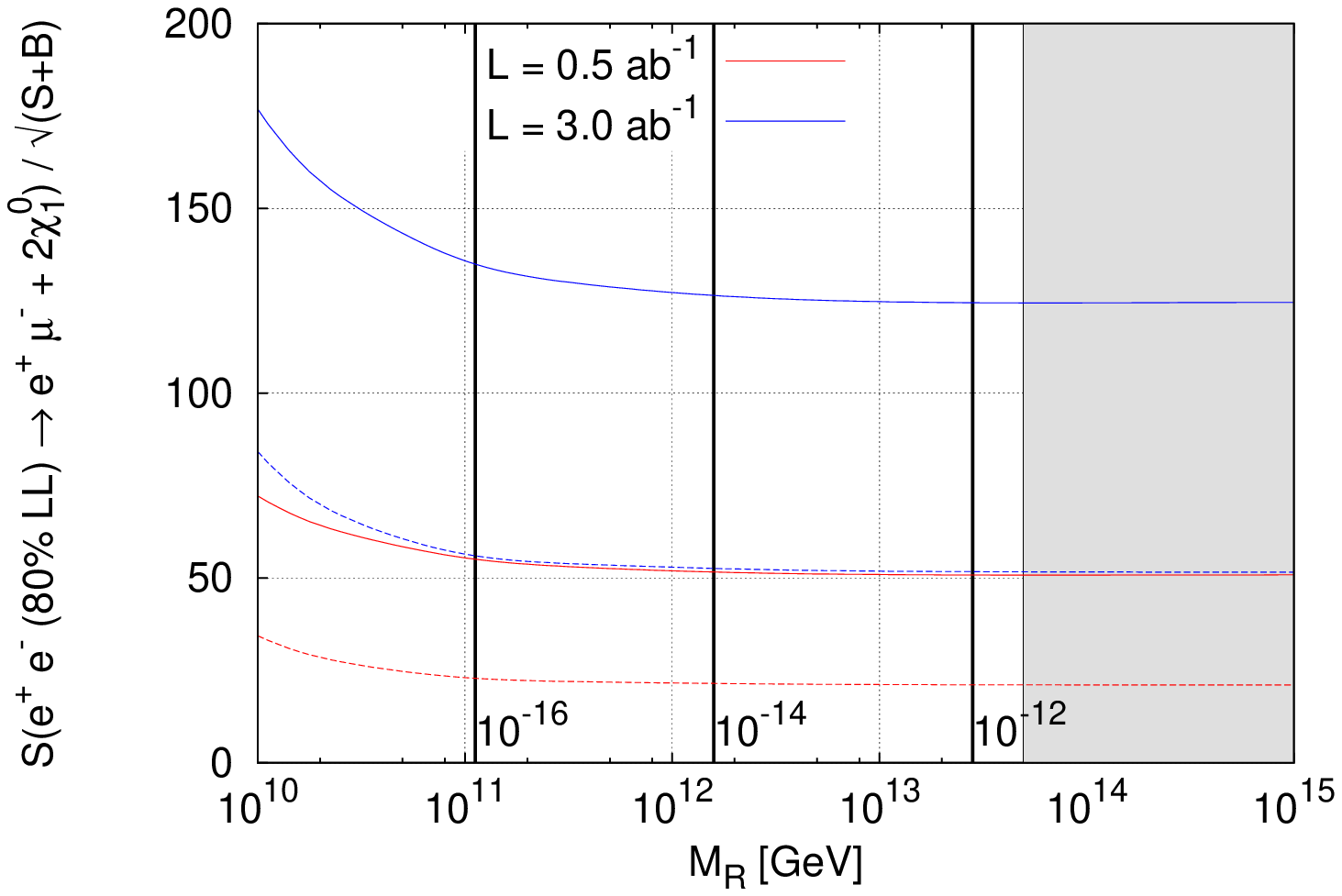,clip=, angle=0, width=75mm}
&
\epsfig{file=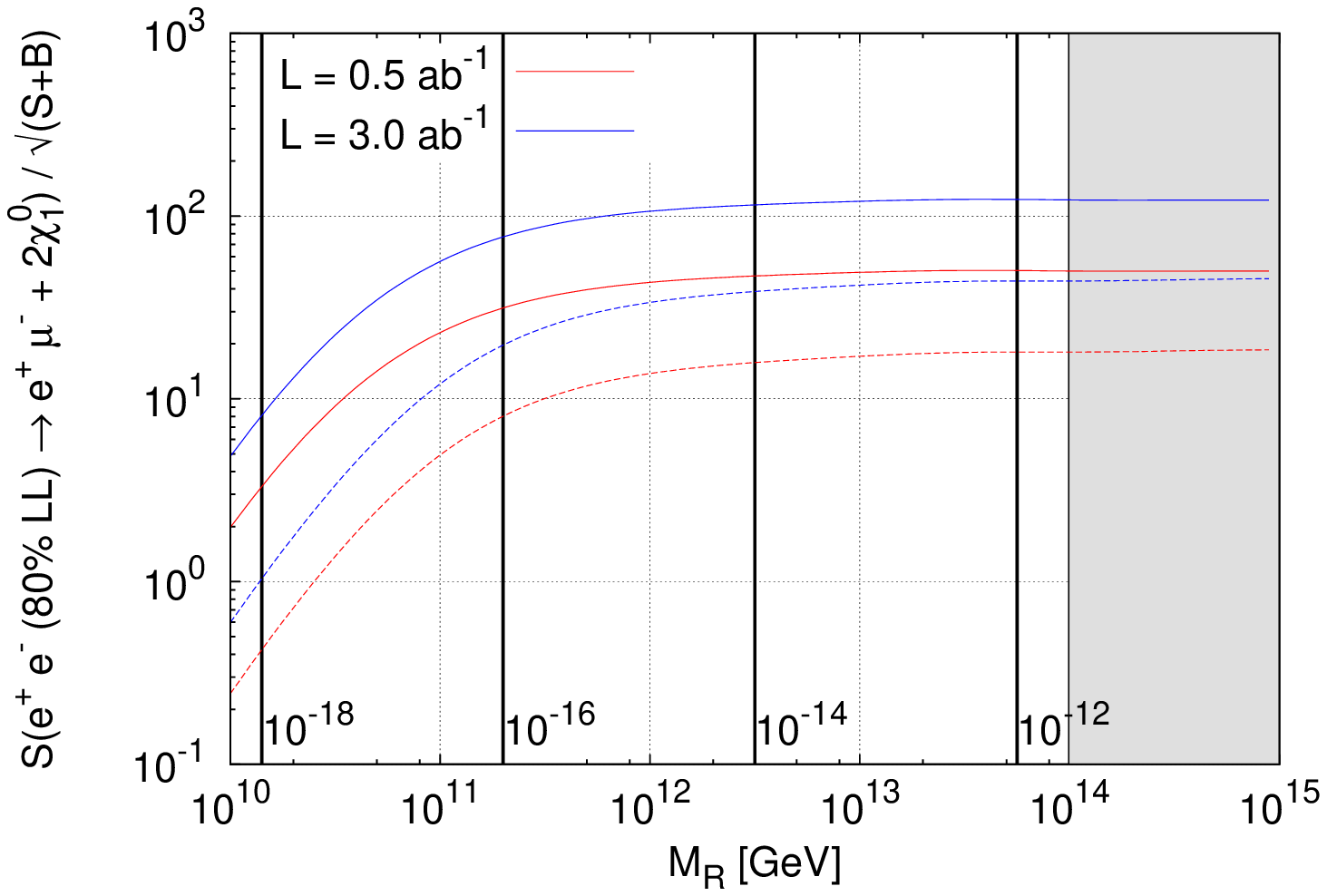,
  clip=, angle=0, width=75mm} 
\\
\epsfig{file=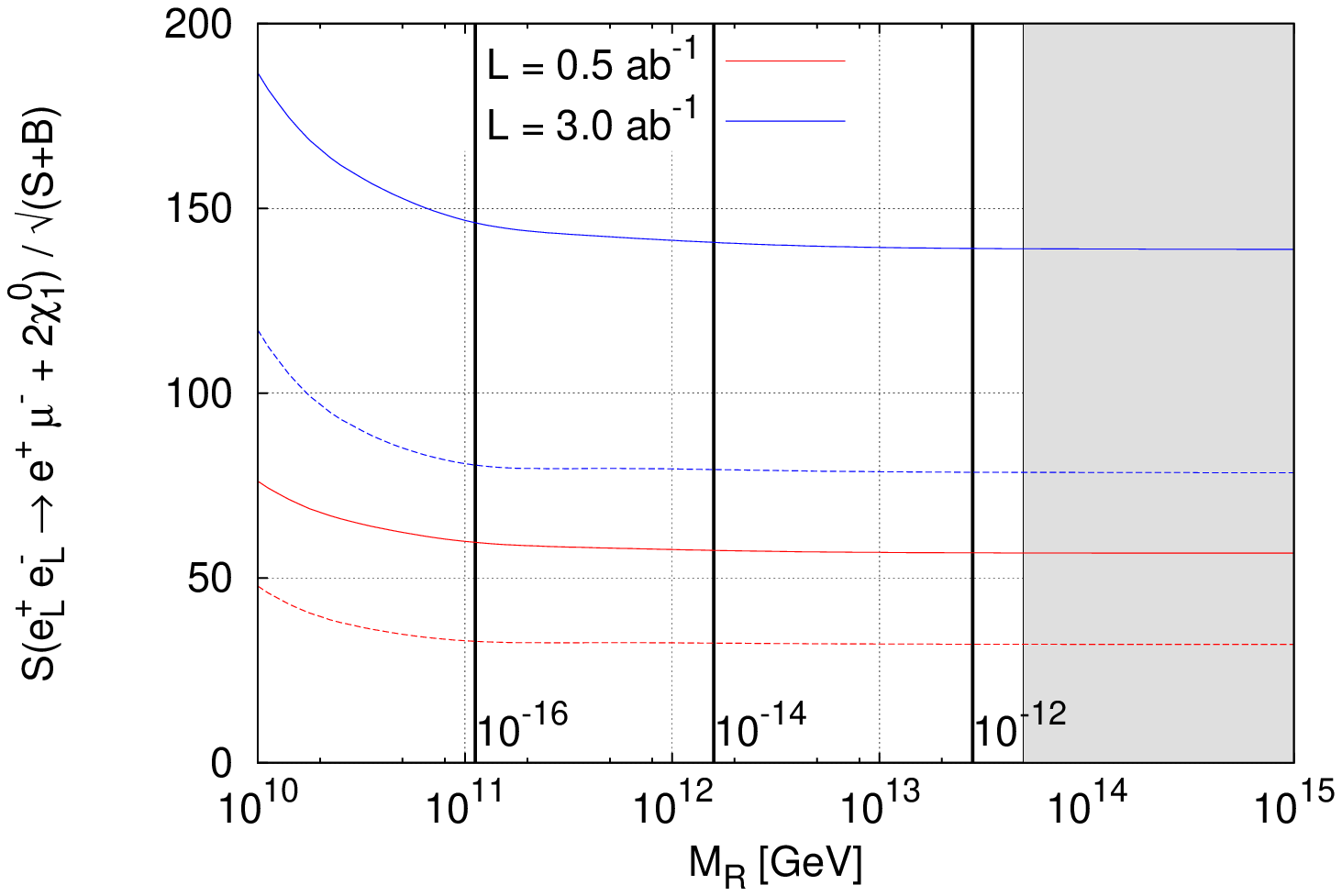,clip=, angle=0, width=75mm}
&
\epsfig{file=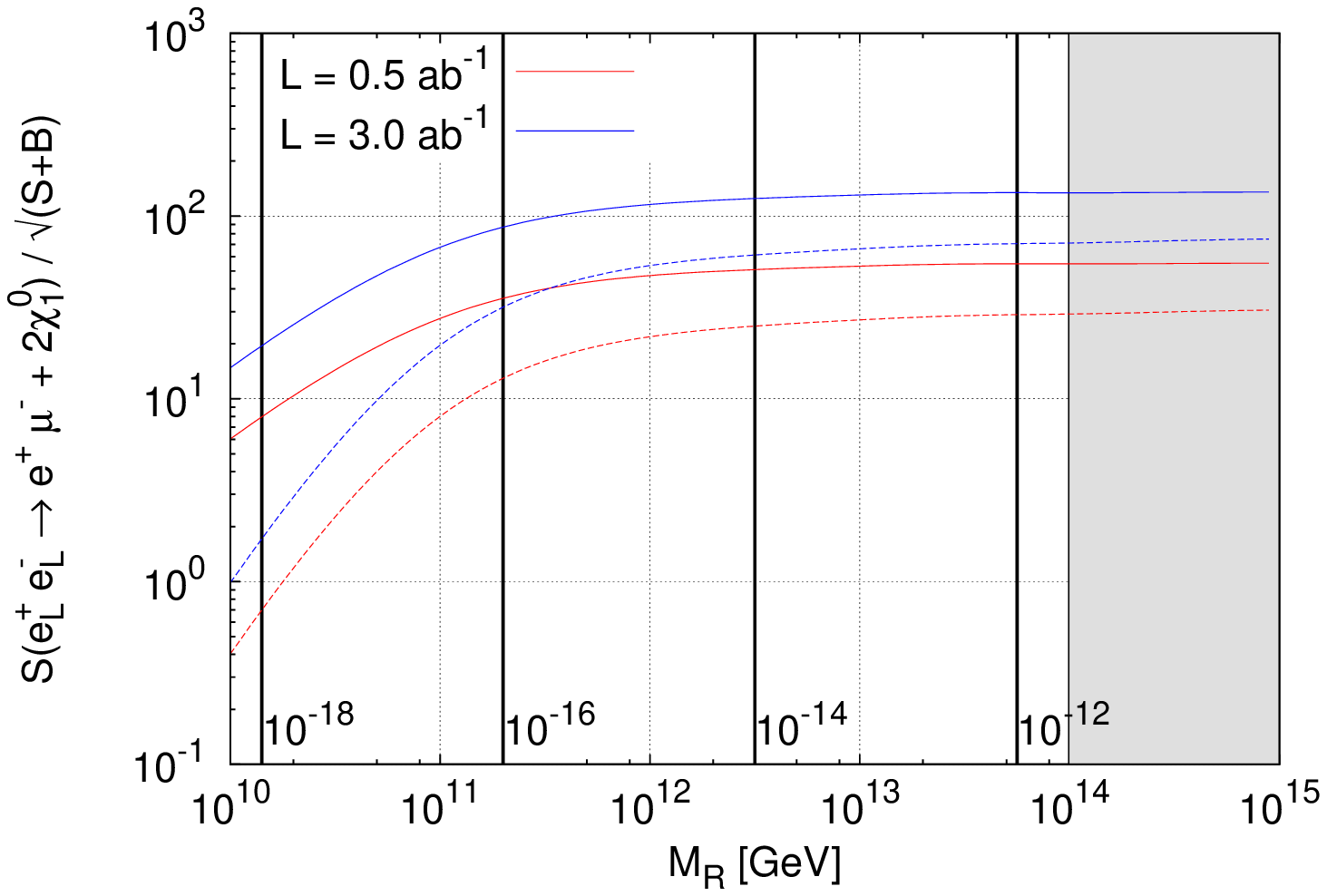,
  clip=, angle=0, width=75mm} 
\end{tabular}
\caption{Significance of the signal for points C-light (left) and
  C-heavy (right), as a function of the seesaw scale ($M_R$), 
  for $\sqrt s$=2 TeV, with 80\% (upper) and 100\% (lower panels), 
  beam polarisation. Line and colour
  code as in Fig.~\ref{fig:Clh:Sig}.}
\label{fig:Clh:80pol:Sig}
\end{center}
\end{figure}

Again, we complete this analysis by displaying in
Fig.~\ref{fig:Clh:80pol:Sig} 
the significance of the
signal for both points C in the case of an 80\%
 and in the ``ideal'' 100\% case beam
polarisation.
The results thus obtained further suggest that
cLFV will be indeed observable at a Linear Collider in $e^+ e^-$
collisions: under an 80\% positron and electron beam polarisation,
the significances are typically above 10, and can be even close to 100.

\subsection{$e^-e^-$ beam option}\label{sec:res:emem}
\begin{figure}[ht!]
\begin{center}
\begin{tabular}{cc}
\epsfig{file=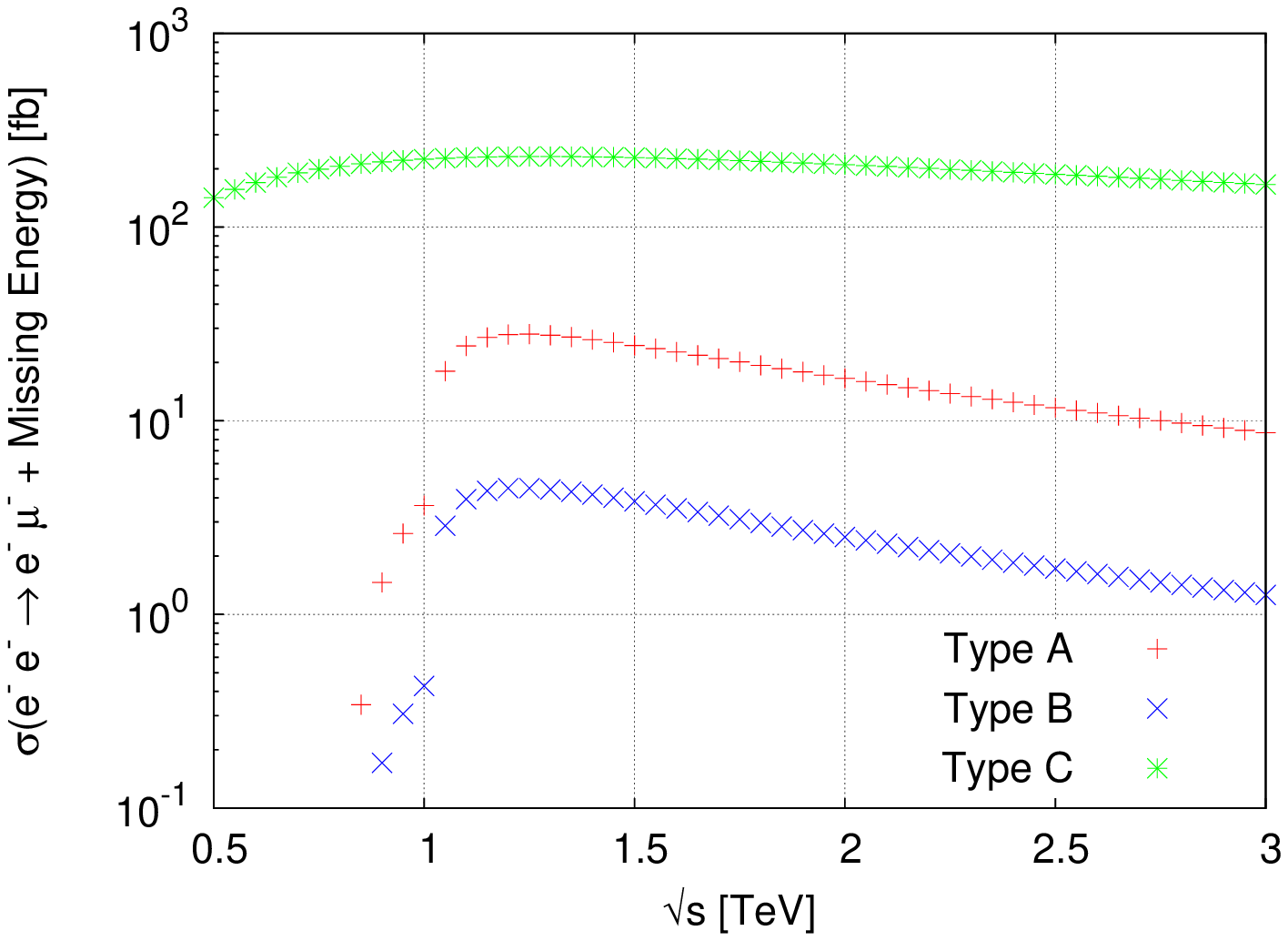, 
clip=, angle=0, width=75mm}
&
\epsfig{file=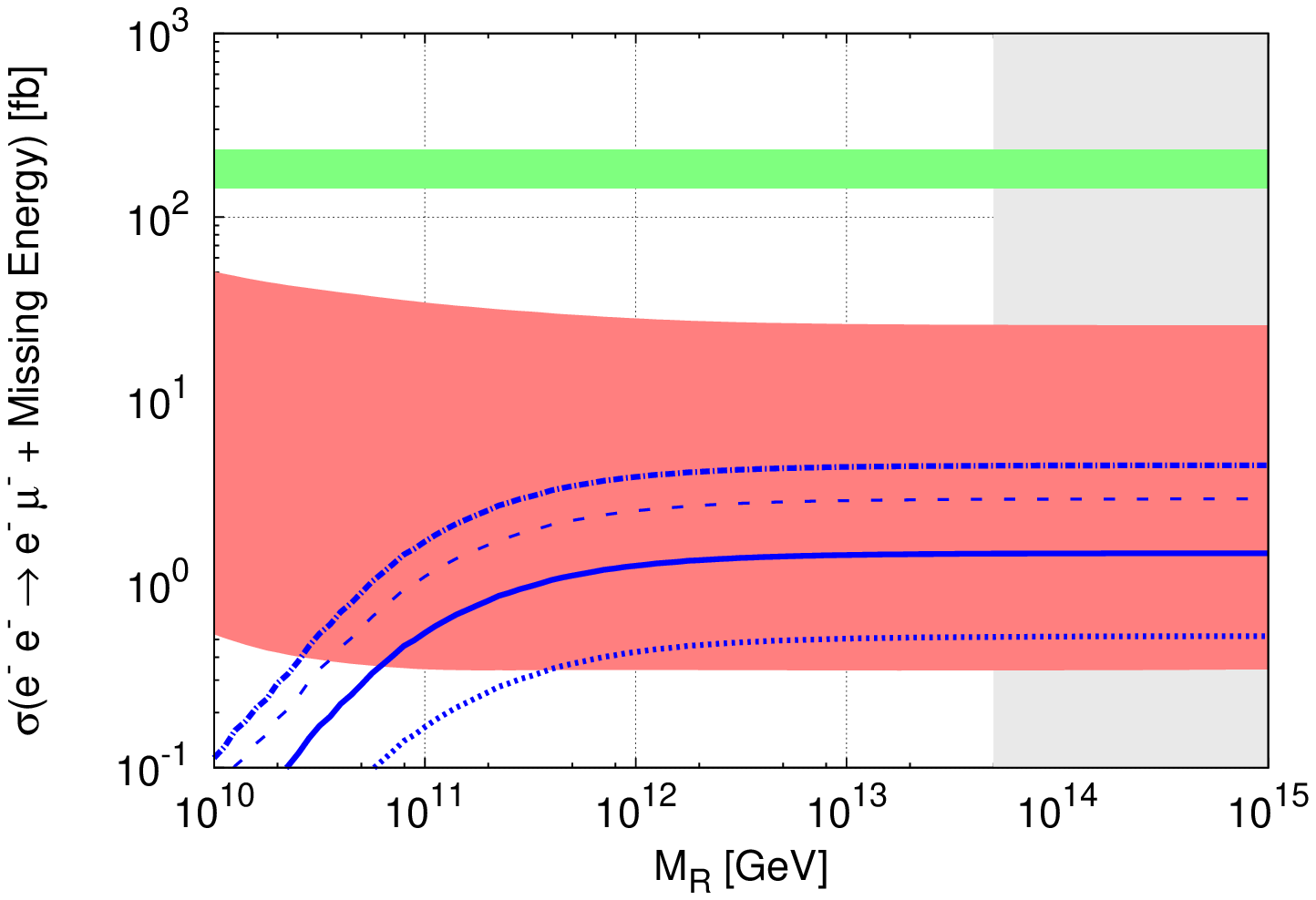, 
clip=, angle=0, width=75mm} 
\\
\epsfig{file=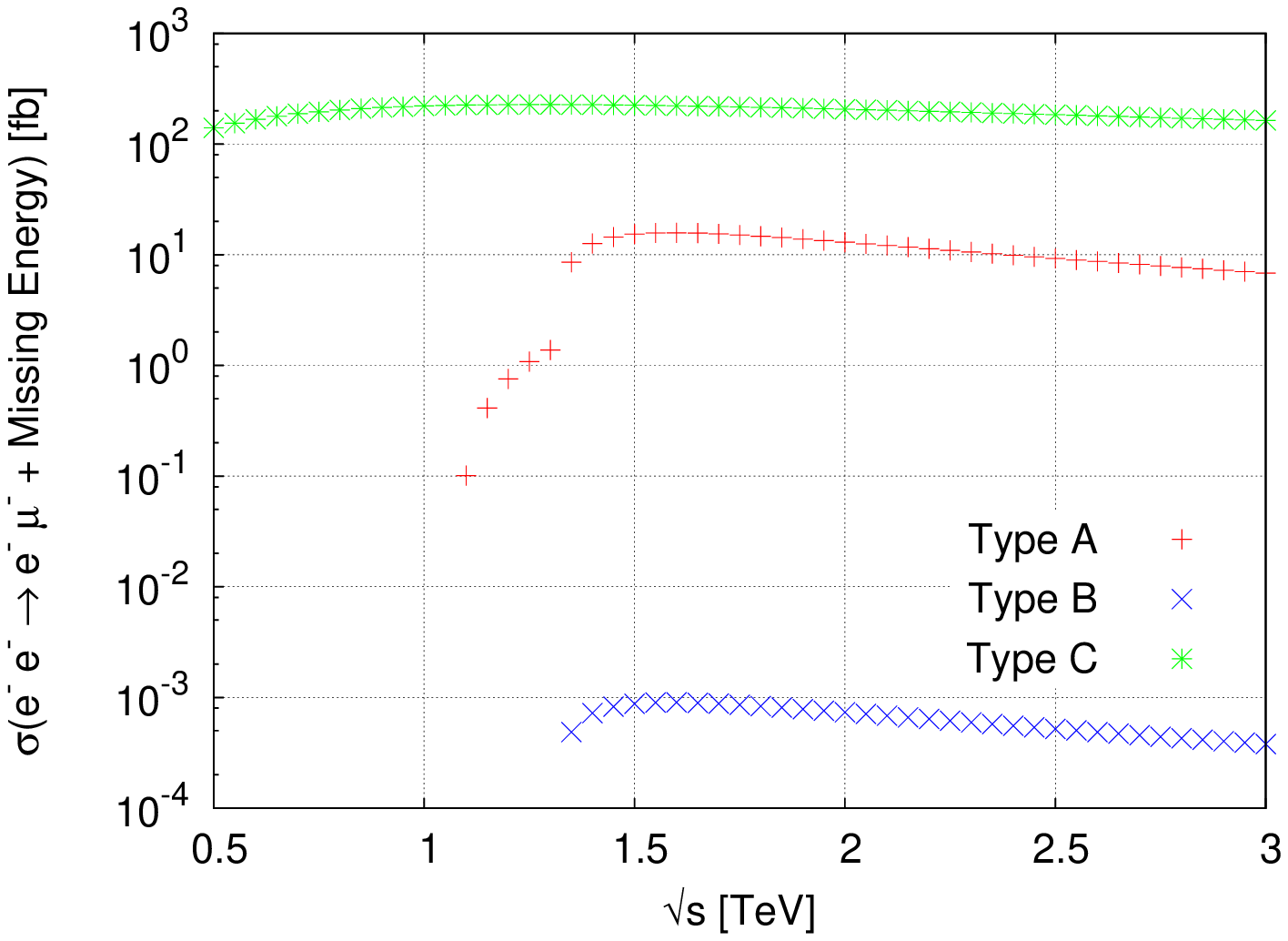, 
clip=, angle=0, width=75mm}
&
\epsfig{file=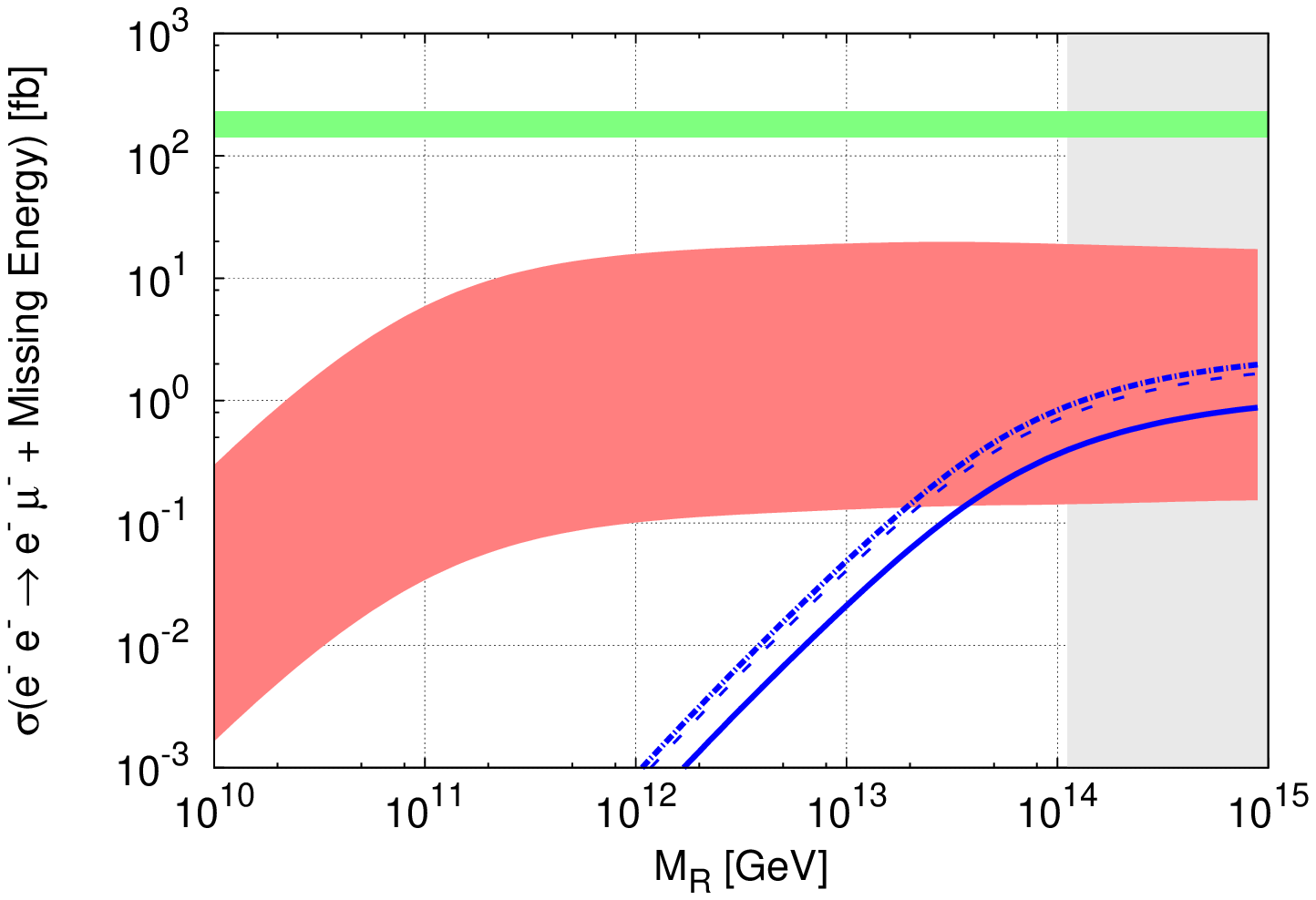, 
clip=, angle=0, width=75mm}
\end{tabular}
\caption{On the left, cross section for $e^- e^- \to e^-
\mu^- + E^T_\text{miss}$ (with { $E^T_\text{miss}=2 \chi_1^0, 2 \chi_1^0
+  (2,4)\nu, (2,4) \nu$}), for points C-light and C-heavy (upper and lower
panels, respectively), as a function of the  centre of mass energy,
$\sqrt{s}$. On the right, cross section for $e^- e^- \to e^-
\mu^- + E^T_\text{miss}$ ({with $E^T_\text{miss}=2 \chi_1^0, 2 \chi_1^0
+  (2,4)\nu, (2,4) \nu$}), for points C-light and C-heavy (upper and lower
panels, respectively), as a function of the right-handed
neutrino mass ($M_R$).
In both cases we have taken a degenerate right-handed neutrino spectrum,
and set $\theta_{13}=10^\circ$.
Line and colour code as in Fig.~\ref{fig:Clh:Msqrts}.}
\label{fig:Clh:Msqrts:emem}
\end{center}
\end{figure}

We now proceed to explore the very interesting LC feature of 
$e^-e^-$ collisions. In Fig.~\ref{fig:Clh:Msqrts:emem}, we display the 
cross section for $e^- e^- \to e^-
\mu^- + E^T_\text{miss}$ (with $E^T_\text{miss}=2 \chi_1^0, 2 \chi_1^0
+  (2,4)\nu, (2,4) \nu$), for points C-light and C-heavy (upper and lower
panels, respectively), as a function of the  centre of mass energy,
$\sqrt{s}$, and of the seesaw scale $M_R$.

\begin{figure}[ht!]
\begin{center}
\begin{tabular}{cc}
\epsfig{file=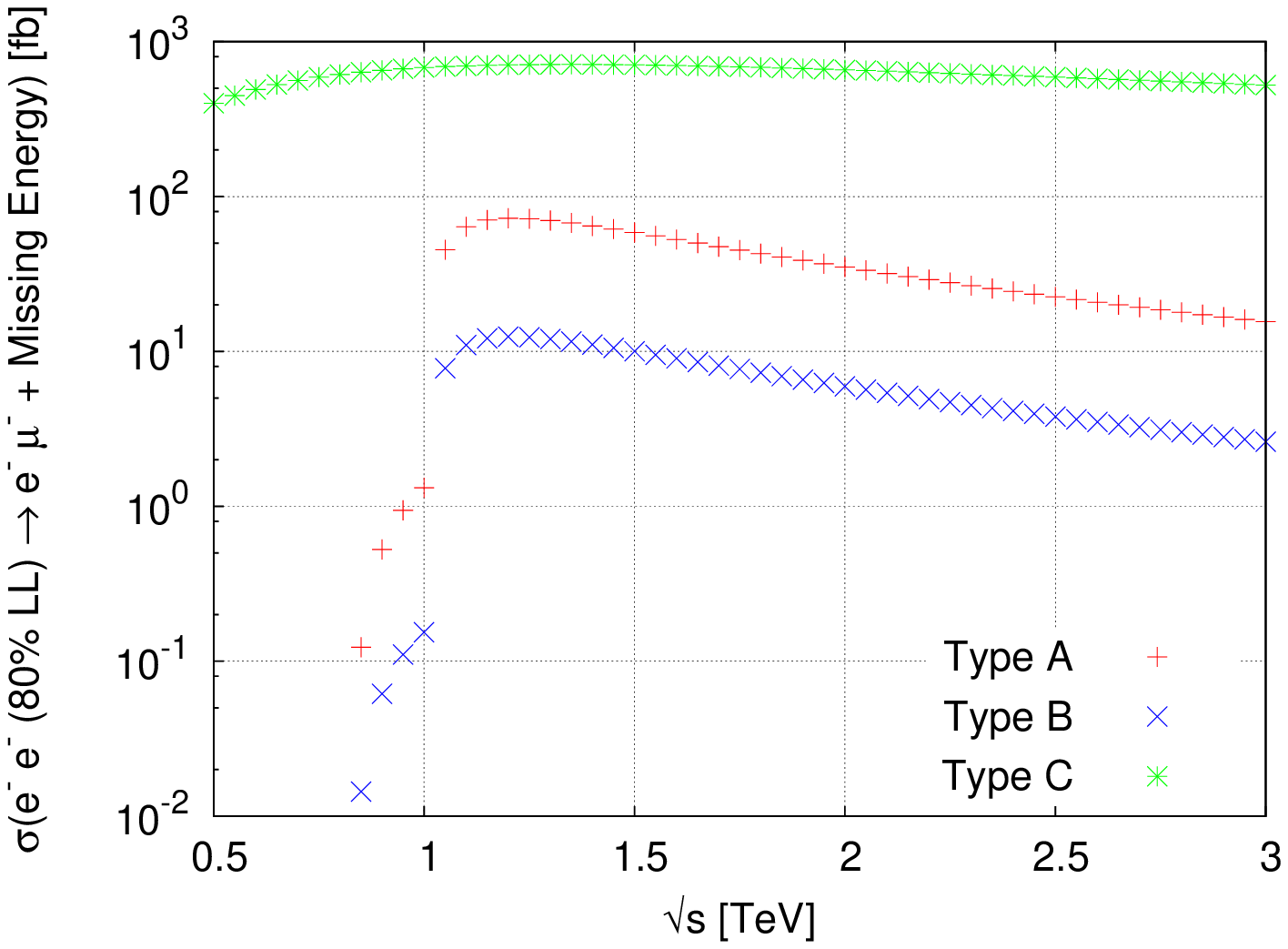, 
clip=, angle=0, width=75mm}
&
\epsfig{file=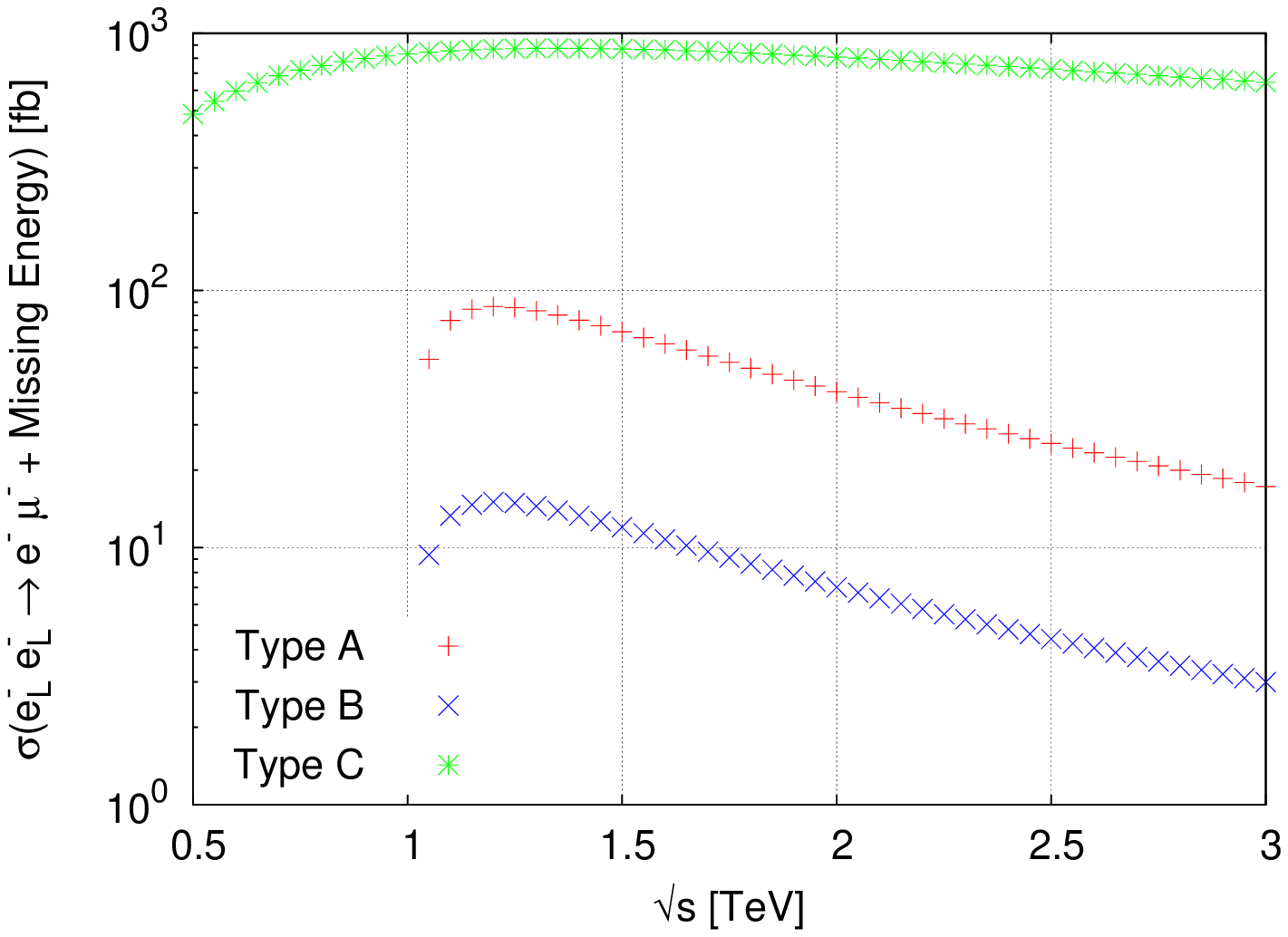, 
clip=, angle=0, width=75mm}\\
\epsfig{file=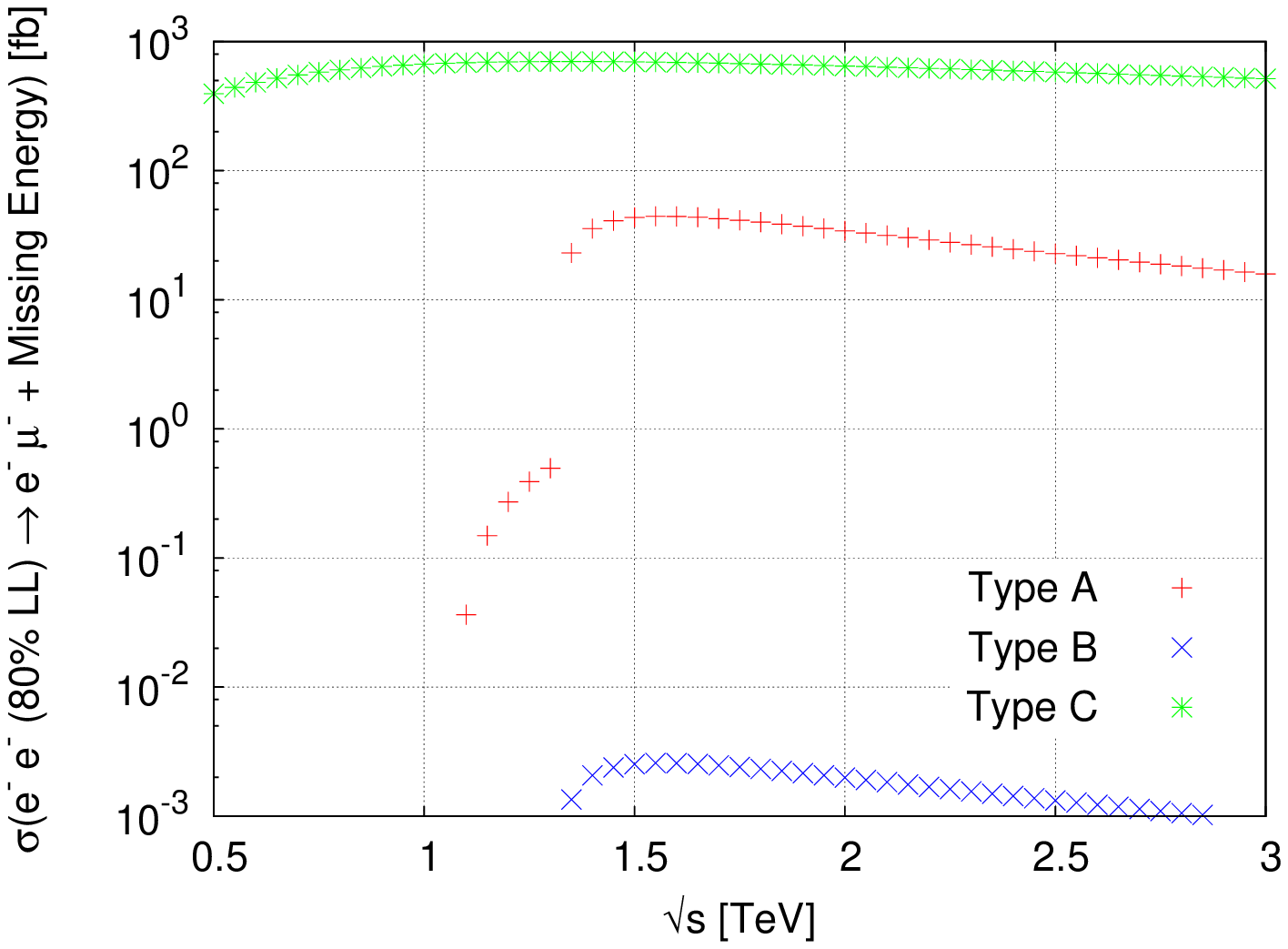, 
clip=, angle=0, width=75mm}
&
\epsfig{file=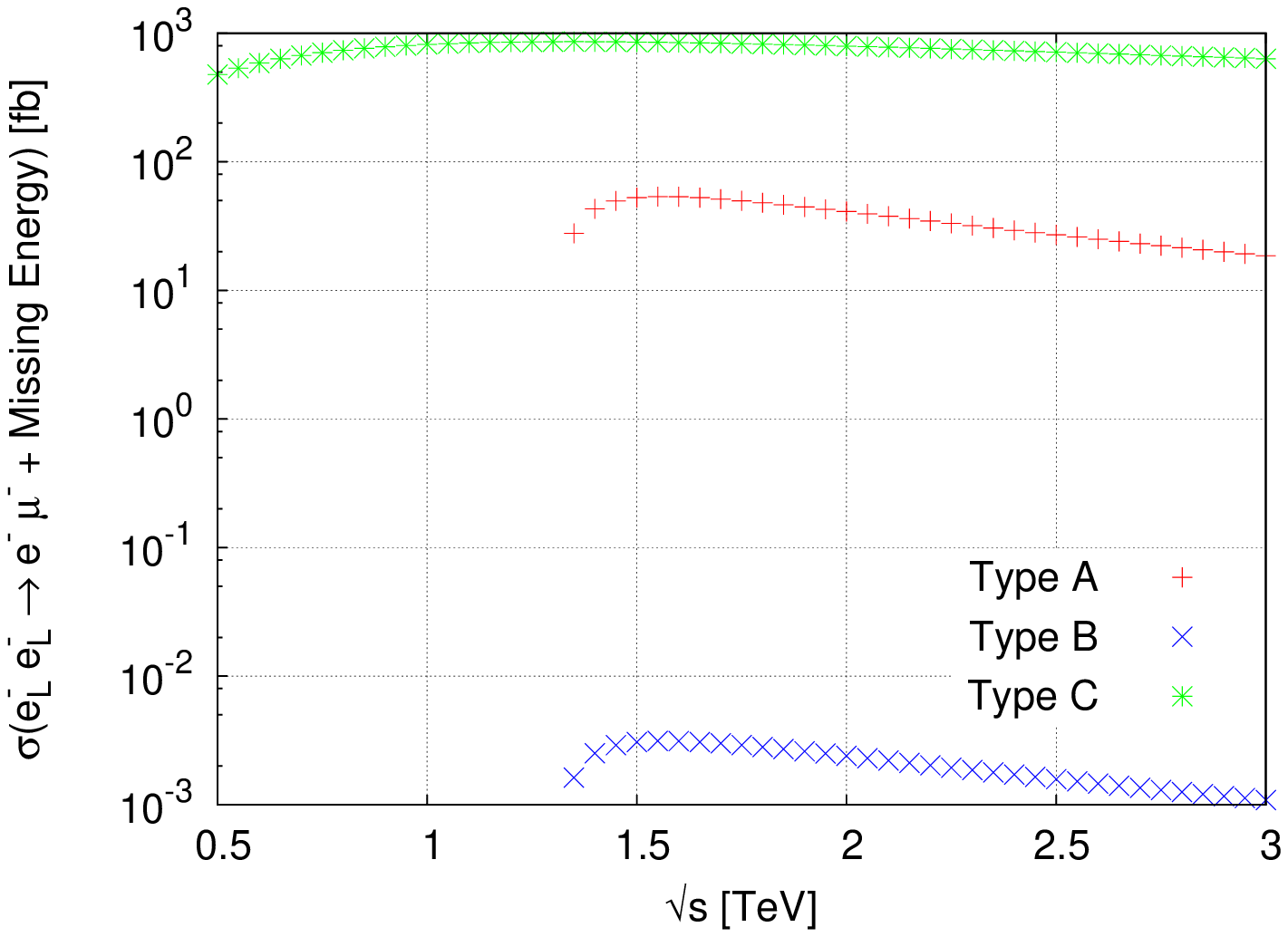, 
clip=, angle=0, width=75mm}
\end{tabular}
\caption{Cross section for $e^- e^- \to e^-
\mu^- + E^T_\text{miss}$ (with 
{$E^T_\text{miss}=2 \chi_1^0, 2 \chi_1^0
+  (2,4)\nu, (2,4) \nu$}), for points C-light and C-heavy
(upper and lower panels, respectively), 
as a function of the  centre of mass energy,
$\sqrt{s}$, for polarised beams. On the left:
$(P_{e^-},P_{e^-})=(-80\%,-80\%)$; on the right:
$(P_{e^-},P_{e^-})=(-100\%,-100\%)$. 
We have taken a degenerate right-handed neutrino spectrum  ($M_R = 10^{12}$ GeV) 
with $\theta_{13}=10^\circ$.}
\label{fig:Clh:Msqrts:pol80-100:emem}
\end{center}
\end{figure}

Although the cross sections for $e^- e^- \to W^ - W^-$ production
are extremely tiny, the SM does still contribute with a dominant
background due to $e^- e^- \to e^- \, \nu\, W^-$ processes (notice that 
the $\sigma_\text{prod}(e^- e^- \to (\mu^-,\tau^-)\,W^- \,\nu) \simeq 0$).
As discussed
in Section~\ref{sec:lfv:lc:emem}, processes contributing to the
non-$\tau$ SUSY background are only present if the spectrum is such
that $m_{\tilde \ell} \gtrsim m_{\chi_1^\pm}$, which is not the case for
points C.
Thus, and as argued in the case of $e^+ e^-$
beams, for SUSY spectra along the co-annihilation region, and 
should appropriate cuts allow to reduce the SM $W$-background, the
prospects of disentangling the cLFV effects of a potential type I SUSY  
seesaw at a Linear Collider are very promising for the $e^-e^-$
beam option. 
As previously discussed in Section~\ref{sec:res:epem}, due to the nature
of the dominant SM background,  
polarising the beams only translates into a small increase of the total (signal) cross section.
 This is  illustrated in Fig.~\ref{fig:Clh:Msqrts:pol80-100:emem}.

Still, notice that the effect of the LL polarisation is clearly visible in
the disappearance of the SUSY (signal and background) 
processes occurring for the lower values of
$\sqrt s$: as can be inferred from the associated spectrum shown in
Table~\ref{table:spectrum}, these would correspond to the production of (at
least) one right-handed slepton.  
For a c.o.m. energy of 2 TeV, the
expected number of events for C-light is around $3\times 10^4$ (2$\times 10^5$) for
$\mathcal{L}=0.5$ (3) ab$^{-1}$; in the case of C-heavy the
expected number of events varies between 300 and 2$\times 10^4$, for 
$\mathcal{L}=0.5$ ab$^{-1}$,  and between 2000 and $10^5$ for 
$\mathcal{L}=3$ ab$^{-1}$.

\begin{figure}[ht!]
\begin{center}
\begin{tabular}{cc}
\epsfig{file=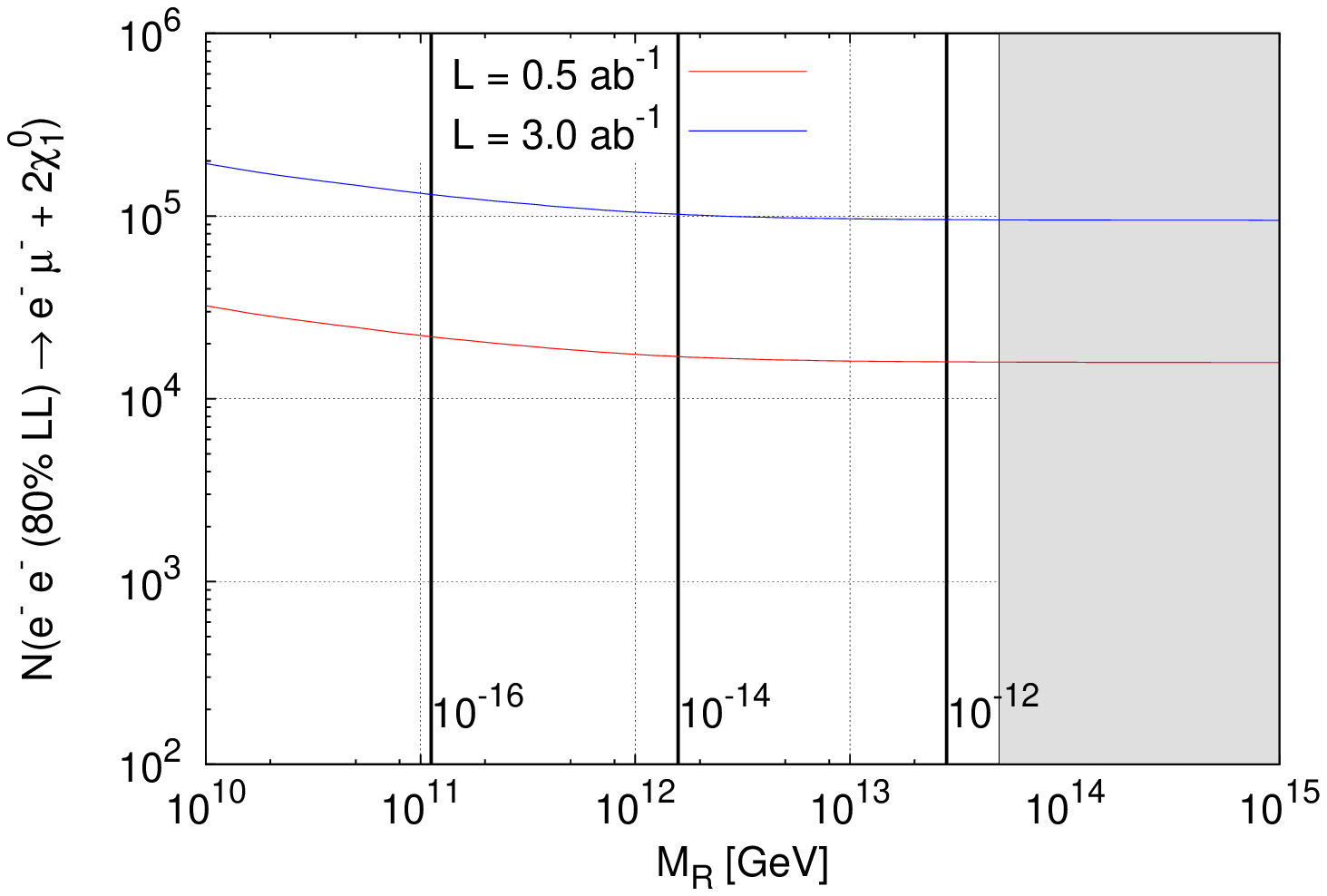, 
clip=, angle=0, width=75mm}
&
\epsfig{file=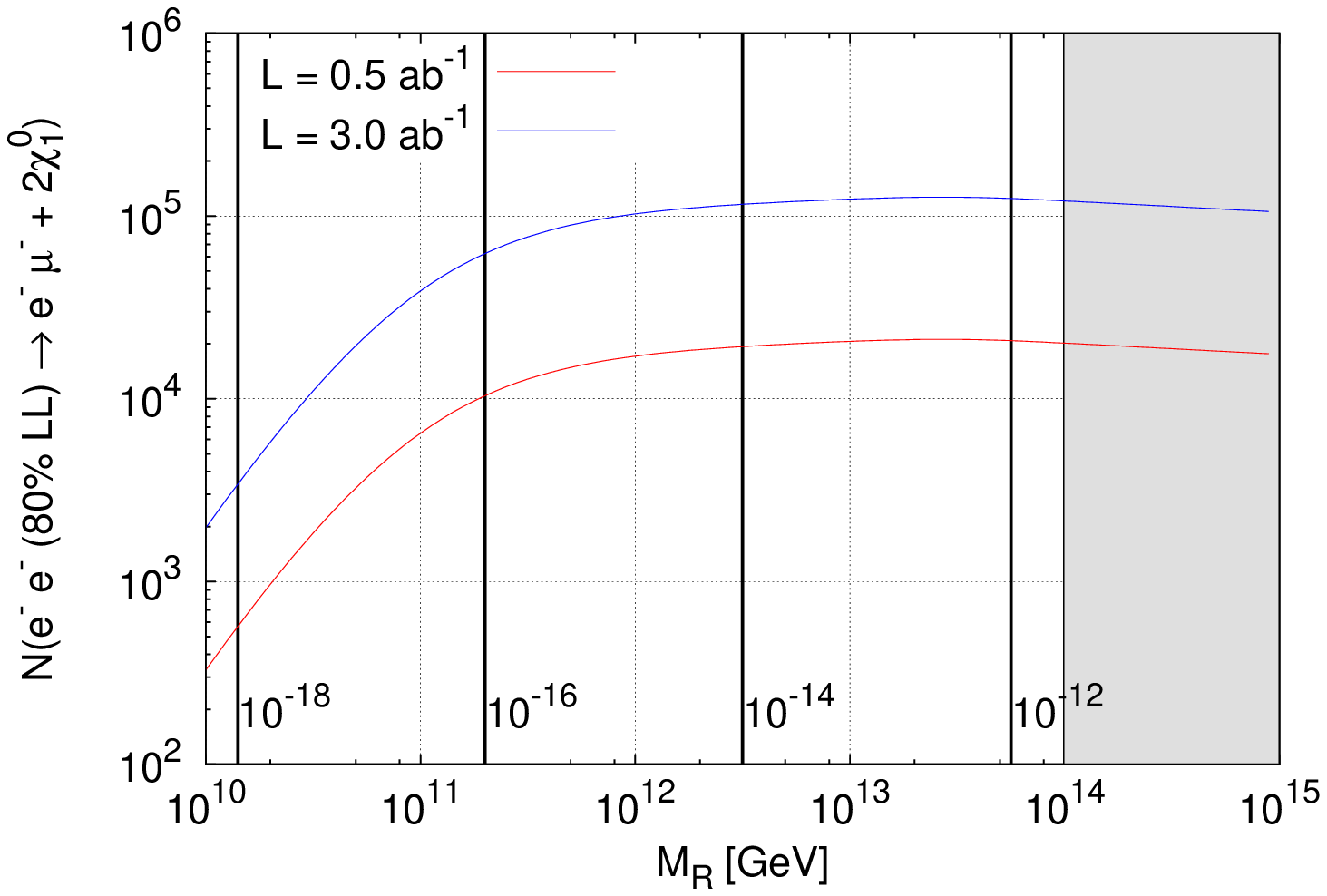, 
clip=, angle=0, width=75mm}
\end{tabular}
\caption{Number of events for $e^- e^- \to e^-
\mu^- + E^T_\text{miss}$ (with $E^T_\text{miss}=2 \chi_1^0$), 
for points C-light (left) and C-heavy (right), 
as a function of the seesaw scale $M_R$, for 
$(P_{e^-},P_{e^-})=(-80\%,-80\%)$
polarised beams. In both cases, we fix $\sqrt s= 2$ TeV, 
and we have taken a degenerate right-handed neutrino spectrum,
with $\theta_{13}=10^\circ$. 
Vertical lines denote the $M_R$-corresponding value of BR($\mu \to e \gamma$)
while the (grey) shaded region represents values of $M_R$ 
already excluded by the present experimental bound on BR($\mu \to e \gamma$).}
\label{fig:Clh:Msqrts:pol80:emem:discussion}
\end{center}
\end{figure}

\bigskip
In order to fully explore the potential of the results so far
obtained, let us  assume that SUSY has been discovered, with a
spectrum resembling one of the points in Table~\ref{table:spectrum}, 
and that a type I seesaw is
indeed the unique source of LFV at work. We focus on points C-light and
C-heavy, as these are associated with a potentially large number of
events, as can be seen in
Fig.~\ref{fig:Clh:Msqrts:pol80:emem:discussion}. 
 
To illustrate our discussion, 
we consider the expected number of
events  $e^- e^- \to e^-
\mu^- + E^T_\text{miss}$ (with $E^T_\text{miss}=2 \chi_1^0$), 
in the case of a 
realistic scenario for the beam polarisation, i.e.
$(P_{e^-},P_{e^-})=(-80\%,-80\%)$.  
 We further assume that dedicated
cuts will have allowed to significantly reduce the ($W$-strahlung) SM
background. Notice that it is quite likely that the cuts needed to reduce the background will also somewhat reduce the signal; this would naturally imply a rescaling of the corresponding curves of Fig.~\ref{fig:Clh:Msqrts:pol80:emem:discussion}. 

Firstly, if at the LC, no cLFV event of the type $e^- e^- \to e^-
\mu^- + E^T_\text{miss}$ is observed (for any possible luminosity and
c.o.m. energy), then a high-scale type I SUSY seesaw should be clearly
disfavoured as an hypothesis for the (unique) underlying source of LFV. 
Secondly, let us assume that a sizable number of events is indeed
collected: if such a number can be accommodated by the predictions (as
illustrated by the curves of
Fig.~\ref{fig:Clh:Msqrts:pol80:emem:discussion}), then one can derive
information about 
the seesaw scale. For instance, if for a C-heavy-like spectrum, more
than $10^5$ events are observed (for $\mathcal{L} \sim 3$ ab$^{-1}$),
then the seesaw scale ($M_R$, or in the hierarchical case, $M_{N_3}$)
should be above $10^{12}$ GeV. Furthermore, compatibility with BR($\mu
\to e \gamma$) bounds also puts an upper bound on this scale, so that
in this case one would be led to $10^{12} \text{ GeV }\lesssim M_R \lesssim
10^{14}$ GeV. 
The observation of a $\mu \to e \gamma$ decay at MEG can be
instrumental: let us assume that a BR($\mu \to e \gamma$)$\sim \mathcal{O}
(10^{-12})$ is indeed found. In this case, any number of $e^- e^- \to e^-
\mu^- + E^T_\text{miss}$ events below $10^{5}$ renders the two cLFV
manifestations incompatible with the assumption of a unique source of
LFV: either some (unaccounted for) destructive interference occurred in
the high-energy processes, which lowered the number of events, or then
there are additional sources of LFV, only manifest in the
low-energy observables, which account for the enhancement of the BR($\mu
\to e \gamma$). 
However, if the number of events corroborates the theoretical
expectations for a seesaw scale that would indeed account for such a BR($\mu
\to e \gamma$), then this interplay between high- and low-energy
observables strengthens the hypothesis of a type I
seesaw as the unique source of LFV. 

Should there be additional data on cLFV from the LHC (for example, in
association with neutralino decays into sleptons), then one can fully
explore the synergy between a large array of high- and
low-energy cLFV observables.

\begin{figure}[ht!]
\begin{center}
\begin{tabular}{cc}
\epsfig{file=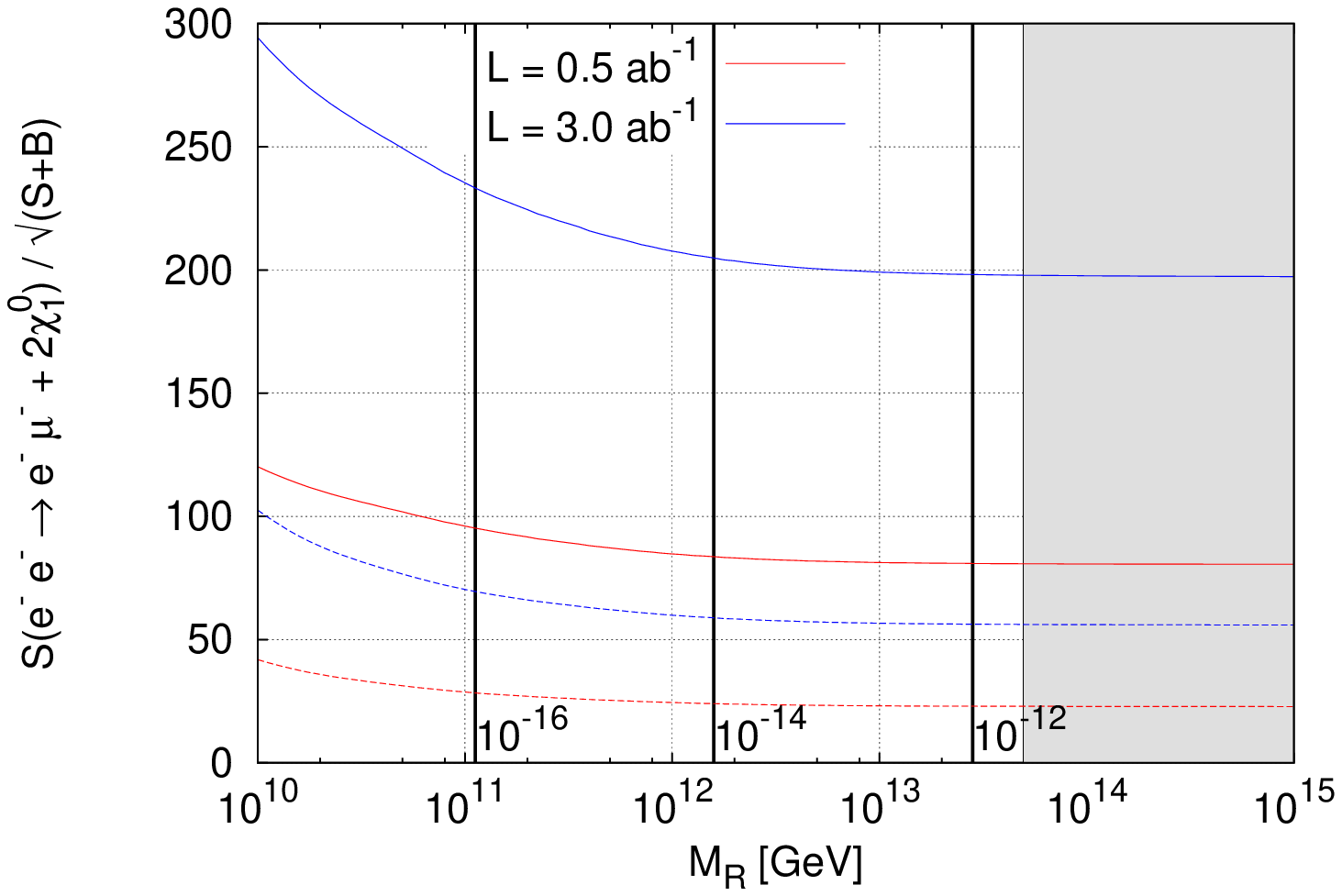,clip=, angle=0, width=75mm}
&
\epsfig{file=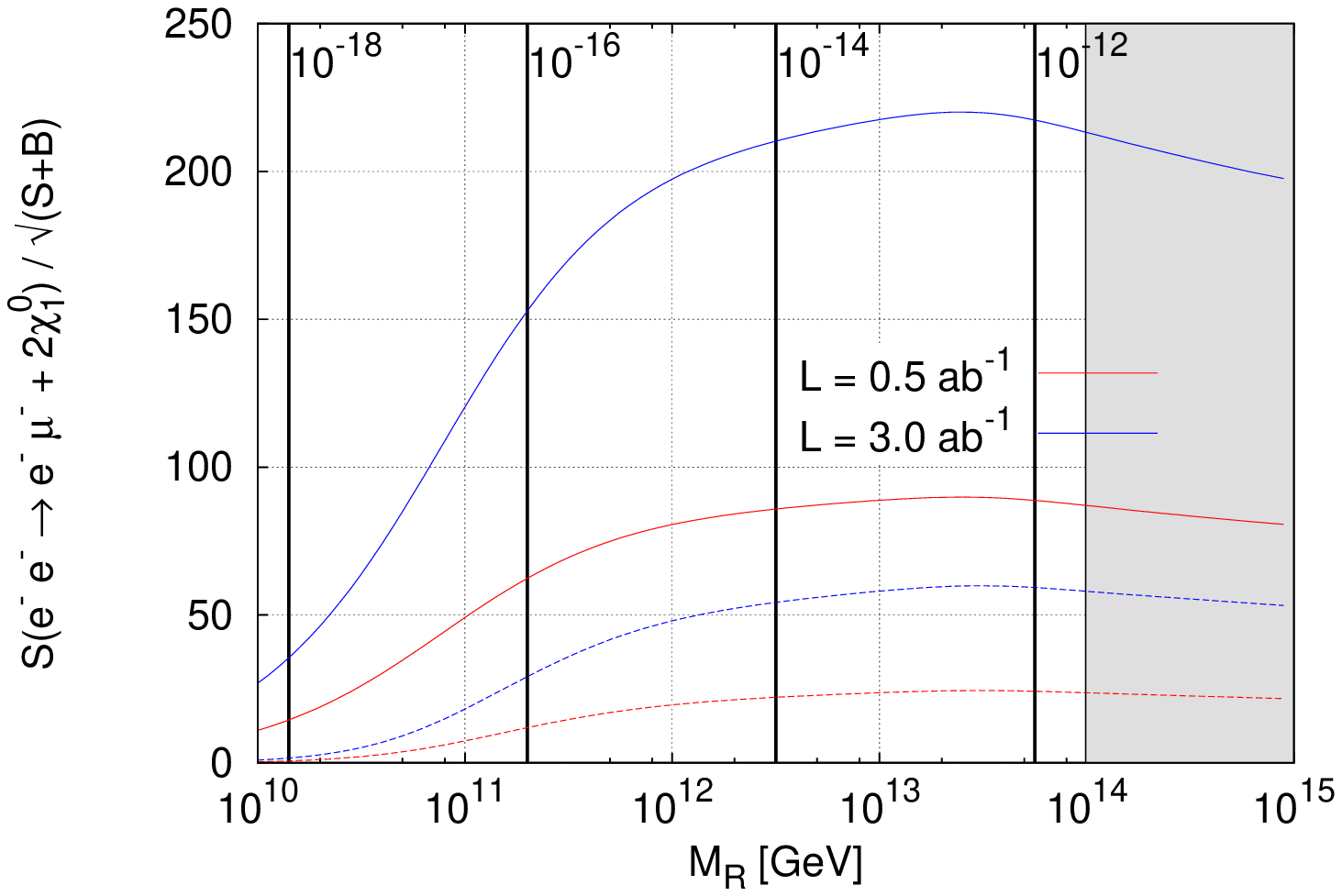,clip=, angle=0, width=75mm}
\\
\epsfig{file=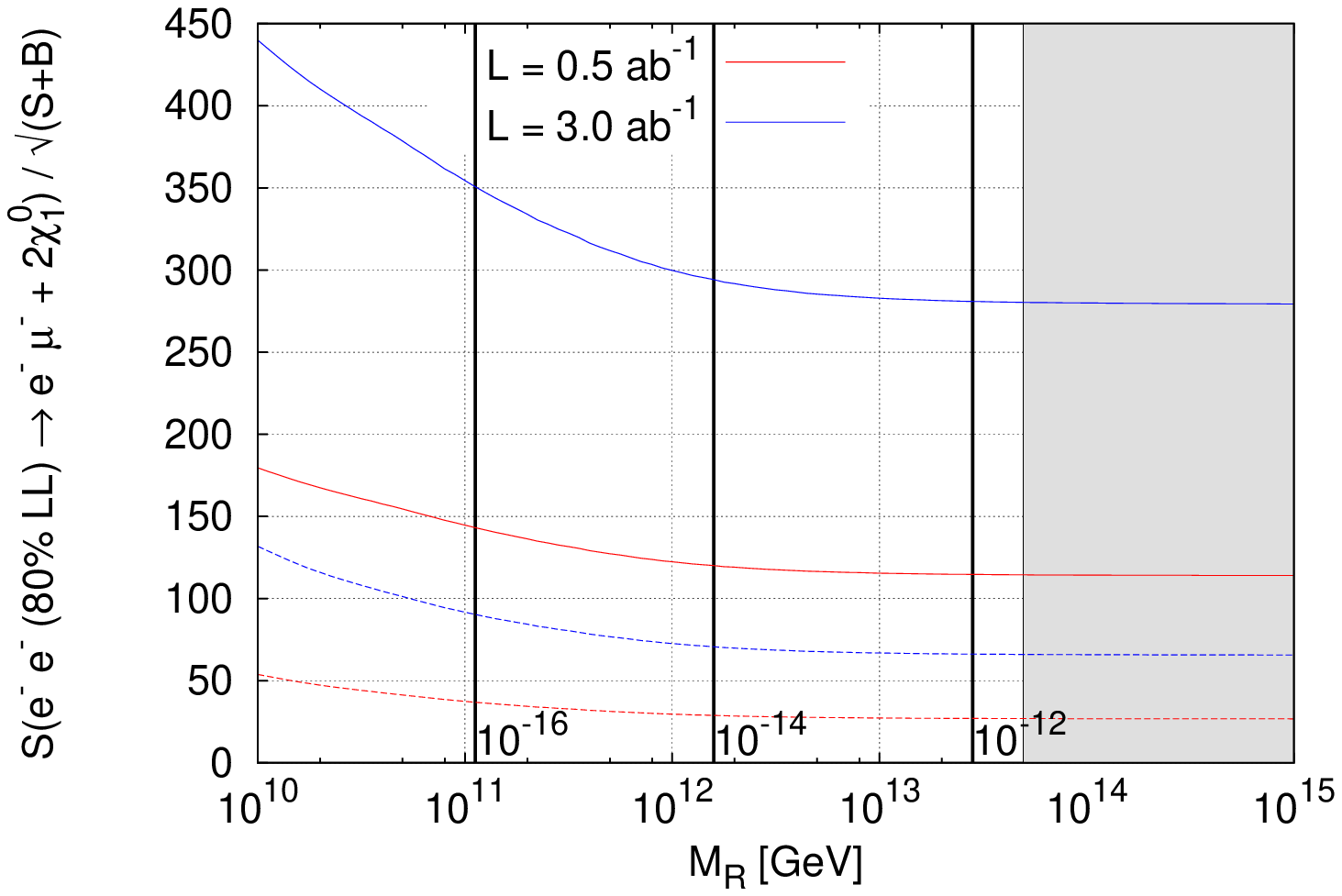,clip=, angle=0, width=75mm}
&
\epsfig{file=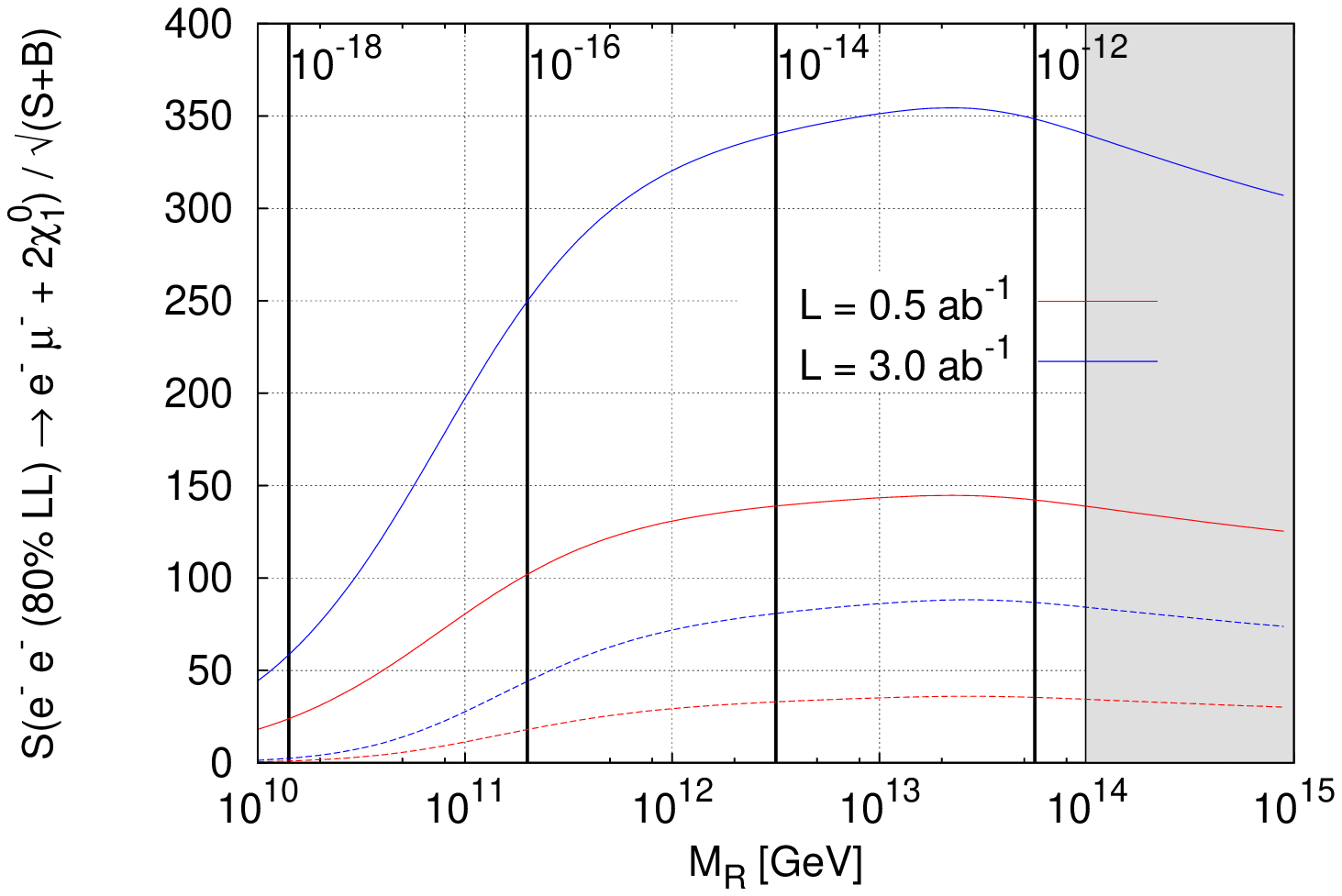, clip=, angle=0, width=75mm}
\\
\epsfig{file=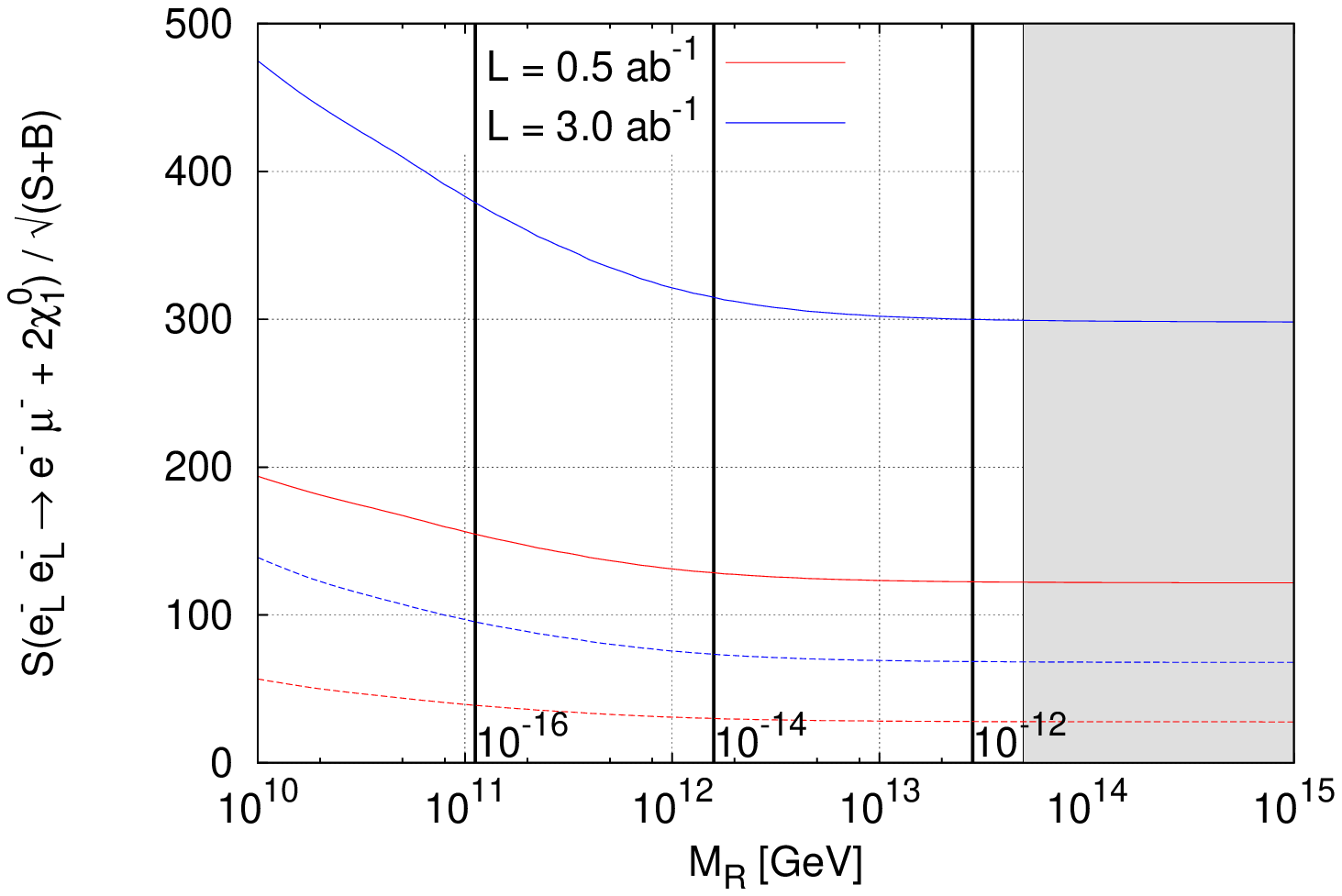,clip=, angle=0, width=75mm}
&
\epsfig{file=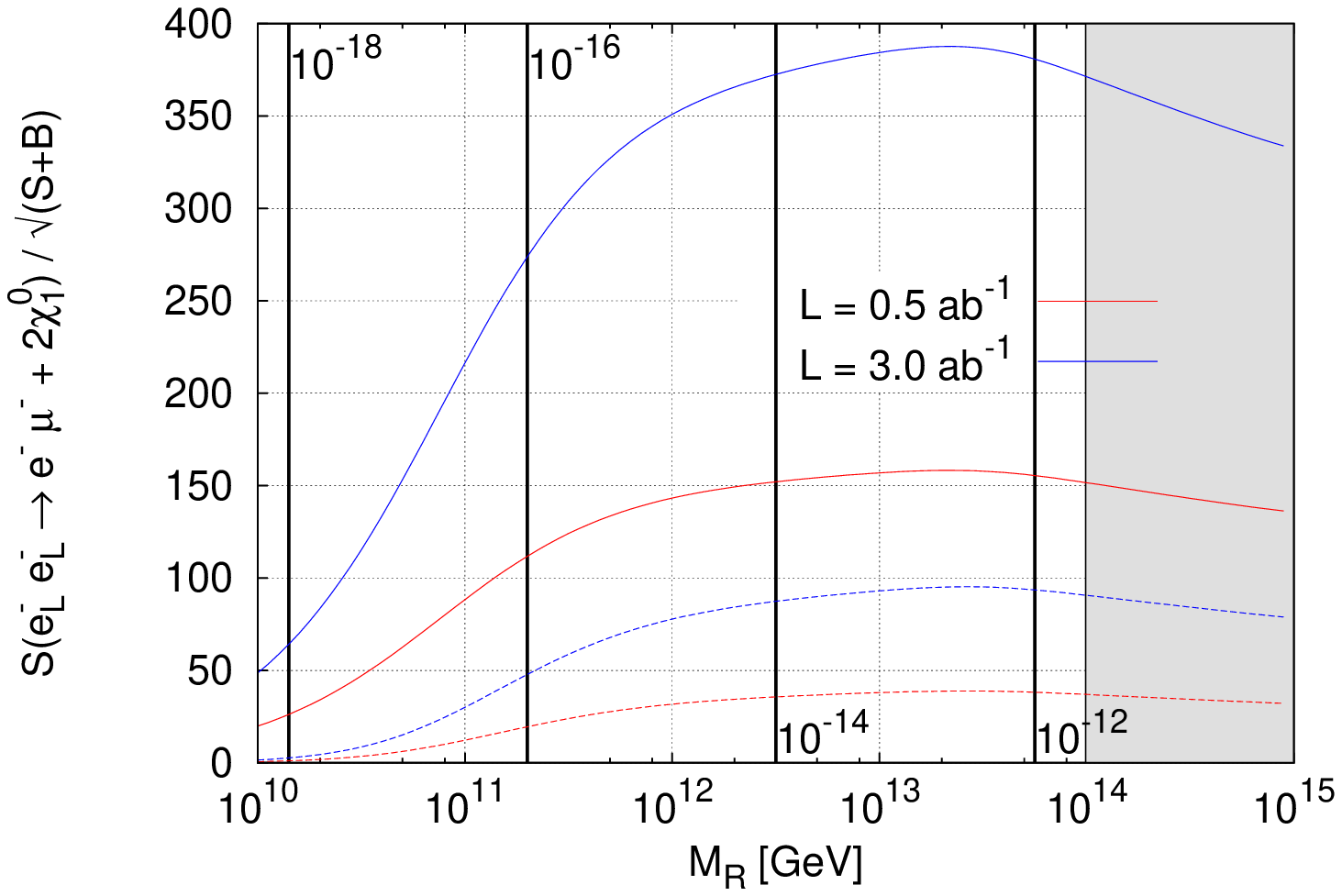, clip=, angle=0, width=75mm} 
\end{tabular}
\caption{Significance of the signal $e^- e^- \to e^- \mu^- +
  E^T_\text{miss}$ for points C-light (left) and
  C-heavy (right), as a function of the seesaw scale ($M_R$), 
  for $\sqrt s$=2 TeV, and for unpolarised (upper), 80\% (mid),  and
  fully polarised (lower panels)  $e^-$ beams. Line and colour
  code as in Fig.~\ref{fig:Clh:Sig}.}
\label{fig:Clh:Msqrts:allpol:emem:sign}
\end{center}
\end{figure}

\medskip
Finally, we present in
Fig.~\ref{fig:Clh:Msqrts:allpol:emem:sign}, the expected significance
of the signal, for unpolarised, 80\% polarised and fully polarised
electron beams. 
As in the case of $e^+ e^- \to e^+ \mu^- +
E^T_\text{miss}$ collisions, an 80\% polarised electron beam
configuration would already allow to have a significance of the
signal around 10, for both points C-light and C-heavy, even for 
an integrated luminosity of 0.5 ab$^{-1}$ (in the case of C-heavy,
only for $M_R \gtrsim 10^{11}$ GeV). We again stress that dedicated
cuts could even further improve these values.

\subsection{cLFV from $\mu^-\mu^-$ final states}\label{golden}
To complete our discussion, we finally comment on what might possibly
be a ``golden channel'' for the detection of cLFV at a Linear
Collider.  

\begin{figure}[ht!]
\begin{center}
\begin{tabular}{cc}
\epsfig{file=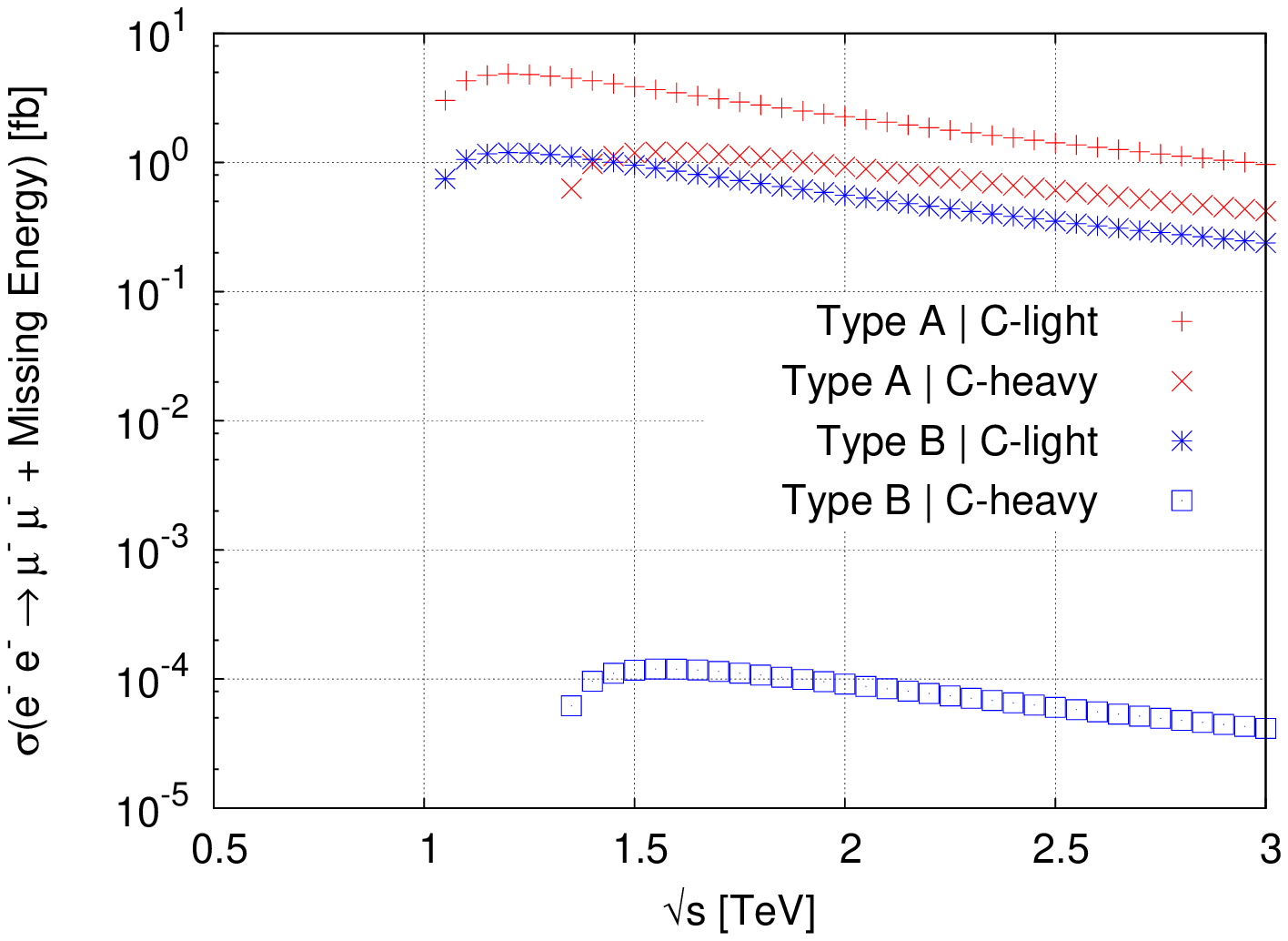, 
clip=, angle=0, width=75mm}
&
\epsfig{file=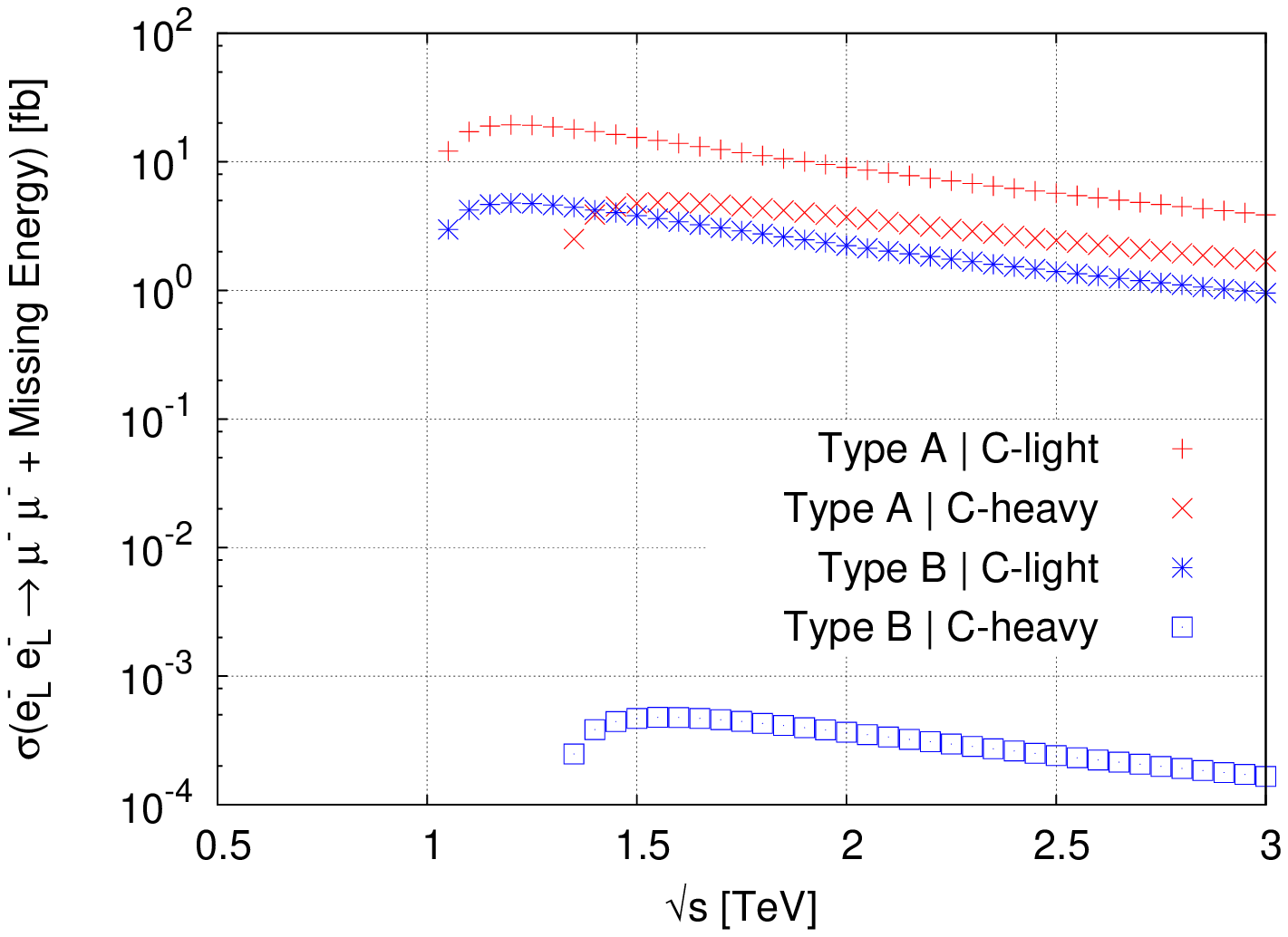, 
clip=, angle=0, width=75mm} 
\end{tabular}
\caption{Cross section for $e^- e^- \to \mu^-
\mu^- + E^T_\text{miss}$ ({with $E^T_\text{miss}=2 \chi_1^0, 2 \chi_1^0
+  (2,4)\nu, (2,4) \nu$}), for points C-light and 
C-heavy as a function of the  centre of mass energy, 
$\sqrt{s}$, in the unpolarised beam case (left) and for fully 
polarised beams $(P_{e^-},P_{e^-})=(-100\%,-100\%)$ (right). 
C-light: the signal (SUSY background) is denoted by red crosses 
(blue asterisks); C-heavy: the signal (SUSY background) is represented by red times (blue squares). 
We have  taken a degenerate right-handed neutrino spectrum ($M_R =
10^{12}$ GeV) and set $\theta_{13}=10^\circ$.} 
\label{fig:Cl:Msqrts:mumuminus}
\end{center}
\end{figure}

$e^- e^- \to \mu^-\mu^- + E^T_\text{miss}$ events turn out
to be extremely clean probes of cLFV, from an experimental and
a theoretical point of view:
\textcolor{black}{firstly, 
the efficiency of the muon detectors can be fully explored
when looking for $e^- e^- \to \mu^-\mu^- + E^T_\text{miss}$ signals;}
secondly, and especially when compared to the other (already
discussed) signals - i.e. $e^{+(-)} e^- \to e^{+(-)} \mu^- +
E^T_\text{miss}$ - the SM model background is extremely
tiny in this case. SUSY background processes are still present, but are
subdominant when compared to the signal, as the corresponding cross
sections differ by around one (four) order of magnitude for C-light
(C-heavy), as is illustrated on Fig.~\ref{fig:Cl:Msqrts:mumuminus}. 

For the case of fully polarised electron beams, and a 
c.o.m. energy of 2 TeV, the
expected number of events for C-light is $\sim$ 1000 (6000) for
$\mathcal{L}=0.5$ (3) ab$^{-1}$; in the case of C-heavy the
expected number of events is $\sim$ 500 (3000).
Should this be the case, one would clearly identify the presence of a new
physics scenario, such as the SUSY seesaw - always under the 
assumption of having a unique source of Lepton Flavour Violation
present. Naturally, one would be also confirming the Majorana nature
of the exchanged particles in the $t$-channel.

\section{Concluding remarks}\label{sec:concs}
Following recent developments in lepton physics and the new physics
searches at the LHC, we have revisited the potential of
a Linear Collider concerning the  
study of lepton flavour violation: in particular, we investigated cLFV 
in $e^+e^-$ and $e^-e^-$ collisions in the framework of a type I SUSY seesaw. 
Due to the leptonic mixing associated with charged current interactions 
($U_\text{MNS}$), cLFV will always occur in any model that can 
accommodate neutrino oscillations, 
independently from any mechanism of mass generation or 
new physics model.

If it is to explain neutrino masses and mixings with natural values
for the Yukawa couplings, 
the type I seesaw is impossible to probe directly (at high scale);
however, and when embedded into supersymmetric theories, it might
radiatively induce several cLFV phenomena, which can be manifest 
in both low-energy observables and in a number of
processes at high-energy colliders.  

In this work, and following early analyses~\cite{Deppisch:2003wt,Deppisch:2004pc},
we have considered $e^{\pm}\,e^-\, \to
\,\ell_i^{\pm}\,\ell_j^- + E_T^\text{miss}$ and discussed how the
signatures of a type I SUSY seesaw can be
disentangled from the SM and/or the MSSM, when the latter are 
effectively enlarged to accommodate low-energy neutrino data
(i.e. $U_\text{MNS} \neq 1$). To fully explore the LC beam
potentialities and maximise the existent synergie with very sensitive 
low-energy observables such as $\mu\to e\gamma$ decay and $\mu - e$
conversion in nuclei, we selected as ``signal'' the processes  
$e^{\pm}\,e^-\, \to \,e^{\pm}\,\mu^- + 2\,\chi_1^0 $.
We considered the
possibility of electron and positron beam polarisation, investigating
the prospects for  c.o.m. energies in the interval $[500$ GeV, $3$
TeV$]$ and for nominal  integrated
luminosities of $\mathcal{L}=0.5$ and 3 ab$^{-1}$.

Although the SM background typically dominates, 
we have shown that for an $e^+ e^-$ Linear Collider, 
running for instance at $\sqrt s=$ 2 TeV, and  prior to any selection cuts,  
80\% polarised beams would provide as much as 
$\sim10^3$ $(10^4)$ $e^+\,e^-\, \to \,e^+\,\mu^- + 2\,\chi_1^0 $ events 
for the above values of  $\mathcal{L}$ 
(in the case of a relatively light SUSY spectrum, and in the low $\tan
\beta$ regime). 

Furthermore the  $e^- e^-$ beam option provides one of the cleanest 
experimental setups to probe not only lepton number violation 
but also lepton flavour violation. 
Even without beam polarisation, $e^- e^-$ collisions offer excellent 
prospects for disentangling the specific cLFV effects 
of a type-I SUSY seesaw, 
in particular for SUSY spectra along the so-called 
co-annihilation region. 
For instance, for $\sqrt s=$ 2 TeV, one could expect as much as $\sim10^4$ $(10^5)$ 
$e^-e^-\to \,e^-\mu^- + 2\chi_1^0 $ events for 
an integrated luminosity of $\mathcal{L}=0.5$ (3) ab$^{-1}$.

In view of these promising prospects, the confrontation of the
theoretical estimations for a given reconstructed SUSY spectrum with
future data from dedicated lepton flavour violating studies at a LC
would allow to derive information about the 
scale of the underlying neutrino mass generation mechanism. 
Moreover, in the framework of a type I SUSY seesaw, 
there would be a strict correlation between the processes $e^-e^-
\to e^-\mu^- + 2\chi_1^0 $ and low-energy observables such as BR
($\mu\to e\gamma$).  
As discussed in this paper, the interplay of FV at a 
LC (especially for the $e^-e^-$ beam option)  
and a signal at MEG would strongly 
substantiate the hypothesis of a type-I SUSY seesaw. 

The present study only addressed cLFV at a Linear Collider from a
phenomenological point of view. The results obtained here (signal and
background cross
sections, expected number of events and associated statistical significance)
are just a first step of a larger study~\cite{inprogress}:  
the analysis can and should be improved by a full fledged simulation, generating
the events, applying kinematic cuts (see examples above), as well as acceptance
cuts of a generic detector.

\section*{Acknowledgements}
We are indebted to O. Panella for many
valuable exchanges and suggestions, and for a critical reading of
the manuscript. A. M. T. is grateful to M. Battaglia for interesting
discussions, and input on LC benchmarks.
This work has been done partly under the ANR project CPV-LFV-LHC {NT09-508531}. 
The work of  A. J. R. F.  has been supported by {\it Funda\c c\~ao
  para a Ci\^encia e a 
Tecnologia} through the fellowship SFRH/BD/64666/2009. 
A. A., A. J. R. F. and A. M. T. acknowledge partial support from the
European Union FP7  ITN INVISIBLES (Marie Curie Actions, PITN-
GA-2011- 289442). 
A. J. R. F. and J. C. R. also acknowledge the financial support from
the EU Network grant UNILHC PITN-GA-2009-237920 and from {\it
Funda\c{c}\~ao para a Ci\^encia e a Tecnologia} grants CFTP-FCT UNIT
777, CERN/FP/116328/2010 and PTDC/FIS/102120/2008. 

\appendix 
\section{Appendix: Particle production and decay}
\label{sec:appendix}

In our numerical analysis we have considered all relevant
contributions to 
\begin{align}
e^\pm e^- \,\to \,X \, \to \, \{Y\} \, \to \,e^\pm \mu^- + (0,2,4)\,\chi_1^0
+ (1,2,3) \, \bar \nu \,\nu\,,
\end{align}
with final state particle number no greater than 8. In the above, $X$ denotes any element of the set displayed in
Eq.~(\ref{eq:app:X1}), and $\{Y\}$ the possible two-body decays, as
listed below, in Eqs.~(\ref{eq:app:Y1}-\ref{eq:app:Y7}). 
The primary production modes (on-shell) are as follows\footnote{Notice that we have excluded 
$e^+ e^- \to H Z$, $e^+ e^- \to h A$ and $e^- e^- \to W^- W^-$ since, 
throughout the parameter space explored, 
we have found the associated cross sections to be highly 
suppressed: $\sigma_{H Z} < 1.1 \times 10^{-4}$ fb, 
$\sigma_{h A} < 2 \times 10^{-5}$ fb and $\sigma_{W^- W^-} < 1.1 \times 10^{-21}$ fb.}:
\begin{equation}\label{eq:app:X1}
e^+ \,e^- \,\to \left\{ \begin{array}{l}
			\chi^0_A \,\chi^0_B \\
			\chi^\pm_A\, \chi^\mp_B \\
			\tilde \ell^+_i\, \tilde \ell^-_j \\
			\tilde \nu_i \,\tilde \nu^*_j \\
			h \,Z,\, H\, A,\, H^+\, H^- \\
			W^+\, W^- \\
			\tau^+ \tau^-\\
			W^+ \,\ell^-_i \,\bar\nu + \text{c.c.}
		\end{array} \right. \quad \quad
e^- \,e^- \,\to \left\{ \begin{array}{l}
			\tilde \ell^-_i\, \tilde \ell^-_j \\
			W^-\, \ell^-_i\, \nu
		\end{array} \right. \,.
\end{equation}
In view of the final states under study, the above produced states
were allowed to decay, leading to a two-body decay cascade; among the
final states will be those of the type (A), (B) and (C) processes (see  Eqs.~(\ref{eq:epem:ABC,eq:emem:ABC})), corresponding to the cLFV
signal and to the SUSY and SM backgrounds. 
The following two-body decays (plus the 3-body $\tau$ decays) were considered, obeying kinematical and
simple selection rules (denoted by superscripts in the decay
products). The former are a natural consequence of the hierarchy of
the different considered spectra\footnote{For cases of nearly-degenerate LSP and NLSP,
  typically associated to the dark matter co-annihilation region, the
  NLSP (the lightest stau) becomes extremely long-lived, potentially
  decaying outside the detector. This is the case of point C-heavy, 
  for which the $\tilde \tau_1$'s lifetime is $\mathcal{O}(100 \mu$s).}
\begin{align}\label{eq:app:Y1}
& 
\tilde e^-_L \, \to \left\{ \begin{array}{l}
			( \chi^0_1, \, \chi^0_2 )\,  \ell^-_i \\
			h \, \tilde \tau^-_1 \\
			Z \, (\tilde \tau^-_2,\,  \tilde \tau^-_1) \\
			( \chi^-_1, \, \chi^-_2 )\,  \nu ~{}^\text{\scriptsize (1)} \\
			W^- \, \tilde \nu_\tau
	\end{array} \right. 
\quad
\tilde \mu^-_L \, \to \left\{ \begin{array}{l}
			( \chi^0_1, \, \chi^0_2 )\,  \ell^-_i \\
			h \, \tilde \tau^-_1 \\
			Z \, (\tilde \tau^-_2, \, \tilde \tau^-_1) \\
			( \chi^-_1,\,  \chi^-_2 ) \, \nu ~{}^\text{\scriptsize (1)} \\
			W^- \, \tilde \nu_\tau
	\end{array} \right. 
\quad
\tilde \tau^-_2 \, \to \left\{ \begin{array}{l}
			( \chi^0_1, \, \chi^0_2 ) \, \ell^-_i \\
			h \, \tilde \tau^-_1 \\
			Z \, (\tilde e^-_L, \, \tilde \mu^-_L,\,  \tilde \tau^-_1) \\
			( \chi^-_1,\,  \chi^-_2 )\,  \nu ~{}^\text{\scriptsize (1)} \\
	\end{array} \right.
\end{align}

\begin{align}\label{eq:app:Y2}
& 
\tilde e^-_R \, \to \left\{ \begin{array}{l}
			( \chi^0_1,\,  \chi^0_2 ) \, \ell^-_i \\
			( \chi^-_1,\,  \chi^-_2 ) \, \nu ~{}^\text{\scriptsize (1)}
	\end{array} \right. 
\quad
\tilde \mu^-_R \, \to \left\{ \begin{array}{l}
			( \chi^0_1,\,  \chi^0_2 ) \, \ell^-_i \\
			( \chi^-_1, \, \chi^-_2 )\,  \nu ~{}^\text{\scriptsize (1)}
	\end{array} \right. 
\quad
\tilde \tau^-_1 \, \to \left\{ \begin{array}{l}
			( \chi^0_1, \, \chi^0_2 )\,  \ell^-_i \\
			( \chi^-_1, \, \chi^-_2 )\,  \nu ~{}^\text{\scriptsize (1)}
	\end{array} \right.
\end{align}

\begin{align}\label{eq:app:Y3}
&
\tilde \nu_e \, \to \left\{ \begin{array}{l}
		W^+ \, ( \tilde \tau^-_2, \, \tilde e^-_R, \, \tilde \mu^-_R, \, \tilde \tau^-_1 ) \\
		( \chi^0_1, \, \chi^0_2 ) \, \nu \\
		( \chi^+_1, \, \chi^+_2 ) \, \ell^-_i~{}^\text{\scriptsize (1)}
	\end{array} \right. 
\quad
\tilde \nu_\mu \, \to \left\{ \begin{array}{l}
		W^+ \, ( \tilde \tau^-_2, \, \tilde e^-_R, \, \tilde \mu^-_R, \, \tilde \tau^-_1 ) \\
		( \chi^0_1,\,  \chi^0_2 )\,  \nu \\
		( \chi^+_1,\,  \chi^+_2 )\,  \ell^-_i~{}^\text{\scriptsize (1)}
	\end{array} \right. 
\quad
\tilde \nu_\tau \, \to \left\{ \begin{array}{l}
		W^+ \, ( \tilde e^-_R, \, \tilde \mu^-_R,\,  \tilde \tau^-_1 ) \\
		( \chi^0_1, \, \chi^0_2 )\,  \nu \\
		( \chi^+_1,\,  \chi^+_2 )\,  \ell^-_i~{}^\text{\scriptsize (1)}
	\end{array} \right.
\end{align}

\begin{align}\label{eq:app:Y4}
&
\chi^+_1 \, \to \left\{ \begin{array}{l}
		W^+ \, \chi^0_1 \\
		\tilde \ell^+_i\,  \nu~{}^\text{\scriptsize (2)} \\
		\tilde \nu_i\,  \ell^+_j ~{}^\text{\scriptsize (3)}
\end{array} \right. 
\quad
\chi^+_2 \, \to \left\{ \begin{array}{l}
		W^+ \, ( \chi^0_1, \, \chi^0_2 ) \\
		( Z, \, h )\,  \chi^+_1 \\
		\tilde \ell^+_i\,  \nu ~{}^\text{\scriptsize (2)} \\
		\tilde \nu_i \, \ell^+_j ~{}^\text{\scriptsize (3)}
\end{array} \right.
\end{align}

\begin{align}\label{eq:app:Y5}
&
\chi^0_2 \, \to \left\{ \begin{array}{l}
		\tilde \nu_i \, \bar\nu + \text{c.c.} ~{}^\text{\scriptsize (3)} \\
		( Z, \, h ) \, \chi^0_1 \\
		\tilde \ell^+_i \, \ell^-_j + \text{c.c.} ~{}^\text{\scriptsize (2*)}
\end{array} \right. 
\quad
\chi^0_3 \, \to \left\{ \begin{array}{l}
		( Z, \, h ) \, ( \chi^0_1, \, \chi^0_2 ) \\
		\chi^+_1\,  W^- + \text{c.c.} 
\end{array} \right. 
\quad
\chi^0_4 \, \to \left\{ \begin{array}{l}
		( Z, \, h )\,  ( \chi^0_1,\,  \chi^0_2 ) \\
		\chi^+_1 \, W^- + \text{c.c.} 
\end{array} \right. 
\end{align}

\begin{align}\label{eq:app:Y6}
&
\tau^\pm \, \to \,(e^\pm, \,\mu^\pm) \,\bar\nu \,\nu\, ; \quad 
W^- \,/\, W^+ \,\to\, \ell^-_i \,\bar\nu / \ell^+_i \,\nu \,;\quad 
Z \,\to \left\{ \begin{array}{l}
			\ell^+_i \,\ell^-_i \\
			\bar\nu \,\nu
	\end{array} \right.
\quad
h^0 \,\to \,\tau^+ \,\tau^- 
\end{align}

\begin{align}\label{eq:app:Y7}
&
H \,\to \left\{ \begin{array}{l}
			\tilde e^+_L\, \tilde e^-_L \\
			\tilde e^+_R\, \tilde e^-_R \\
			\tilde \mu^+_L\, \tilde \mu^-_L \\
			\tilde \mu^+_L \,( \tilde \tau^-_2,\, \tilde
                        \mu^-_R, \,\tilde \tau^-_1 ) + \text{c.c.}  \\
			\tilde \tau^+_2 \, \tilde \tau^-_2 \\
			\tilde \tau^+_2\, ( \tilde \mu^-_R, \,\tilde \tau^-_1 ) + \text{c.c.} \\
			\tilde \mu^+_R \,\tilde \mu^-_R \\
			\tilde \tau^+_1 \,\tilde \tau^-_1 \\
			\tilde \nu^*_i\, \tilde \nu_i \\
			\chi^0_1\,( \chi^0_1,\, \chi^0_2 ) \\
			\chi^0_2\, \chi^0_2 \\
			\chi^+_1\, \chi^-_1 \\
			\tau^+\, \tau^- \\
			Z\, Z \\
			W^+\, W^- \\
			h\, h
	\end{array} \right. 
\quad
A \to \left\{ \begin{array}{l}
			\tilde \mu^+_L ( \tilde \mu^-_R, \tilde \tau^-_1 ) + \text{c.c.} \\
			\tilde \tau^+_2 ( \tilde \mu^-_R, \tilde \tau^-_1 ) + \text{c.c.} \\
			\chi^0_1( \chi^0_1, \chi^0_2 ) \\
			\chi^0_2 \chi^0_2 \\
			\chi^+_1 \chi^-_1 \\
			\tau^+ \tau^- \\
			Z h
	\end{array} \right.
\quad
H^+ \to \left\{ \begin{array}{l}
			\tilde e^+_L \tilde \nu_e \\
			\tilde \mu^+_L ( \tilde \nu_\mu, \tilde \nu_\tau ) \\
			\tilde \tau^+_2 ( \tilde \nu_\mu, \tilde \nu_\tau ) \\
			\tilde \mu^+_R \tilde \nu_\mu \\
			\tilde \tau^+_1 ( \tilde \nu_\mu, \tilde \nu_\tau ) \\
			\chi^+_1 ( \chi^0_1, \chi^0_2 ) \\
			( \tau^+, \mu^+ ) \nu \\
			W^+ h
	\end{array} \right. 
\end{align}\\

\noindent The above decay chains are subject to the following rules: a given
decay marked with a superscript (1-3) is \\
\noindent 
(1) avoided, if a $\chi^0_2$ was produced prior in the
          decay chain branch (i.e., assuming $m_{\chi^0_2} \simeq
          m_{\chi^\pm_1})$;  
\\
\noindent 	
(2) considered, if no prior slepton is present in the decay chain
          branch or if any prior slepton is mostly left-handed (2*:
          if the prior slepton is left-handed, this further restricts the decay
          into only right-handed sleptons); 
\\
\noindent	
(3) considered, if no prior sneutrino appears in the corresponding
branch of the decay chain.\\

\noindent Thus, for each final state, all possible allowed decay chains  (i.e.
present in the aforementioned list of decays) were generated,
taking as a starting point the considered primary production modes. 
For the case of supersymmetric $e^+ e^-$ cross sections, as well as
SUSY branching fractions, we used the public code
SPheno~\cite{Porod:2003um}. We developed dedicated routines for the
case of SUSY $e^- e^-$  and SM cross sections, which were implemented onto SPheno.

\end{document}